\documentclass[a4paper,11pt,oneside]{article}
\usepackage[english]{babel}
\usepackage[dvips]{graphicx}
\usepackage{amsthm}
\usepackage{exscale}
\usepackage{amsfonts}
\usepackage{amssymb}
\usepackage[intlimits,sumlimits]{amsmath}
\usepackage{amsmath,cases}
\usepackage[latin1]{inputenc}
\usepackage{mathrsfs}
\usepackage{fancyhdr}

\theoremstyle{definition}

\newtheorem{thm}{Theorem}[section]
\newtheorem{prop}[thm]{Proposition}

\newcommand{\vartot}{\mathrm{d}_{\mathrm{TV}}}
\newcommand{\as}{a_{\ast}}
\newcommand{\WW}{\mathrm{W}}
\newcommand{\XX}{\mathrm{X}}
\newcommand{\YY}{\mathrm{Y}}
\newcommand{\ZZ}{\mathrm{Z}}
\newcommand{\RR}{\mathrm{R}}
\newcommand{\MM}{\mathrm{M}}
\newcommand{\SSS}{\mathrm{S}}
\newcommand{\IIr}{\mathrm{I}_1(\rho, \ub)}
\newcommand{\IIs}{\mathrm{I}_2(\rho, \ub)}
\newcommand{\Mrand}{\mathcal{M}}
\newcommand{\Nrand}{\mathcal{N}}
\newcommand{\xb}{\mathbf{x}}

\newcommand{\vb}{\mathbf{v}}
\newcommand{\wb}{\mathbf{w}}
\newcommand{\ub}{\mathbf{u}}

\newcommand{\qb}{\mathbf{q}}
\newcommand{\hb}{\mathbf{h}}
\newcommand{\xib}{\boldsymbol{\xi}}
\newcommand{\psib}{\boldsymbol{\psi}}

\newcommand{\omb}{\boldsymbol{\omega}}
\newcommand{\ud}{\mathrm{d}}
\newcommand{\dx}{\mathrm{d} \mathbf{x}}
\newcommand{\dxi}{\mathrm{d} \boldsymbol{\xi}}
\newcommand{\unifSo}{u_{S^2}(\ud \boldsymbol{\omega})}
\newcommand{\unifSu}{u_{S^2}(\ud \ub)}

\newcommand{\intethree}{\int_{\mathbb{R}^3}}
\newcommand{\lnorm}{\mid\mid \!}
\newcommand{\rnorm}{\! \mid\mid}
\newcommand{\Lnorm}{\Big{|}\Big{|} \!}

\newcommand{\rnormm}{\! \mid\mid_{\ast}}
\newcommand{\Rnormm}{\! \Big{|}\Big{|}_{\ast}}
\newcommand{\ind}{1 \! \textrm{l}}
\newcommand{\prob}{\mathcal{P}(\mathbb{R}^3)}

\newcommand{\borelthree}{\mathscr{B}(\mathbb{R}^3)}
\newcommand{\rone}{\mathbb{R}}
\newcommand{\rthree}{\mathbb{R}^3}
\newcommand{\rthreeb}{(\mathbb{R}^3, \mathscr{B}(\mathbb{R}^3))}

\newcommand{\BL}{\mathrm{BL}(\mathbb{R}^3)}

\newcommand{\pp}{\textsf{P}}
\newcommand{\pt}{\textsf{P}_t}
\newcommand{\et}{\textsf{E}_t}
\newcommand{\mut}{\mu(\cdot, t)}
\newcommand{\treen}{\mathfrak{t}_n}
\newcommand{\ptreen}{\mathbb{P}(\mathfrak{t}_n)}
\newcommand{\tree}{\mathfrak{t}}

\newcommand{\treesn}{\mathfrak{s}_{n-1}}
\newcommand{\treenk}{\mathfrak{t}_{n, k}}
\newcommand{\mq}{\mathfrak{m}_4}
\newcommand{\Mtre}{\mathfrak{L}_{3}}
\newcommand{\Blnorm}{\Big{|}\Big{|} \!}
\newcommand{\Brnormg}{\! \Big{|}\Big{|}_{S^2}}
\newcommand{\gradient}{\nabla_{S^2}}
\newcommand{\hess}{\mathrm{Hess}_{S^2}}
\newcommand{\diff}{\mathscr{D}}
\newcommand{\Log}{\textrm{Log}}

\author{\textsc{By Emanuele Dolera and Eugenio Regazzini} \footnote{Also
affiliated with CNR-IMATI, Milano, Italy.} \\
\emph{Universit\`a di Modena e Reggio Emilia and Universit\`a di Pavia}}

\title{\textbf{Proof of a McKean conjecture on the rate of convergence of Boltzmann-equation solutions} \footnote{Supported in part by MIUR-2008MK3AFZ}}

\date{}

\begin{document}

\maketitle
\begin{abstract}
{\footnotesize The present work provides a definitive answer to the problem of quantifying relaxation to equilibrium of the solution to the spatially homogeneous Boltzmann equation for Maxwellian molecules. The beginning of the story dates back to a pioneering work by Hilbert, who first formalized the concept of linearization of the collision operator and pointed out the importance of its eigenvalues with respect to a certain asymptotic behavior of the Boltzmann equation. Under really mild conditions on initial data -- close to being necessary -- and a weak, physically consistent, angular cutoff hypothesis, our main result (Theorem \ref{thm:CDGR}) contains the first precise statement that the total variation distance between the solution and the limiting Maxwellian distribution admits an upper bound of the form $C e^{\Lambda_b t}$, $\Lambda_b$ being the least negative of the aforesaid eigenvalues and $C$ a constant which depends only on a few simple numerical characteristics (e.g. moments) of the initial datum. The validity of this quantification was conjectured, about fifty years ago, in a paper by Henry P. McKean but, in spite of several attempts, the best answer known up to now consists in a bound with a rate which can be made arbitrarily close to $\Lambda_b$, to the cost of the ``explosion'' of the constant $C$. Moreover, its deduction is subject to restrictive hypotheses on the initial datum, besides the Grad angular cutoff condition. As to the proof of our results, we have taken as point of reference an analogy between the problem of convergence to equilibrium and the central limit theorem of probability theory, highlighted by
McKean. Our work represents in fact a confirmation of this analogy, since the techniques we develop here crucially rely on certain formulations of the central limit theorem. The proof of Theorem \ref{thm:CDGR} starts by assuming the Grad angular cutoff and proceeds with these steps: 1) A new representation, in Theorem \ref{thm:representation}, for the solution of the Boltzmann equation as expectation of a random probability distribution of a weighted random sum of independent and identically distributed random vectors. 2) An upper bound for the total variation distance of interest expressed as sum of expectations of the total variation distance, between the aforesaid random probability distribution and the limiting Maxwellian law, over two appropriate events $U$ and $U^c$. 3) The proof that the probability of $U$ approaches zero, as time goes to infinity, at an exponential rate equal to $\Lambda_b$. 4) An extension of a classical Beurling inequality which, combined with new Berry-Esseen-like inequalities, leads to the validity of the desired exponential rate $\Lambda_b$ of decay also for the expectation over $U^c$. Then, the conclusion can be extended to the case of weak cutoff hypothesis by a standard truncation argument. To complete this description of the paper, we mention the use of the aforesaid representation to characterize, in Theorem \ref{thm:CLT}, the domain of attraction of the Maxwellian limit.}
\end{abstract}


\tableofcontents

\section{Introduction and new results} \label{sect:intro}

On the basis of an analogy pointed out by McKean in \cite{mck6, mck7}, a few years ago we started a program which aims at studying the long-time behavior of solutions of some kinetic equations, by means of representations which connect these solutions to probability laws of certain weighted sums of independent and identically distributed (i.i.d.) random variables. The discovery of the right representation is comparatively simple for the solution of the spatially homogeneous one-dimensional Kac equation. This fact has produced both new results and improvements on the existing ones
concerning the Kac equation. See \cite{bl, blm, blr, cgrUNI, dgr, dore, gr8, gr10}.

Our goal in the present paper is to go back to the original kinetic model, the \emph{spatially homogeneous Boltzmann equation for Maxwellian molecules} (SHBEMM), which had inspired the aforesaid one-dimensional model. The reason for having deferred its treatment is connected, on the one hand, with the mathematical complexity of the subject and, on the other hand, with the hope that useful insights could be derived from the study of simpler allied cases. More specifically, we discuss here the problem of quantifying the ``best'' rate of relaxation to equilibrium. The starting point of the argument is the new probabilistic representation exhibited in Subsection \ref{sect:representation} of the present paper.

The last part of the program, to be developed in forthcoming papers, is concerned with the inhomogeneous
Boltzmann equation for Maxwellian molecules. Although the assumption of spatial homogeneity adopted here may seem a strong restriction, it is nonetheless proving an interesting and inspiring basis for studying qualitative properties of the complete model.

\subsection{The equation} \label{sect:equation}

In classical kinetic theory, a gas is thought of as a system of a very large number $N$ of like particles, described by means of a time-dependent statistical distribution $\mut$ on the phase space $X \times \rthree$, where $X$ stands for the spatial domain. Then, for any subset $A$ of $X \times \rthree$, $\mu(A, t)$ provides an approximation, independent of $N$, of the statistical frequency of particles in $A$, at time $t$. It is worth noting that $\mut$ can be also interpreted, consistently with its statistical meaning, as a genuine probability distribution (p.d.) by arguing about $\mu(A, t)$ as probability that the position-velocity of a randomly selected particle, at time $t$, belongs to $A$. See the discussion in Subsection 2.1 in Chapter 2A of \cite{vil}. The basic assumptions for the derivation of the classical equation which governs the evolution of $\mut$ are that the gas is dilute (\emph{Boltzmann-Grad limit}) and that the particles interact via binary, elastic and microscopically reversible collisions. Particles which are just about to collide are viewed as stochastically independent (\emph{Boltzmann's Stosszahlansatz}). See \cite{cerS, cip, trumu} for a comprehensive treatment. In this work, we also assume spatial homogeneity, so that the phase space reduces to $\rthree$ and the SHBEMM can be written as
\begin{eqnarray}
\frac{\partial}{\partial t} f(\vb, t) &=& \intethree\int_{S^2} [f(\vb_{\ast}, t)
f(\wb_{\ast}, t) \ - \ f(\vb, t) f(\wb, t)] \times \nonumber \\
&\times& b\left(\frac{\wb - \vb}{|\wb - \vb|} \cdot \omb \right) \unifSo \ud \wb \label{eq:boltzmann}
\end{eqnarray}
where $(\vb, t)$ varies in $\rthree \times (0, +\infty)$, $f(\cdot, t)$ stands for a density function of $\mut$ and $u_{S^2}$ for the uniform p.d. (normalized Riemannian measure) on the unit sphere $S^2$, embedded in $\rthree$. The symbols $\vb_{\ast}$ and $\wb_{\ast}$ denote post-collisional velocities which, according to the conservation laws of momentum and kinetic energy, must satisfy
$$
\vb + \wb = \ \vb_{\ast} + \wb_{\ast} \ \ \ \ \ \ \ \ \text{and} \ \ \ \ \ \ \ \
|\vb|^2 + |\wb|^2 = \ |\vb_{\ast}|^2 + |\wb_{\ast}|^2 \ .
$$
Throughout the paper, $\vb_{\ast}$ and $\wb_{\ast}$ are parametrized according to the $\omb$-representation, i.e.
$$
\vb_{\ast} = \vb + [(\wb - \vb) \cdot \omb] \ \omb \ , \ \ \ \ \ \ \ \wb_{\ast} = \wb - [(\wb - \vb) \cdot \omb] \ \omb
$$
where $\cdot$ denotes the standard scalar product. The \emph{angular collision kernel} $b$ is a non-negative measurable function on $[-1, 1]$. Henceforth, for the sake of mathematical convenience, it will be assumed that $b$ meets the \emph{symmetry condition}
\begin{equation} \label{eq:bsymm}
b(x) = b(\sqrt{1 - x^2}) \frac{|x|}{\sqrt{1 - x^2}} = b(-x)
\end{equation}
for all $x$ in $(-1, 1)$, an assumption which does not reduce the generality of (\ref{eq:boltzmann}), as explained in Subsection 4.1 in Chapter 2A of \cite{vil}. \footnote{It should be noted that condition (\ref{eq:bsymm}) is tantamount to assuming that the counterpart of $b$ in the $\boldsymbol{\sigma}$-representation is an even function.} In presence of a general interaction potential governing the mechanism of binary collisions, $b$ is replaced by a more complex function called \emph{collision kernel}. See Section 3 in Chapter 2A of \cite{vil}. Maxwell \cite{max} was the first to study particles which repel each other with a force inversely proportional to the fifth power of their distance, named \emph{Maxwellian molecules} after him. In this particular circumstance, the resulting collision kernel turns out to be a specific function only of $\frac{\wb - \vb}{|\wb - \vb|} \cdot \omb$, as in (\ref{eq:boltzmann}), with a non-summable singularity near 0. It is customary, as we do here, to call \emph{Maxwellian} any collision kernel which is a function only of $\frac{\wb - \vb}{|\wb - \vb|} \cdot \omb$, and to distinguish Maxwellian kernels depending on whether they are summable or not. The former case corresponds to a SHBEMM with \emph{Grad (angular) cutoff}. Without any loss of generality, this condition can be formalized assuming that
\begin{equation} \label{eq:cutoff}
\int_{0}^{1} b(x) \ud x = 1
\end{equation}
since any SHBEMM with cutoff can be reduced, via a time-scaling, to a SHBEMM with a kernel satisfying (\ref{eq:cutoff}). The case when $b$ is not summable corresponds to a SHBEMM of the \emph{non-cutoff type}. We shall confine ourselves to considering the \emph{weak (angular) cutoff}, i.e.
\begin{equation} \label{eq:wcutoff}
\int_{0}^{1} x b(x) \ud x < +\infty \ .
\end{equation}
This condition is actually satisfied by the explicit form of $b$ given by Maxwell, namely the only form of $b$ that has been justified from a physical standpoint.

The first rigorous results on existence and uniqueness, given a probability density function $f_0$ on $\rthree$
as initial datum, were obtained in \cite{mor, wil} under the validity of (\ref{eq:cutoff}). To discuss this question about the SHBEMM with or without cutoff within a unitary framework, one needs a reformulation of the problem. Accordingly, the weak version of (\ref{eq:boltzmann}) used throughout this paper reads
\begin{eqnarray}
\frac{\ud}{\ud t} \intethree \psi(\vb) \mu(\ud \vb, t) &=& \intethree \intethree \int_{S^2} [\psi(\vb_{\ast}) - \psi(\vb)] \times \nonumber \\
&\times& b\left(\frac{\wb - \vb}{|\wb - \vb|} \cdot \omb \right)
\unifSo \mu(\ud \vb, t) \mu(\ud \wb, t) \label{eq:wboltzmann}
\end{eqnarray}
where $\psi$ varies in $\BL$, the space of all bounded and Lipschitz-continuous functions defined on $\rthree$. This formulation enables us to consider any p.d. $\mu_0$ on $\rthreeb$ as initial datum, $\borelthree$ standing for the Borel class on $\rthree$. The term \emph{weak solution} designates any family $\{\mut\}_{t \geq 0}$ of p.d.'s on $\rthreeb$ such that
\begin{enumerate}
\item[i)] $\mu(\cdot, 0) = \mu_0(\cdot)$;
\item[ii)] $t \mapsto \intethree \psi(\vb) \mu(\vb, t)$ belongs to $\mathrm{C}([0, +\infty)) \cap \mathrm{C}^1((0, +\infty))$ for all $\psi$ in $\BL$;
\item[iii)] $\intethree |\vb| \mu(\vb, t) < +\infty$ for all $t \geq 0$, if $b$ is not summable but obeys (\ref{eq:wcutoff});
\item[iv)] $\mut$ satisfies (\ref{eq:wboltzmann}) for all $t > 0$ and for all $\psi$ in $\BL$.
\end{enumerate}
From now on, the term \emph{solution of} (\ref{eq:boltzmann}) has to be meant as weak solution, according to the above definition, of the Cauchy problem with initial datum $\mu_0$. Tanaka \cite{ta} gave a rigorous result of existence and uniqueness for weak solutions by probabilistic arguments. The sole assumption required on $\mu_0$, only  if $b$ is not summable but obeys (\ref{eq:wcutoff}), is the finiteness of the absolute first moment. Apropos of the uniqueness, see also \cite{tovil}.

It should be recalled that, in the non-cutoff case, existence can be recovered from the existence of the solution to the SHBEMM with cutoff, via a truncation procedure originally introduced in \cite{ark}. More precisely, given a non-summable $b$ satisfying (\ref{eq:wcutoff}) and a p.d. $\mu_0$ on $\rthreeb$ with finite first absolute moment, consider the sequence of collision kernels $\{[b(x) \wedge n]/B_n\}_{n \geq 1}$, with $B_n := \int_{0}^{1} [b(x) \wedge n] \ud x$. Since $[b(x) \wedge n]/B_n$ satisfies (\ref{eq:cutoff}), one can find the solution $\mu_n(\cdot, t)$ to (\ref{eq:boltzmann}), with $b$ replaced by $[b(x) \wedge n]/B_n$ and initial datum $\mu_0$. Following \cite{ark, Dbasic}, it can be shown that $\mu_n(\cdot, B_n t)$ \emph{converges weakly} as $n$ goes to infinity to some limit $\mut$, for every $t \geq 0$, and that $\mut$ turns out to be the solution to the original Cauchy problem. Recall that a sequence $\{P_n\}_{n \geq 1}$ of p.d.'s on some topological space $\mathrm{T}$, endowed with its Borel $\sigma$-algebra, converges weakly to a p.d. $P$ on the same space if and only if $\lim_{n \rightarrow \infty} \int_{\mathrm{T}} h \ud P_n = \int_{\mathrm{T}} h \ud P$, for every bounded and continuous function $h$ on $\mathrm{T}$. Henceforth, this kind of convergence will be denoted with $P_n \Rightarrow P$.

Apropos of the long-time behavior of $\mut$, a well-known fact is the \emph{macroscopic conservation of momentum and kinetic energy}, i.e.
\begin{equation} \label{eq:Vu0}
\intethree \vb \mu(\ud \vb, t) = \intethree \vb \mu_0(\ud \vb) \ \ \ \text{and} \ \ \ \intethree |\vb|^2 \mu(\ud \vb, t) = \intethree |\vb|^2 \mu_0(\ud \vb)
\end{equation}
for every $t \geq 0$, which hold true when the hypothesis
\begin{equation} \label{eq:secondmoment}
\intethree |\vb|^2 \mu_0(\ud \vb) < +\infty
\end{equation}
is in force. Section 8 of \cite{ta} is a reference also for the non-cutoff case. Another fundamental fact is that the equilibrium corresponds to the so-called \emph{Maxwellian distribution}
\begin{equation} \label{eq:maxwellians}
\gamma_{\vb_0, \sigma^2}(\ud \vb) = M_{\vb_0, \sigma^2}(\vb) \ud\vb = \left(\frac{1}{2\pi \sigma^2}\right)^{3/2} \exp\{-\frac{1}{2 \sigma^2} |\vb - \vb_0|^2\} \ud \vb
\end{equation}
which is characterized by the first two moments $\vb_0 = \intethree \vb \mu_0(\ud \vb)$ and $\sigma^2 = \frac{1}{3} \intethree |\vb - \vb_0|^2 \mu_0(\ud \vb)$. Note that $\gamma_{\vb_0, 0}$ stands for the unit mass $\delta_{\vb_0}$ at $\vb_0$. The already quoted paper \cite{ta} proves that, under (\ref{eq:wcutoff}) and (\ref{eq:secondmoment}), $\mut \Rightarrow \gamma_{\vb_0, \sigma^2}$ as $t$ goes to infinity.

\subsection{The conjecture and its motivations} \label{sect:motivations}

Relaxation to equilibrium of solutions to the Boltzmann equation is at the core of kinetic theory ever since the works of Boltzmann himself. The importance of accurate estimates of the rate of convergence is tightly connected with the issue on the physical value of any convergence statement of Boltzmann-equation solutions w.r.t. the time scale on which the Boltzmann description may be relevant. See, for example, Section 2 in Chapter 2C of \cite{vil}. Within this framework, a first preliminary question arises apropos of the choice of the topology in which this convergence ought to take place, keeping in mind that one is dealing with convergence of probability measures (p.m.'s). In fact, the literature has dealt with a variety of probability metrics, but no doubt the \emph{total variation distance} (t.v.d.) continues to be a formidable reference for the study of relaxation to equilibrium in kinetic models. Recall that, for any pair $(\alpha, \beta)$ of p.d.'s on some measurable space $(S, \mathscr{S})$, such a distance is defined by
$$
\vartot(\alpha , \beta) := \sup_{B \in \mathscr{S}} | \alpha(B) - \beta(B)|
$$
and that it can be written as
$$
\vartot(\alpha , \beta) = \frac{1}{2} \int_S |p(x) - q(x)| \lambda(\ud x)
$$
when $\lambda$ is any $\sigma$-finite measure dominating both $\alpha$ and $\beta$, and $p$, $q$ are probability density functions w.r.t. $\lambda$ of $\alpha$ and $\beta$, respectively. See Chapter III of \cite{stro} for more information. Once the right metric has been singled out, the problem of convergence to equilibrium is greatly enhanced by the knowledge of the rate of approach to the limiting distribution and even more so by a precise bound on the error in approximating the limit for each fixed instant. To introduce the reader to the essential part of the problem, we recall that, for an entire class  $\mathcal{I}$ of initial data $\mu_0$, one can prove that
\begin{equation} \label{eq:boundsotto}
\vartot(\mu(\cdot, t), \gamma_{\vb_0, \sigma^2}) \geq C_{\ast} e^{\Lambda_b t}
\end{equation}
is met with a suitable constant $C_{\ast}$ and
\begin{equation} \label{eq:spectralgap}
\Lambda_b := -2\int_{0}^{1} x^2 (1 - x^2) b(x) \ud x \ .
\end{equation}
This result can be reached from a well-known statement by Ikenberry and Truesdell \cite{iktr}, according to which
\begin{equation} \label{eq:iktr}
\Big{|} \intethree \vb^{\boldsymbol{\alpha}} \mu(\ud\vb, t) - \intethree \vb^{\boldsymbol{\alpha}} \gamma_{\vb_0, \sigma^2}(\ud\vb) \Big{|} \leq C_{\boldsymbol{\alpha}} e^{\Lambda_b t}
\end{equation}
holds true with suitable constants $C_{\boldsymbol{\alpha}}$, for any multi-index $\boldsymbol{\alpha}$ such that $\intethree |\vb|^{|\boldsymbol{\alpha}|} \mu_0(\ud\vb) < +\infty$. Recently, it has been proved that $\mathcal{I}$ contains all the p.d.'s $\mu_0$ satisfying $\intethree e^{i \xib \cdot \vb} \mu_0(\ud\vb) = \int_{\rone} e^{i |\xib| x} \zeta_0(\ud x)$ for every $\xib$ in $\rthree$, where $\zeta_0$ is a symmetric p.d. on $(\rone, \mathscr{B}(\rone))$ with \emph{non-zero kurtosis coefficient}. See \cite{dore}. Such being the case, inequality (\ref{eq:boundsotto}) is conducive to checking whether it is possible to establish also the reverse relation
\begin{equation} \label{eq:boundsopra}
\vartot(\mu(\cdot, t), \gamma_{\vb_0, \sigma^2}) \leq C^{\ast} e^{\Lambda_b t} \ .
\end{equation}
Actually, when (\ref{eq:boundsotto}) and (\ref{eq:boundsopra}) are in force simultaneously, $\Lambda_b$ can be viewed as \emph{the best rate of exponential convergence} of $\mut$ to equilibrium. The characterization of the largest class of initial data for which (\ref{eq:boundsopra}) is valid is commonly referred to as \emph{McKean's conjecture}. The reference to McKean is due to the fact that, relative to the solution $\mut$ of the well-known Kac's simplification of the SHBEMM, he was the first to prove rigorously, in \cite{mck6}, that $\vartot(\mu(\cdot, t), \gamma_{\vb_0, \sigma^2}) \leq C^{'} e^{\lambda t}$ holds true with $\lambda \approx -0.016$ and for a suitable constant $C^{'}$. However, this value of $\lambda$ is strictly greater than $\Lambda_b$, equal to $-1/4$ in the case of Kac's equation. See \cite{dgr}.

As a completion of the argument, it is interesting to point out the meaning of $\Lambda_b$ w.r.t. the asymptotic behavior of $\mut$. Besides the important role played in (\ref{eq:iktr}), $\Lambda_b$ represents also the least negative eigenvalue of the \emph{linearized collision operator}
\begin{eqnarray}
L_b[h](\vb) &:=& \intethree\int_{S^2} \ [h({\vb_{\ast}}) +
h({\wb_{\ast}}) \ - \ h(\vb) - h(\wb)] \times \nonumber \\
&\times& b\left(\frac{\wb - \vb}{|\wb - \vb|} \cdot \omb \right) \unifSo \gamma_{\vb_0, \sigma^2}(\ud \wb) \nonumber
\end{eqnarray}
defined on $\mathcal{H} := \mathrm{L}^2(\rthree, \gamma_{\vb_0, \sigma^2}(\ud \xb))$. Hilbert \cite{hil} was the first to derive this operator from a linearization of (\ref{eq:boltzmann}) and to highlight the opportunity of choosing the domain $\mathcal{H}$ with a view to carrying out the spectral analysis. In the Hilbert setting, $L_b$ turns out to be self-adjoint and negative with discrete spectrum and $|\Lambda_b|$ represents the \emph{spectral gap}. See \cite{do}. Finally, it is worth recalling that $\Lambda_b$ arises also in Kac-like derivations of the SHBEMM \cite{kac56}, based on a stochastic evolution of an $N$-particle system. See \cite{ccl, cgl}.

\subsection{A glance at the literature on McKean's conjecture} \label{sect:literature}

The formulation of the Boltzmann H-theorem originated a significant mathematical research, aimed at studying the convergence to equilibrium in total variation, whose first rigorous outcomes are in \cite{car, mor55}. In any case, in spite of the huge literature on this subject, the number of works which expressly pursued the validation of the conjecture is small. Essentially, four lines of research have been followed to achieve the goal, based on: 1) use of contractive functionals or probability metrics; 2) entropy methods; 3) linearization; 4) central limit theorem. 1) As for the first line of research, the papers \cite{cgt, gtw, muta, ta, vil98} are worth mentioning. In particular, Theorem 1.1 in \cite{cgt} constitutes the closest result to the McKean conjecture obtained so far. It is valid only under (\ref{eq:cutoff}) and states that, for every $\varepsilon > 0$, there is $C_{\varepsilon}(\mu_0, b)$ such that
\begin{equation} \label{eq:CGT}
\vartot(\mu(\cdot, t), \gamma_{\vb_0, \sigma^2}) \leq C_{\varepsilon}(\mu_0, b) e^{(1 - \varepsilon)\Lambda_b t}
\end{equation}
holds for every $t \geq 0$, but this $C_{\varepsilon}$ \emph{goes to infinity as} $\varepsilon$ \emph{goes to zero}. Therefore, the presence of $\varepsilon$, together with such a behavior of $C_{\varepsilon}(\mu_0, b)$, defeats the hope of extending (\ref{eq:CGT}) to the solution of the SHBEMM of non-cutoff type through the truncation argument explained in Subsection \ref{sect:equation}: A strong motivation for the pursuit of a bound with $\varepsilon = 0$ and of a constant $C(\mu_0)$ depending only on $\mu_0$, in the place of $C_{\varepsilon}(\mu_0, b)$. Moreover, (\ref{eq:CGT}) has been deduced thanks to rather strong conditions on $\mu_0(\ud \xb) = f_0(\xb) \ud \xb$, such as finiteness of all absolute moments, Sobolev regularity and finiteness of the Linnik functional. 2) Entropy methods aim at proving quantitative H-theorems, on the basis of the seminal ideas introduced in \cite{cc92, cc94}. An attempt to improve this strategy, towards the achievement of the McKean conjecture, was represented by the \emph{Cercignani conjecture} which, however, proved to be false in the case of Maxwellian molecules. See \cite{bobcer, vilcerc}. Anyway, quantitative H-theorems are still considered as conducive to the most powerful strategy to study relaxation to equilibrium in non-homogeneous frameworks. See \cite{devil}. 3) The strategy of the linearization is outlined in \cite{doLom, gru}. It gives general positive answers to the problem of quantifying the relaxation to equilibrium only when the solution enters a small neighborhood of the equilibrium itself, so that the spectral analysis of $L_b$, as an operator on $\mathcal{H}$, becomes relevant to the study of the nonlinear problem. It is only recently that, in the case of the homogeneous Boltzmann equation with \emph{hard potentials}, the linearization has been used successfully to prove the conjecture. See \cite{mouh}. However, the radical difference between the situation of hard potentials and that of Maxwellian molecules hampers a direct extension of the positive conclusion from the former to the latter. 4) Finally, the link with the central limit theorem discovered by McKean in \cite{mck6, mck7} has been taken into serious consideration only recently in \cite{ccg0, ccg5}, two works which have strongly inspired and motivated our program.

\subsection{The main result} \label{sect:mainresult}

A precise and complete formulation is encapsulated in the following theorem where $\hat{\mu}$ stands for the Fourier transform of the p.d. $\mu$ on $\rthreeb$, namely $\hat{\mu}(\xib) := \intethree e^{i \xib \cdot \vb} \mu(\ud \vb)$ for $\xib$ in $\rthree$.
\begin{thm} \label{thm:CDGR}
\emph{Assume that} (\ref{eq:bsymm}) \emph{and} (\ref{eq:wcutoff}) \emph{are in force and that the initial datum} $\mu_0$ \emph{satisfies}
\begin{equation}\label{eq:fourthmom}
\mathfrak{m}_4 := \intethree |\vb|^4 \mu_0(\ud \vb) < +\infty
\end{equation}
\emph{and}
\begin{equation}\label{eq:tailsCDGR}
|\hat{\mu}_0(\xib)| = o(|\xib|^{-p}) \ \ \ \ \ \ \ \ \ \ (|\xib| \rightarrow +\infty)
\end{equation}
\emph{for some strictly positive} $p$. \emph{Then, the solution} $\mu(\cdot, t)$ \emph{meets}
\begin{equation} \label{eq:CDGR}
\vartot(\mu(\cdot, t), \gamma_{\vb_0, \sigma^2}) \leq C(\mu_0) e^{\Lambda_b t}
\end{equation}
\emph{for every} $t \geq 0$, \emph{where} $\Lambda_b$ \emph{is given by} (\ref{eq:spectralgap}) \emph{and} $C(\mu_0)$ \emph{is a positive constant which depends only on} $\mu_0$.
\end{thm}
Indications for numerical evaluation of $C(\mu_0)$ can be derived from specific passages of the proof, in Subsection \ref{sect:proof}. With reference to the SHBEMM with cutoff, this theorem represents the first direct validation of the McKean conjecture, without unnecessary extra-conditions. Moreover, as far as the non-cutoff case is concerned, the same theorem is, at the best of our knowledge, the only existing sharp quantification of the speed of convergence to equilibrium. A detailed explanation of these points is given in the following \\

\noindent \emph{Remarks}
\begin{enumerate}
\item[1.] The proof of Theorem \ref{thm:CDGR} will be developed, in Subsection \ref{sect:proof}, under the cutoff condition (\ref{eq:cutoff}). Indeed, once (\ref{eq:CDGR}) has been established under (\ref{eq:cutoff}), one can resort to the truncation procedure described in Subsection \ref{sect:equation} to write, for every $n$ in $\mathbb{N}$,
$$
\vartot(\mu_n(\cdot, B_n t), \gamma_{\vb_0, \sigma^2}) \leq C(\mu_0) \exp\Big{\{}-2 t \int_{0}^{1} x^2(1 - x^2) [b(x) \wedge n] \ud x\Big{\}} \ .
$$
Now, the combination of this inequality with
$$
\vartot(\mu(\cdot, t), \gamma_{\vb_0, \sigma^2}) \leq \liminf_{n \rightarrow \infty} \vartot(\mu_n(\cdot, B_n t), \gamma_{\vb_0, \sigma^2})
$$
leads to the desired conclusion.
\item[2.] Let us now discuss assumption (\ref{eq:fourthmom}). It is interesting to recall that, under the cutoff condition, convergence in the total variation metric to the Maxwellian holds under (\ref{eq:secondmoment}). See \cite{cl}. The necessity of this condition, in a cutoff setting, is stated both in \cite{cgr} and in Theorem \ref{thm:CLT} of the present paper. In \cite{cl} it is also shown that convergence to equilibrium, under the sole assumption of finiteness of the second moment of $\mu_0$, could be arbitrarily slow, whereas the finiteness of the $(2+\delta)$-th absolute moment, for some $\delta > 0$, is enough to get exponentially decreasing bounds. Nevertheless, if $\delta < 2$, these bounds can be worse than that conjectured by McKean. Here is an example which shows that, even if the tail condition (\ref{eq:tailsCDGR}) is fulfilled, the desired bound is not achieved because of ``infinitesimal'' deviations from hypothesis (\ref{eq:fourthmom}). Consider the class of initial data $\mu^{(q)}_{0}(\ud \vb) = f_{0, q}(\vb) \ud \vb$ with
$$
f_{0, q}(\vb) = \frac{q}{4\pi \ |\vb|^{3 + q}} \ind_{\{|\vb| \geq 1\}}
$$
for $q$ in $(3, 4)$. The Fourier transform of this density at $\xib$ is
$$
1 - \frac{q}{6 (q - 2)} |\xib|^2 - \frac{\Gamma(1 - q) \cos(q \pi/2)}{1+q} |\xib|^q  - q \sum_{m \geq 2} \frac{(-1)^m |\xib|^{2m}}{(2m + 1)! (2m - q)}
$$
which meets $\hat{\mu}^{(q)}_{0}(\xib) = O(|\xib|^{-1})$ when $|\xib|$ goes to infinity. Then, $\mu^{(q)}_{0}$ satisfies (\ref{eq:tailsCDGR}) and has finite absolute moment of order $(3 + \delta)$ for every $\delta$ in $(0, q - 3)$, but has infinite absolute fourth moment. Denoting by $\mu^{(q)}(\cdot, t)$ the solution of (\ref{eq:boltzmann}) relative to $\mu^{(q)}_{0}$, one can mimic the argument explained in \cite{dore} to prove that
$$
\vartot(\mu^{(q)}(\cdot, t), \gamma_{\mathbf{0}, \sigma^2}) \geq C_q \exp\{-(1 - 2l_q(b)) t\}
$$
holds for every $t \geq 0$, where $3\sigma^2 = q/(q - 2)$, $C_q$ is a strictly positive constant independent of $b$,
$l_q(b) := \int_{0}^{1} (1 - x^2)^{q/2} b(x) \ud x$ and $\Lambda_b < -(1 - 2l_q(b)) < 0$.
\item[3.] As far as the tail assumption (\ref{eq:tailsCDGR}) is concerned, it is worth noting that it is implied by the finiteness of the Linnik functional, according to Lemma 2.3 in \cite{cgt}. Also the relationship between (\ref{eq:tailsCDGR}) and certain regularity conditions adopted to guarantee the validity of classical local limit theorems of probability theory are worth noting. See, for example, Theorem 19.1 in \cite{barao}.
\end{enumerate}

\subsection{A probabilistic representation of the solution} \label{sect:representation}

The proof of Theorem \ref{thm:CDGR} relies on a representation of the solution $\mut$ -- already proposed and studied in \cite{dophd} -- which is valid under the cutoff condition (\ref{eq:cutoff}). The motivation for this representation is twofold. On the one hand, it leads us to study the problem of convergence to equilibrium from the standpoint of the central limit problem of probability theory. On the other hand, it lends itself to computability of certain derivatives of the Fourier transform of $\mut$ involved in the first steps of the proof of Theorem \ref{thm:CDGR}. See, for example, (\ref{eq:appliedbeurling}) below. It should be mentioned that the existing representations, essentially based on the Bobylev identity (see Section 3 of \cite{bob88}), turn out to be unfit for the aforesaid computations.

In a nutshell, the probabilistic representation at issue states that
\begin{equation} \label{eq:main}
\mu(B, t) = \textsf{E}_t \left[\mathcal{M}(B)\right]
\end{equation}
for every $t \geq 0$ and every $B$ in $\borelthree$, where $\et$ is an expectation and $\Mrand$ is a \emph{random p.m.} connected with a distinguished weighted sum of random vectors, to be defined below. Here and in the rest of the paper we use the term random p.m. to designate any measurable function from some measurable space into the space $\prob$ of all p.m.'s on $\rthreeb$, endowed with the Borel $\sigma$-algebra of weak convergence of p.m.'s. See, e.g., Chapters 11-12 of \cite{ka} for further details. Then, to carry out our programme, it remains to provide the reader with those definitions and preliminary results which are necessary to understand (\ref{eq:main}). In this way, we shall present also the core of the notation used in the rest of the paper.

The starting point is the introduction of the sample space
$$
\Omega := \mathbb{N} \times \mathbb{T} \times [0, \pi]^{\infty} \times (0, 2\pi)^{\infty} \times (\rthree)^{\infty}
$$
where: For any nonempty set $X$, $X^{\infty}$ stands for the set of all sequences $(x_1, x_2, \dots)$ whose elements belong to $X$; $\mathbb{T} := \textsf{X}_{\substack{n \geq 1}} \mathbb{T}(n)$ and $\mathbb{T}(n)$ is the (finite) set of all \emph{McKean binary trees} with $n$ leaves. See, e.g., \cite{gr6} for details. On writing $\treen$ to denote an element of $\mathbb{T}(n)$, $\treenk$ indicates the \emph{germination} of $\treen$ at its $k$-th leaf (cf. Fig. 1), while $\treen^l$ and $\treen^r$ symbolize the two trees, of $n_l$ and $n_r$ leaves respectively, obtained by a \emph{split-up} of $\treen$ (cf. Fig. 2). A recent and comprehensive treatment of random trees can be found in \cite{drm}.

%%%%%%%%%%%%%%%%%%%%%%%%%%%%%
%????DISEGNO????
%%%%%%%%%%%%%%%%%%%%%%%%%%%%%%%%

Then, associate with $\Omega$ the $\sigma$-algebra
$$
\mathscr{F} := 2^{\mathbb{N}} \otimes  2^{\mathbb{T}} \otimes  \mathscr{B}([0, \pi]^{\infty}) \otimes  \mathscr{B}((0, 2\pi)^{\infty}) \otimes  \mathscr{B}((\rthree)^{\infty})
$$
where $2^X$ stands for the power set of $X$ and $\mathscr{B}(X)$ for the Borel class on $X$. Define
$$
\nu,\ \{\tau_n\}_{n \geq 1},\ \{\phi_n\}_{n \geq 1},\ \{\vartheta_n\}_{n \geq 1}, \{\mathbf{V}_n\}_{n \geq 1}
$$
to be the coordinate random variables of $\Omega$ and, by them, generate the $\sigma$-algebras
\begin{eqnarray}
\mathscr{G} &:=& \sigma\big{(}\nu,\ \{\tau_n\}_{n \geq 1},\ \{\phi_n\}_{n \geq 1} \big{)} \nonumber \\
\mathscr{H} &:=& \sigma\big{(}\nu,\ \{\tau_n\}_{n \geq 1},\ \{\phi_n\}_{n \geq 1},\ \{\vartheta_n\}_{n \geq 1} \big{)} \ . \nonumber
\end{eqnarray}
Now, for every $t \geq 0$, consider the unique p.d. $\pt$ on $(\Omega, \mathscr{F})$ which makes the random coordinates
\emph{stochastically independent}, consistently with the following marginal p.d.'s:
\begin{enumerate}
\item[a)] \begin{equation} \label{eq:nu}
\pt[ \nu = n ] = e^{-t}(1 - e^{-t})^{n-1} \ \ \ \ \ \ \ (n = 1, 2, \dots)
\end{equation}
with the proviso that $0^0 := 1$.
\item[b)] $\{\tau_n\}_{n \geq 1}$ is a Markov sequence driven by
\begin{equation} \label{eq:markov}
\begin{array}{lll}
\pt[\tau_1 = \tree_1] &= 1 & {} \\
\pt[ \tau_{n+1} = \treenk \ | \ \tau_n = \treen ] &= \frac{1}{n} & \ \text{for}\ k = 1, \dots, n \\
\pt[ \tau_{n+1} = \tree_{n+1} \ | \ \tau_n = \treen ] &= 0 & \ \text{if}\ \tree_{n+1} \not\in \mathbb{G}(\treen)
\end{array}
\end{equation}
for every $n$ in $\mathbb{N}$ and $\treen$ in $\mathbb{T}(n)$, where, for a given $\treen$, $\mathbb{G}(\treen)$ is the subset of $\mathbb{T}(n+1)$ containing all the germinations of $\treen$.
\item[c)] The elements of $\{\phi_n\}_{n \geq 1}$ are i.i.d. random numbers with p.d.
\begin{equation}\label{eq:beta}
\beta(\ud \varphi) := \frac{1}{2} b(\cos\varphi) \sin\varphi \ud \varphi \ , \ \ \ \ \ \ (\varphi \in [0, \pi]) \ .
\end{equation}
\item[d)] The elements of $\{\vartheta_n\}_{n \geq 1}$ are i.i.d. with uniform p.d. on $(0, 2\pi)$, $u_{(0, 2\pi)}$.
\item[e)] The elements of $\{\mathbf{V}_n\}_{n \geq 1}$ are i.i.d. with p.d. $\mu_0$, the initial datum of the Cauchy problem relative to (\ref{eq:boltzmann}).
\end{enumerate}
According to the above notation, $\et$ denotes expectation w.r.t. $\pt$.

A constituent of the representation under study is $\boldsymbol{\pi} := \{\pi_{j, n}\ | \ j = 1, \dots, n; n \in \mathbb{N}\}$, an array of $[-1,1]$-valued random numbers. They are obtained by setting
\begin{equation} \label{eq:pijn}
\pi_{j, n} := \pi_{j, n}^{\ast}(\tau_n, (\phi_1, \dots, \phi_{n-1}))
\end{equation}
for $j = 1, \dots, n$ and $n$ in $\mathbb{N}$. The $\pi_{j, n}^{\ast}$'s are functions on $\mathbb{T}(n) \times [0, \pi]^{n-1}$ defined by putting $\pi_{1, 1}^{\ast} \equiv 1$ and, for $n \geq 2$,
\begin{equation} \label{eq:pijnast}
\pi_{j, n}^{\ast}(\treen, \boldsymbol{\varphi}) := \left\{ \begin{array}{ll}
\pi_{j, n_l}^{\ast}(\mathfrak{t}_{n}^{l}, \boldsymbol{\varphi}^l) \cos\varphi_{n-1} & \text{for} \ j = 1, \dots, n_l \\
\pi_{j - n_l, n_r}^{\ast}(\mathfrak{t}_{n}^{r}, \boldsymbol{\varphi}^r) \sin\varphi_{n-1} & \text{for} \ j = n_l + 1, \dots, n
\end{array} \right.
\end{equation}
for every $\boldsymbol{\varphi} = (\boldsymbol{\varphi}^l, \boldsymbol{\varphi}^r, \varphi_{n-1})$ in $[0, \pi]^{n-1}$, with
$$
\boldsymbol{\varphi}^l := (\varphi_1, \dots, \varphi_{n_l-1})\ \ \ \ \ \text{and}\ \ \ \ \ \boldsymbol{\varphi}^r := (\varphi_{n_l}, \dots, \varphi_{n-2}) \ .
$$
An induction argument shows that
\begin{equation} \label{eq:sumpijn}
\sum_{j=1}^{n} \pi_{j, n}^{2} = 1
\end{equation}
for every $n$ in $\mathbb{N}$. It is also worth recalling the identity
\begin{equation} \label{eq:gare}
\et\left[\sum_{j = 1}^{\nu} |\pi_{j, \nu}|^s \right] = e^{-(1 - 2 l_s(b))t}
\end{equation}
valid for every $t, s > 0$, with $l_s(b) := \int_{0}^{1} (1 - x^2)^{s/2} b(x) \ud x$. The original derivation is in \cite{gr6} but, for the sake of completeness, we have included its proof in \ref{a:gare}. Throughout the paper, A.$n$ designates the $n$-th subsection of the Appendix. With a view to the proof of Theorem \ref{thm:CDGR}, it is interesting to point out that
\begin{equation} \label{eq:l4spectral}
-(1 - 2l_4(b)) = \Lambda_b \ .
\end{equation}

Another constituent of the desired representation is the array $\boldsymbol{\mathrm{O}} := \{\mathrm{O}_{j, n}\ | \ j = 1, \dots, n; n \in \mathbb{N}\}$ of random matrices $\mathrm{O}_{j, n}$, taking values in the Lie group $\mathbb{SO}(3)$ of orthogonal matrices with positive determinant, defined by
\begin{equation} \label{eq:Ojn}
\mathrm{O}_{j, n} := \mathrm{O}_{j, n}^{\ast}(\tau_n, (\phi_1, \dots, \phi_{n-1}), (\vartheta_1, \dots, \vartheta_{n-1}))
\end{equation}
for $j = 1, \dots, n$ and $n$ in $\mathbb{N}$. The $\mathrm{O}_{j, n}^{\ast}$'s are $\mathbb{SO}(3)$-valued functions obtained by putting $\mathrm{O}_{1, 1}^{\ast} \equiv \mathrm{Id}_{3 \times 3}$ and, for $n \geq 2$,
\begin{eqnarray} \label{eq:Ojnast}
&{}& \mathrm{O}_{j, n}^{\ast}(\treen, \boldsymbol{\varphi}, \boldsymbol{\theta}) \nonumber \\
&:=& \left\{ \begin{array}{ll}
\mathrm{M}^l(\varphi_{n-1}, \theta_{n-1}) \mathrm{O}_{j, n_l}^{\ast}(\treen^l, \boldsymbol{\varphi}^l, \boldsymbol{\theta}^l) & \text{for} \ j = 1, \dots, n_l \\
\mathrm{M}^r(\varphi_{n-1}, \theta_{n-1}) \mathrm{O}_{j - n_l, n_r}^{\ast}(\treen^r, \boldsymbol{\varphi}^r, \boldsymbol{\theta}^r) & \text{for} \ j = n_l + 1, \dots, n
\end{array} \right.
\end{eqnarray}
for every $\treen$ in $\mathbb{T}(n)$, $\boldsymbol{\varphi}$ in $[0, \pi]^{n-1}$ and $\boldsymbol{\theta}$ in $(0, 2\pi)^{n-1}$. Here
$$
\boldsymbol{\theta}^l := (\theta_1, \dots, \theta_{n_l-1})\ \ \ \ \ \text{and} \ \ \ \ \ \boldsymbol{\theta}^r := (\theta_{n_l}, \dots, \theta_{n-2})
$$
and, finally,
$$
\mathrm{M}^l(\varphi, \theta) := \left( \begin{array}{ccc} -\cos\theta \cos\varphi & \sin\theta & \cos\theta \sin\varphi \\
- \sin\theta \cos\varphi & -\cos\theta & \sin\theta \sin\varphi  \\
\sin\varphi & 0 & \cos\varphi \\
\end{array} \right)
$$
$$
\mathrm{M}^r(\varphi, \theta) := \left( \begin{array}{ccc} \sin\theta & \cos\theta \sin\varphi & -\cos\theta \cos\varphi \\
-\cos\theta & \sin\theta \sin\varphi & - \sin\theta \cos\varphi \\
0 & \cos\varphi & \sin\varphi \\
\end{array} \right) \ .
$$
Working out the recursion formula (\ref{eq:Ojn}) gives
\begin{equation} \label{eq:pathOijn}
\mathrm{O}_{j, n}^{\ast}(\treen, \boldsymbol{\varphi}, \boldsymbol{\theta}) = \prod_{h = 1}^{\delta_j(\treen)} \mathrm{M}^{\epsilon_h(\treen, j)}(\varphi_{m_h(\treen, j)}, \theta_{m_h(\treen, j)})
\end{equation}
where $\prod_{h=1}^{n} A_h := A_1 \times \dots \times A_n$ and $\delta_j(\treen)$ indicates the \emph{depth} of the $j$-th leaf of $\treen$, that is the number of generations separating this leaf from the root (the top node of the tree). The $\epsilon_h(\treen, j)$'s take values in $\{l, r\}$ and, in particular, $\epsilon_1(\treen, j)$ equals $l$ ($r$, respectively) if $j \leq n_l$ ($j > n_l$, respectively). Then,
$$
\epsilon_h(\treen, j) := \left\{ \begin{array}{ll} \epsilon_{h-1}(\treen^l, j) & \text{for} \ j = 1, \dots, n_l \\
\epsilon_{h-1}(\treen^r, j - n_l) & \text{for} \ j = n_l + 1, \dots, n
\end{array} \right.
$$
when $h \geq 2$. Each $m_h$ belongs to $\{1, \dots, n-1\}$ and $m_1 \neq \dots \neq m_{\delta_j(\treen)}$. In fact, $m_1(\treen, j) := n-1$ for every $\treen$ in $\mathbb{T}(n)$, $j = 1, \dots, n$, and
$$
m_h(\treen, j) := \left\{ \begin{array}{ll} m_{h-1}(\treen^l, j) & \text{for} \ j = 1, \dots, n_l \\
m_{h-1}(\treen^r, j - n_l) & \text{for} \ j = n_l + 1, \dots, n
\end{array} \right.
$$
when $h \geq 2$.

Now, choose a \emph{non-random} measurable function $\mathrm{B}$ from $S^2$ onto $\mathbb{SO}(3)$ such that $\mathrm{B}(\ub) \mathbf{e}_3 = \ub$ for every $\ub$ in $S^2$, and define the random functions $\psib_{j, n} : S^2 \rightarrow S^2$ through the relation
\begin{equation} \label{eq:psijn}
\psib_{j, n}(\ub) := \mathrm{B}(\ub) \mathrm{O}_{j, n} \mathbf{e}_3
\end{equation}
for $j = 1, \dots, n$ and $n$ in $\mathbb{N}$, with $\mathbf{e}_3 :=\  ^t(0, 0, 1)$. It should be noticed that such a $\mathrm{B}$ actually exists and that it cannot be continuous. See, e.g., Chapter 5 of \cite{hir}.

The central object of our construction is the random sum
\begin{equation} \label{eq:Su}
S(\ub) := \sum_{j = 1}^{\nu} \pi_{j, \nu} \psib_{j, \nu}(\ub) \cdot \mathbf{V}_j
\end{equation}
whose characteristic function (c.f.) serves the new representation according to
\begin{thm}\label{thm:representation}
\emph{Assume that} (\ref{eq:bsymm})-(\ref{eq:cutoff}) \emph{are in force. Then, the function}
\begin{gather}
\hat{\Mrand}(\xib) := \textsf{E}_t \left[e^{i \rho S(\ub)} \ | \ \mathscr{G} \right] \nonumber \\
= \left\{ \begin{array}{ll}
\hat{\mu}_0(\xib) & \text{if} \ \nu = 1 \\
\int_{(0, 2\pi)^{\nu - 1}} \left[\prod_{j = 1}^{\nu} \hat{\mu}_0\big{(}\rho \pi_{j, \nu}\mathrm{B}(\ub)\mathrm{O}_{j, \nu}^{\ast}(\tau_{\nu}, \boldsymbol{\phi}, \boldsymbol{\theta}) \mathbf{e}_3\big{)}\right] u_{(0, 2\pi)}^{\otimes_{\nu - 1}}(\ud \boldsymbol{\theta}) & \text{if} \ \nu \geq 2\ ,
\end{array} \right. \label{eq:M}
\end{gather}
\emph{with} $\xib$ \emph{in} $\rthree\setminus\{\mathbf{0}\}$, $\rho := |\xib|$, $\ub := \xib/|\xib|$ \emph{and} $\boldsymbol{\phi} := (\phi_1, \dots, \phi_{\nu - 1})$, \emph{is the Fourier transform of a random p.d. on} $\rthreeb$, \emph{denoted by} $\Mrand$. \emph{This} $\Mrand$ \emph{turns out to be independent of the choice of} $\mathrm{B}$, \emph{and satisfies} (\ref{eq:main})
\emph{for every} $t \geq 0$ \emph{and} $B$ \emph{in} $\borelthree$.
\end{thm}
The proof of the theorem is contained in Subsection \ref{sect:proofrepresentation}. Many relevant properties of $\Mrand$ rely on the analysis of the random function
\begin{equation} \label{eq:N}
\hat{\Nrand}(\rho; \ub) := \textsf{E}_t \left[e^{i \rho S(\ub)} \ | \ \mathscr{H} \right]  = \prod_{j = 1}^{\nu} \hat{\mu}_0(\rho \pi_{j, \nu} \psib_{j, \nu}(\ub))
\end{equation}
which, as a function of $\xib$, \emph{is not} c.f. and \emph{depends} on the choice of $\mathrm{B}$.

One of the merits of representation (\ref{eq:main}) is that it allows the formulation of a central limit-like theorem for the asymptotic behavior of the solution of the SHBEMM with cutoff, condensed in the following
\begin{thm} \label{thm:CLT}
\emph{When} (\ref{eq:bsymm})-(\ref{eq:cutoff}) \emph{are in force}, $\mu(\cdot, t)$ \emph{converges weakly as} $t$ \emph{goes to infinity if and only if} (\ref{eq:secondmoment}) \emph{holds true. Moreover, in case this condition is satisfied, the limiting distribution is given by} (\ref{eq:maxwellians}).
\end{thm}
As already mentioned at the beginning of Remark 2, this theorem is well-known. In fact, the ``if part'' was proved in \cite{cl, ta}, while the ``only if'' part was proved, in a quite different way, in \cite{cgr}. What is new is the proof we develop in Subsection \ref{sect:CLT} on the basis of (\ref{eq:main}).

\section{Proofs} \label{sect:proof2}

In this section, we present the skeleton of the proofs of Theorems \ref{thm:CDGR}, \ref{thm:representation} and \ref{thm:CLT}. Some technical issues are deferred to the Appendix and to \cite{doBE, Dbasic}. We start from the basic representation formulated in Theorem \ref{thm:representation}.

\subsection{Proof of Theorem \ref{thm:representation}}\label{sect:proofrepresentation}

When (\ref{eq:bsymm})-(\ref{eq:cutoff}) are in force, recall that $\mut$ can be expressed by means of the so-called \emph{Wild-McKean sum} \cite{mck7, wil}, namely
$$
\mu(B, t) = \sum_{n = 1}^{+\infty} e^{-t} (1 - e^{-t})^{n-1} \sum_{\treen \in \mathbb{T}(n)} p_n(\treen) \mathcal{Q}_{\treen}[\mu_0](B)
$$
for every $t \geq 0$ and $B$ in $\borelthree$. According to McKean, the weights $p_n(\treen)$ are defined inductively starting from $p_1(\tree_1) := 1$ and then putting
\begin{equation} \label{eq:pntn}
p_n(\treen) := \frac{1}{n-1} p_{n_l}(\treen^l) p_{n_r}(\treen^r)
\end{equation}
for every $n \geq 2$ and $\treen$ in $\mathbb{T}(n)$. These $p_n$'s are connected with the p.d. of $\{\tau_n\}_{n \geq 1}$ through the identity
\begin{equation} \label{eq:pntntaun}
p_n(\treen) = \pt[\tau_n = \treen]
\end{equation}
valid for every $n$ in $\mathbb{N}$ and $\treen$ in $\mathbb{T}(n)$. See \ref{a:pntn} for the proof. As far as the $\mathcal{Q}_{\treen}$'s are concerned,
$$
\begin{array}{lll}
\mathcal{Q}_{\tree_1}[\mu_0] &:= \mu_0 & {} \\
\mathcal{Q}_{\treen}[\mu_0] &:= \mathcal{Q}\left[\mathcal{Q}_{\treen^l}[\mu_0], \mathcal{Q}_{\treen^r}[\mu_0]\right] & \ \ \ \ \text{for} \ n \geq 2
\end{array}
$$
where $\mathcal{Q}$ is an operator which sends a pair $(\zeta, \eta)$ belonging to $\prob \times \prob$ into a new element
$\mathcal{Q}[\zeta, \eta]$ of $\prob$ according to the following rule. First, take two sequences $\{\zeta_n\}_{n\geq 1}$ and $\{\eta_n\}_{n\geq 1}$ of absolutely continuous p.m.'s such that $\zeta_n \Rightarrow \zeta$ and $\eta_n \Rightarrow \eta$, and denote with $p_n$ ($q_n$, respectively) the density of $\zeta_n$ ($\eta_n$, respectively). Then, denoting limit w.r.t. weak convergence by $\text{w-lim}$, put
\begin{equation} \label{eq:extendedQ}
\mathcal{Q}[\zeta, \eta](\ud \vb) := \text{w-lim}_{n \rightarrow \infty} Q[p_n, q_n](\vb) \ud \vb
\end{equation}
where
$$
Q[p, q](\vb) := \intethree\int_{S^2} p(\vb_{\ast})
q(\wb_{\ast}) b\left(\frac{\wb - \vb}{|\wb - \vb|} \cdot \omb \right) \unifSo \ud \wb \ .
$$
Note that $Q[p, q] = Q[q, p]$, as a consequence of (\ref{eq:bsymm}). As shown in \cite{Dbasic}, the limit in (\ref{eq:extendedQ}) exists and is independent of the choice of the approximating sequences $\{\zeta_n\}_{n\geq 1}$ and $\{\eta_n\}_{n\geq 1}$.

To carry on with the proof, consider the Fourier transform and apply the well-known Bobylev formula, as in \cite{Dbasic}, to get
$$
\hat{\mathcal{Q}}[\zeta, \eta](\xib) = \int_{S^2} \hat{\zeta}((\xib \cdot \omb)\omb) \hat{\eta}(\xib - (\xib \cdot \omb)\omb)  \ b\left(\frac{\xib}{|\xib|} \cdot \omb \right) \unifSo
$$
for every $\xib$ in $\rthree\setminus\{\mathbf{0}\}$. This, by the change of variable $\omb = \omb(\varphi, \theta, \xib) = \sin\varphi \cos\theta \mathbf{a}(\ub) + \sin\varphi \sin\theta \mathbf{b}(\ub) +
\cos\varphi \ub$, becomes
\begin{equation} \label{eq:parametrizedbobylev}
\hat{\mathcal{Q}}[\zeta, \eta](\xib) = \int_{0}^{\pi}\int_{0}^{2\pi} \hat{\zeta}(\rho \cos\varphi \boldsymbol{\psi}^l) \hat{\eta}(\rho \sin\varphi \boldsymbol{\psi}^r) u_{(0, 2\pi)}(\ud \theta) \beta(\ud \varphi)
\end{equation}
where $\rho = |\xib|$, $\ub = \xib/|\xib|$ and $\boldsymbol{\psi}^l$, $\boldsymbol{\psi}^r$ are abbreviations for the quantities
\begin{equation} \label{eq:psi}
\begin{array}{lll}
\psib^l(\varphi, \theta, \ub) &:= &{}\cos\theta \sin\varphi \mathbf{a}(\ub) + \sin\theta \sin\varphi \mathbf{b}(\ub) + \cos\varphi \ub \\
\psib^r(\varphi, \theta, \ub) &:= &-\cos\theta \cos\varphi \mathbf{a}(\ub) - \sin\theta \cos\varphi \mathbf{b}(\ub) + \sin\varphi \ub
\end{array}
\end{equation}
which depend on the choice of the orthonormal basis $\{\mathbf{a}(\ub), \mathbf{b}(\ub), \ub\}$ of $\rthree$. The components of this basis are exactly the columns of the matrix $\mathrm{B}$ introduced in (\ref{eq:psijn}). The inner integral in (\ref{eq:parametrizedbobylev}), that is
\begin{equation} \label{eq:internalintegralbobylev}
I(\xib, \varphi) := \left\{ \begin{array}{ll}
\int_{0}^{2\pi} \hat{\zeta}(\rho \cos\varphi \boldsymbol{\psi}^l) \hat{\eta}(\rho \sin\varphi \boldsymbol{\psi}^r) u_{(0, 2\pi)}(\ud \theta) & \text{if} \ \xib \neq \mathbf{0} \\
1 & \text{if} \ \xib = \mathbf{0} \ ,
\end{array} \right.
\end{equation}
has interesting properties, which are at the basis of the new representation (\ref{eq:main}). In particular, $I$ is a measurable function of $(\xib, \varphi)$ independent of the choice of $\{\mathbf{a}(\ub), \mathbf{b}(\ub), \ub\}$. Moreover, for every fixed $\varphi$ in $[0, \pi]$, $I(\cdot, \varphi)$ is the Fourier transform of a p.m. on $\rthreeb$, say $\mathcal{C}[\zeta, \eta; \varphi]$, that is $I(\xib, \varphi) = \hat{\mathcal{C}}[\zeta, \eta; \varphi](\xib)$ for every $\xib$ in $\rthree$. The link with $\mathcal{Q}$ is given by
\begin{equation} \label{eq:QCbis}
\mathcal{Q}[\zeta, \eta](B) = \int_{0}^{\pi} \mathcal{C}[\zeta, \eta; \varphi](B) \beta(\ud \varphi)
\end{equation}
for every $B$ in $\borelthree$. The proof of these facts is contained in \ref{a:C}. At this stage, mimicking the iteration procedure developed for $\mathcal{Q}$ leads to the following definition
$$
\begin{array}{lll}
\mathcal{C}_{\tree_1}[\mu_0; \emptyset] &:= \mu_0 & {} \\
\mathcal{C}_{\treen}[\mu_0; \boldsymbol{\varphi}] &:= \mathcal{C}\left[\mathcal{C}_{\treen^l}[\mu_0; \boldsymbol{\varphi}^l], \mathcal{C}_{\treen^r}[\mu_0; \boldsymbol{\varphi}^r]; \varphi_{n-1}\right] & \ \ \ \ \text{for} \ n \geq 2
\end{array}
$$
for every $\treen$ in $\mathbb{T}(n)$ and $\boldsymbol{\varphi}$ in $[0, \pi]^{n-1}$, with the proviso that $\boldsymbol{\varphi}^l$ ($\boldsymbol{\varphi}^r$, respectively) is void when $n_l$ ($n - n_l$, respectively) is equal to one. For every $n \geq 2$ and $\treen$ in $\mathbb{T}(n)$, the mapping $\boldsymbol{\varphi} \mapsto \mathcal{C}_{\treen}[\mu_0; \boldsymbol{\varphi}]$ is a random p.m. and
\begin{equation} \label{eq:QCtnbis}
\mathcal{Q}_{\treen}[\mu_0](B) = \int_{[0, \pi]^{n-1}} \mathcal{C}_{\treen}[\mu_0; \boldsymbol{\varphi}](B) \beta^{\otimes_{n-1}}(\ud \boldsymbol{\varphi})
\end{equation}
holds true for every $B$ in $\borelthree$, as proved in \ref{a:QCtnbis}. In view of this link, the Wild-McKean can be re-written
as
$$
e^{-t}\mu_0(B) + \sum_{n = 2}^{+\infty} e^{-t} (1 - e^{-t})^{n-1} \sum_{\treen \in \mathbb{T}(n)} p_n(\treen) \int_{[0, \pi]^{n-1}} \mathcal{C}_{\treen}[\mu_0; \boldsymbol{\varphi}](B) \beta^{\otimes_{n-1}}(\ud \boldsymbol{\varphi})
$$
which coincides with $\et\left[\mathcal{C}_{\tau_{\nu}}[\mu_0; (\phi_1, \dots, \phi_{\nu - 1})](B)\right]$. Therefore, to show the validity of (\ref{eq:main}), it is enough to verify that $\mathcal{M}(B) = \mathcal{C}_{\tau_{\nu}}[\mu_0; (\phi_1, \dots, \phi_{\nu - 1})](B)$ for every $B$ in $\borelthree$ or, equivalently, that
\begin{eqnarray}
&{}& \hat{\mathcal{C}}_{\treen}[\mu_0; \boldsymbol{\varphi}](\xib) \nonumber \\
&=& \int_{(0, 2\pi)^{n-1}} \left[\prod_{j=1}^{n} \hat{\mu}_0\big{(} \rho  \pi_{j, n}^{\ast}(\treen, \boldsymbol{\varphi})
\qb_{j, n}(\treen, \boldsymbol{\varphi}, \boldsymbol{\theta}, \ub)
\big{)}\right] u_{(0, 2\pi)}^{\otimes_{n-1}}(\ud \boldsymbol{\theta}) \ \ \ \ \ \ \label{eq:FourierCtnast} \\
&=& \int_{(0, 2\pi)^{n-1}} \left[\prod_{j=1}^{n} \hat{\mu}_0\big{(}\rho \pi_{j, n}^{\ast}(\treen, \boldsymbol{\varphi})
\mathrm{B}(\ub) \mathrm{O}_{j, n}^{\ast}(\treen, \boldsymbol{\varphi}, \boldsymbol{\theta}) \mathbf{e}_3
\big{)}\right] u_{(0, 2\pi)}^{\otimes_{n-1}}(\ud \boldsymbol{\theta}) \ \ \ \ \ \ \label{eq:FourierCtn}
\end{eqnarray}
hold true for every $n \geq 2$, $\treen$ in $\mathbb{T}(n)$, $\boldsymbol{\varphi}$ in $[0, \pi]^{n-1}$ and $\xib \neq \mathbf{0}$. The $\qb_{j, n}$'s are defined inductively starting from $\qb_{1, 1}(\tree_1, \emptyset, \emptyset, \ub) := \ub$ and then putting
\begin{eqnarray} \label{eq:psijnabstract}
&{}& \qb_{j, n}(\treen, \boldsymbol{\varphi}, \boldsymbol{\theta}, \ub) \nonumber \\
&=& \left\{ \begin{array}{ll}
\qb_{j, n_l}(\treen^l, \boldsymbol{\varphi}^l, \boldsymbol{\theta}^l, \psib^l(\varphi_{n-1}, \theta_{n-1}, \ub)) & \text{for} \ j = 1, \dots, n_l \\
\qb_{j - n_l, n_r}(\treen^r, \boldsymbol{\varphi}^r, \boldsymbol{\theta}^r, \psib^r(\varphi_{n-1}, \theta_{n-1}, \ub)) & \text{for} \ j = n_l + 1, \dots, n
\end{array} \right.
\end{eqnarray}
for every $n \geq 2$, $\treen$ in $\mathbb{T}(n)$, $\boldsymbol{\varphi}$ in $[0, \pi]^{n-1}$ and $\boldsymbol{\theta}$ in $(0, 2\pi)^{n-1}$.

To prove (\ref{eq:FourierCtnast}), first consider the case when $n = 2$ and observe that $\pi_{1, 2}^{\ast} = \cos\varphi_1$, $\pi_{2, 2}^{\ast} = \sin\varphi_1$, $\qb_{1, 2} = \psib^l$, $\qb_{2, 2} = \psib^r$. Then, (\ref{eq:FourierCtnast}) reduces to (\ref{eq:internalintegralbobylev}) with $\zeta = \eta = \mu_0$. Next, by mathematical induction, assume $n \geq 3$ and combine (\ref{eq:internalintegralbobylev}) with the definition of $\mathcal{C}_{\treen}$ to write
\begin{eqnarray}
&& \hat{\mathcal{C}}_{\treen}[\mu_0; \boldsymbol{\varphi}](\boldsymbol{\xi}) = \int_{0}^{2\pi} \hat{\mathcal{C}}_{\treen^l}[\mu_0; \boldsymbol{\varphi}^l](\rho \cos\varphi_{n-1} \boldsymbol{\psi}^l(\varphi_{n-1}, \theta_{n-1}, \ub))  \nonumber \\
&\times& \hat{\mathcal{C}}_{\treen^r}[\mu_0; \boldsymbol{\varphi}^r](\rho \sin\varphi_{n-1} \boldsymbol{\psi}^r(\varphi_{n-1}, \theta_{n-1}, \ub)) u_{(0, 2\pi)}(\ud \theta_{n-1}) \ . \label{eq:inductionCtn}
\end{eqnarray}
Thus, assuming that (\ref{eq:FourierCtnast}) holds true for every $m$ in $\{1, \dots, n-1\}$ and every tree $\tree_{m}$ in $\mathbb{T}(m)$, deduce
\begin{gather}
\hat{\mathcal{C}}_{\treen^s}[\mu_0; \boldsymbol{\varphi}^s](x \psib^s(\varphi_{n-1}, \theta_{n-1}, \ub)) = \int_{(0, 2\pi)^{n_s - 1}} \Big{\{}\prod_{j=1}^{n_s} \hat{\mu}_0 \big{[} x \pi_{j, n_s}^{\ast}(\treen^s, \boldsymbol{\varphi}^s) \times \nonumber \\
\times \qb_{j, n_s}(\treen^s, \boldsymbol{\varphi}^s, \boldsymbol{\theta}^s, \boldsymbol{\psi}^s(\varphi_{n-1}, \theta_{n-1}, \ub))\big{]} \Big{\}} u_{(0, 2\pi)}^{\otimes_{n_s - 1}}(\ud \boldsymbol{\theta}^s) \nonumber
\end{gather}
where $(s, x)$ is $(l, \rho \cos\varphi_{n-1})$ or $(r, \rho \sin\varphi_{n-1})$. To complete the argument, combine the last two  equalities with (\ref{eq:pijnast}) and (\ref{eq:psijnabstract}).

As far as the proof of (\ref{eq:FourierCtn}) is concerned, start by noting that $\qb_{j, 2}(\tree_2, \varphi, \theta, \ub)$ equals $\mathrm{B}(\ub)\mathrm{O}_{j, 2}^{\ast}(\tree_2, \varphi, \theta) \mathbf{e}_3$ for $j = 1, 2$, for every $\varphi$ in $[0, \pi]$, $\theta$ in $(0, 2\pi)$ and $\ub$ in $S^2$, provided that the basis $\{\mathbf{a}(\ub), \mathbf{b}(\ub), \ub\}$ in (\ref{eq:psi}) is formed by the three columns of $\mathrm{B}(\ub)$.
Then, assume $n \geq 3$ and argue by induction starting from (\ref{eq:FourierCtnast}), definitions (\ref{eq:Ojn}) and (\ref{eq:psijnabstract}). Whence,
\begin{gather}
\int_{(0, 2\pi)^{n-1}} \left[\prod_{j=1}^{n} \hat{\mu}_0\big{(} \rho  \pi_{j, n}^{\ast}(\treen, \boldsymbol{\varphi}) \qb_{j, n}(\treen, \boldsymbol{\varphi}, \boldsymbol{\theta}, \ub)\big{)}\right] u_{(0, 2\pi)}^{\otimes_{n-1}}(\ud \boldsymbol{\theta}) \nonumber \\
= \int_{0}^{2\pi} \! \int_{(0, 2\pi)^{n_l-1}} \int_{(0, 2\pi)^{n_r-1}} P_{j, n}^{l} P_{j, n}^{r} u_{(0, 2\pi)}^{\otimes_{n_r - 1}}(\ud \boldsymbol{\theta}^r)  u_{(0, 2\pi)}^{\otimes_{n_l - 1}}(\ud \boldsymbol{\theta}^l) u_{(0, 2\pi)}(\ud  \theta_{n-1})\label{eq:intermediatepsiastpsi}
\end{gather}
where
$$
P_{j, n}^{l} := \prod_{j=1}^{n_l} \hat{\mu}_0\Big{(} \rho  \pi_{j, n}^{\ast}(\treen, \boldsymbol{\varphi}) \mathrm{B}(\psib^l(\varphi_{n-1}, \theta_{n-1}, \ub)) \mathrm{O}_{j, n_l}^{\ast}(\treen^l, \boldsymbol{\varphi}^l, \boldsymbol{\theta}^l) \mathbf{e}_3 \Big{)}
$$
and
$$
P_{j, n}^{r} := \prod_{j=n_l + 1}^{n} \hat{\mu}_0\Big{(} \rho  \pi_{j, n}^{\ast}(\treen, \boldsymbol{\varphi})
\mathrm{B}(\psib^r(\varphi_{n-1}, \theta_{n-1}, \ub)) \mathrm{O}_{j-n_l, n_r}^{\ast}(\treen^r, \boldsymbol{\varphi}^r, \boldsymbol{\theta}^r) \mathbf{e}_3 \Big{)} \ .
$$
For the sake of clarity, the integral $\int_{(0, 2\pi)^{n_l-1}}$ ($\int_{(0, 2\pi)^{n_r-1}}$, respectively) in (\ref{eq:intermediatepsiastpsi}) should not be written if $n_l = 1$ ($n_r = 1$, respectively) since $\boldsymbol{\theta}^l$ ($\boldsymbol{\theta}^r$, respectively) corresponds to the empty set. At this stage, it will be proved that
\begin{gather}
\int_{(0, 2\pi)^{n_l-1}} P_{j, n}^{l} u_{(0, 2\pi)}^{\otimes_{n_l - 1}}(\ud \boldsymbol{\theta}^l) \nonumber \\
= \int_{(0, 2\pi)^{n_l-1}} \prod_{j=1}^{n_l} \hat{\mu}_0\Big{(} \rho  \pi_{j, n}^{\ast}(\treen, \boldsymbol{\varphi})
\mathrm{B}(\ub) \mathrm{O}_{j, n}^{\ast}(\treen, \boldsymbol{\varphi}, \boldsymbol{\theta}) \mathbf{e}_3
\Big{)} u_{(0, 2\pi)}^{\otimes_{n_l - 1}}(\ud \boldsymbol{\theta}^l) \label{eq:robben}
\end{gather}
holds for every $\rho$ in $\rone$, $\ub$ in $S^2$, $\boldsymbol{\varphi}$ in $[0, \pi]^{n-1}$ and $\theta_{n-1}$ in $(0, 2\pi)$. If $n_l = 1$, the proof of (\ref{eq:robben}) reduces to verify that
$$
\mathrm{B}(\psib^l(\varphi_{n-1}, \theta_{n-1}, \ub)) \mathbf{e}_3 = \psib^l(\varphi_{n-1}, \theta_{n-1}, \ub) = \mathrm{B}(\ub) \mathrm{M}^l(\varphi_{n-1}, \theta_{n-1}) \mathbf{e}_3\ .
$$
To proceed, since the third column of $\mathrm{B}(\boldsymbol{\psi}^l(\varphi_{n-1}, \theta_{n-1}, \ub))$ is the same as that of $\mathrm{B}(\ub) \mathrm{M}^l(\varphi_{n-1}, \theta_{n-1})$, then there exists an orthogonal matrix
$$
\mathrm{R}(\alpha) := \left( \begin{array}{ccc}
\cos\alpha & -\sin\alpha & 0 \\
\sin\alpha & \cos\alpha & 0 \\
0 & 0 & 1 \\
\end{array} \right)
$$
for which $\mathrm{B}(\boldsymbol{\psi}^l(\varphi_{n-1}, \theta_{n-1}, \ub)) = \mathrm{B}(\ub) \mathrm{M}^l(\varphi_{n-1}, \theta_{n-1}) \mathrm{R}(\alpha)$, where $\alpha$ depends only on $\varphi_{n-1}$, $\theta_{n-1}$ and $\ub$. Now, note that $\mathrm{R}(\alpha) \mathrm{M}^s(\varphi, \theta) = \mathrm{M}^s(\varphi, \theta + \alpha)$ is valid for $s = l, r$ and for every $\varphi$ and $\theta$. Then, when $n_l \geq 2$, consider the definition of $P_{j,n}^{l}$, recall (\ref{eq:pathOijn}) and       take account that the product $\mathrm{R}(\alpha)\mathrm{M}^{\epsilon_{1}(\treen^l, j)}(\varphi_{n_l-1}, \theta_{n_l-1})$ equals $\mathrm{M}^{\epsilon_{1}(\treen^l, j)}(\varphi_{n_l-1}, \theta_{n_l-1} + \alpha)$. The change of variable $\theta_{n_l-1}^{'} = \theta_{n_l-1} + \alpha$ transforms the LHS of (\ref{eq:robben}) into
$$
\int_{(0, 2\pi)^{n_l-1}} \prod_{j=1}^{n_l} \hat{\mu}_0\big{(} \rho  \pi_{j, n}^{\ast}(\treen, \boldsymbol{\varphi})  \mathrm{B}(\ub) \mathrm{M}^l(\varphi_{n-1}, \theta_{n-1}) \mathrm{O}_{j, n_l}^{\ast}(\treen^l, \boldsymbol{\varphi}^l, \boldsymbol{\theta}^l) \mathbf{e}_3 \big{)}  u_{(0, 2\pi)}^{\otimes_{n_l - 1}}(\ud \boldsymbol{\theta}^l)
$$
which, in view of (\ref{eq:Ojn}), turns out to be the same as the RHS of (\ref{eq:robben}). The proof of (\ref{eq:FourierCtn}) is completed using (\ref{eq:intermediatepsiastpsi}), after noting that an equality similar to (\ref{eq:robben}) can be stated by changing subscripts and superscripts from $l$ to $r$, and replacing $\mathrm{O}_{j, n}^{\ast}$ with $\mathrm{O}_{j + n_l, n}^{\ast}$.

Finally, the invariance of $\Mrand$ w.r.t. $\mathrm{B}$ is equivalent to the invariance of representation (\ref{eq:FourierCtn}) when $\mathrm{B}(\ub)$ is replaced by any matrix $\mathrm{B}^{'}(\ub)$ having the same characteristics as $\mathrm{B}(\ub)$. Anyway, such an equivalence follows from the above reasoning.

\subsection{Proof of Theorem \ref{thm:CDGR}} \label{sect:proof}

In the first place, we recall that the entire proof will be developed under hypotheses (\ref{eq:bsymm})-(\ref{eq:cutoff}) on $b$, in view of Remark 1 in Subsection \ref{sect:mainresult}. Then, we set a few conditions on $\mu_0$ to simplify a number of arguments without loss of generality. In this sense, we make use of (\ref{eq:Vu0}) to assume, from now on,
\begin{equation} \label{eq:normalizations}
\intethree \vb \mu_0(\ud \vb) = \mathbf{0} \ \ \ \ \ \text{and} \ \ \ \ \ \intethree |\vb|^2 \mu_0(\ud \vb) = 3
\end{equation}
implying that the limiting Maxwellian is $\gamma := \gamma_{\mathbf{0}, 1}$. We also assume that the covariance matrix $V = V[\mu_0]$ of $\mu_0$ is diagonal. In fact, since for any covariance matrix $V$ there is an orthogonal matrix $Q$ such that $Q V Q^t$ is diagonal, then $\mu_0 \circ f_{Q}^{-1}$ has a diagonal covariance matrix, $f_Q$ standing for the function $\xb \mapsto Q\xb$. At this stage, since $\vartot\big{(} \mu(\cdot, t) \circ f_{Q}^{-1}, \gamma \big{)}$ is equal to
$\vartot\big{(} \mu(\cdot, t), \gamma \big{)}$ for every $t$, we can prove (\ref{eq:CDGR}) by taking $\mu_0 \circ f_{Q}^{-1}$ as initial distribution. Compare \cite{Dbasic} for a more detailed explanation. Hence, we suppose that
\begin{equation} \label{eq:covariancematrix}
\begin{array}{llll}
& \intethree v_{i}^{2} \mu_0(\ud \vb) &= \sigma_{i}^{2} & \ \ \ (i = 1, 2, 3) \\
& \intethree v_i v_j \mu_0(\ud \vb) &= 0 &\ \ \ (1 \leq i < j \leq 3) \\
& \sigma_{1}^{2} + \sigma_{2}^{2} + \sigma_{3}^{2} &= 3 & {}
\end{array}
\end{equation}
are in force. In fact, extra-conditions (\ref{eq:normalizations})-(\ref{eq:covariancematrix}) yield the following
\begin{prop} \label{prop:tails}
\emph{Let} $\mu_0$ \emph{satisfy} (\ref{eq:normalizations})-(\ref{eq:covariancematrix}) \emph{in addition to the hypotheses of Theorem} \ref{thm:CDGR}. \emph{Then, there exists a constant} $\lambda$ \emph{such that}
\begin{equation} \label{eq:globtails}
|\hat{\mu}_0(\xib)| \leq \left(\frac{\lambda^2}{\lambda^2
+ |\xib|^2} \right)^q
\end{equation}
\emph{is valid for every} $\boldsymbol{\xi}$ \emph{in} $\rthree$, \emph{with} $q = 1/(2 \lceil 2/p \rceil)$.
\end{prop}
Here, $\lceil x \rceil$ indicates the least integer not less than $x$, while $p$ is the same as in (\ref{eq:tailsCDGR}). As to the numerical evaluation of $\lambda$, the reader is referred to the proof of the proposition in \ref{a:tails}. \\

As first step of the real proof, an application of (\ref{eq:main}) yields
$$
\vartot(\mu(\cdot, t), \gamma) = \sup_{B \in \borelthree} \big{|} \et[\Mrand(B)] - \gamma(B) \big{|} \leq \et[\vartot(\Mrand, \gamma)] \ .
$$
After introducing the random number
\begin{equation} \label{eq:W}
\WW := \sum_{j = 1}^{\nu} \pi_{j, \nu}^4
\end{equation}
we put
\begin{equation} \label{eq:ra}
r := 11 \lceil 2/p \rceil \ \ \ \ \ \ \ \ \text{and}\ \ \ \ \ \ \ \ \as := (2^r r!)^{-1}
\end{equation}
to define the partition $\{U, U^c\}$ of $\Omega$ by
$$
U := \{\nu \leq r\} \cup \{\prod_{j=1}^{\nu} \pi_{j, \nu} = 0\} \cup \{\WW \geq \as\} \ .
$$
This can be used to write
\begin{equation} \label{eq:split}
\et[\vartot(\Mrand, \gamma)] = \et[\vartot(\Mrand, \gamma); U] + \et[\vartot(\Mrand, \gamma); U^c]
\end{equation}
where $\et[X; S]$ denotes $\int_{S} X \ud \pt$. The former summand on the right of (\ref{eq:split}) will be bounded by utilizing the fact that $U$ has ``asymptotically small'' probability. As to the latter, it will be shown that $\Mrand(\cdot; \omega)$ has nice analytical properties for each $\omega$ in $U^c$, so that a proper bound will be derived from these very same properties. In fact, as $\vartot(\Mrand, \gamma) \leq 1$ entails $\et[\vartot(\Mrand, \gamma); U] \leq \pt(U)$, we get
$$
\begin{array}{ll}
\pt(U) &\leq \pt\{\nu \leq r\} + \pt\{\prod_{j=1}^{\nu} \pi_{j, \nu} = 0\} + \pt\{\WW \geq \as\} \\
&\leq r e^{-t} + \pt\{\WW \geq \as\} \ .
\end{array}
$$
The inequality $\pt\{\nu \leq r\} \leq r e^{-t}$ follows from (\ref{eq:nu}), while $\pt\{\prod_{j=1}^{\nu} \pi_{j, \nu} = 0\}$ equals zero since $\pt\{\prod_{j=1}^{\nu} \pi_{j, \nu} = 0 \ | \ \nu, \tau_{\nu}\} = 0$. This claim is obvious on $\{\nu = 1\}$ while, on $\{\nu \geq 2\}$,
$$
\pt\{\prod_{j=1}^{\nu} \pi_{j, \nu} = 0 \ | \ \nu, \tau_{\nu}\} \leq \sum_{j = 1}^{\nu - 1} \pt\{\phi_j \in \{0, \pi/2, \pi\}\}
$$
and the RHS is equal to zero since each $\phi_j$ has an absolutely continuous law. To complete the evaluation of $\pt(U)$,
it is enough to combine the Markov inequality with (\ref{eq:gare})-(\ref{eq:l4spectral}) to get $\pt\{\WW \geq \as\} \leq (1/\as) \ \et[\WW] = (2^r r!) \ e^{\Lambda_b t}$. Whence,
\begin{equation} \label{eq:bad}
\et[\vartot(\Mrand, \gamma); U] \leq (r + 2^r r!) e^{\Lambda_b t} \ .
\end{equation}

The argument to deduce a bound for the expectation over $U^c$ occupies the rest of this subsection. It is based on the following multidimensional extension of a result by Beurling \cite{beu}.
\begin{prop} \label{prop:beurling}
\emph{Let} $\chi$ \emph{be a finite signed measure on} $\rthreeb$ \emph{such that} $\intethree |\xb|^2 |\chi|(\ud\xb) < +\infty$, $|\chi|$ \emph{standing for the total variation of} $\chi$. \emph{Then,}
$$
\sup_{B \in \borelthree} |\chi(B)| \leq 2^{-5/4}\pi^{-1/2} \left(\int_{\rthree} [|\hat{\chi}(\boldsymbol{\xi})|^2 + |\Delta_{\xib} \hat{\chi}(\boldsymbol{\xi})|^2] \ud \boldsymbol{\xi} \right)^{1/2}
$$
\emph{where} $\Delta_{\xib}$ \emph{denotes the Laplacian operator.}
\end{prop}
The proof is deferred to \ref{a:beurling}. The applicability of this proposition to $\chi = \Mrand - \gamma$ is made possible by
\begin{prop} \label{prop:elmroth}
\emph{If} (\ref{eq:normalizations}) \emph{holds and}
\begin{equation}
\mathfrak{m}_h := \intethree |\vb|^h \mu_0(\ud \vb) < +\infty
\end{equation}
\emph{for} $h = 1, \dots, 2k$ \emph{and some integer} $k \geq 2$, \emph{then there are positive constants} $g_h$ \emph{depending on} $\mu_0$ \emph{only through} $\mathfrak{m}_h$, \emph{such that}
\begin{equation} \label{eq:momlS}
\sup_{\ub \in S^2}\et\left[|S(\ub)|^h \ | \ \mathscr{H} \right] \leq g_h
\end{equation}
\emph{for} $h = 1, \dots, 2k$ \emph{and any choice of} $\mathrm{B}$, $\pt$-\emph{almost surely. Moreover,} $\rho \mapsto \frac{\partial^h}{\partial \rho^h} \hat{\Mrand}(\rho \ub)$ \emph{exists for every} $\ub$ \emph{in} $S^2$ \emph{and}
\begin{equation}\label{eq:boundderivativesMhat}
\sup_{(\rho, \ub) \in [0, +\infty) \times S^2}\Big{|}\frac{\partial^h}{\partial \rho^h} \hat{\Mrand}(\rho \ub)\Big{|} \leq g_h
\end{equation}
$\pt$-\emph{almost surely with} $h = 1, \dots, 2k$, \emph{which entails}
\begin{equation} \label{eq:finmom2M}
\intethree |\vb|^{2k} \Mrand(\ud\vb) < +\infty
\end{equation}
$\pt$-\emph{almost surely and} $\xib \mapsto \hat{\Mrand}(\xib) \in \mathrm{C}^{2k}(\rthree)$.
\end{prop}
See \ref{a:elmroth} for the proof and the numerical evaluation of the constants $g_h$. At this stage, Proposition \ref{prop:beurling} yields
\begin{gather}
\et[\vartot(\Mrand, \gamma); U^c] \leq 2^{-5/4}\pi^{-1/2} \times \nonumber \\
\times \et\left[\left(\intethree \big{|}\hat{\mathcal{M}}(\xib) - e^{-|\xib|^2/2}\big{|}^2 \dxi + \intethree \big{|}\Delta_{\xib}[\hat{\mathcal{M}}(\xib) - e^{-|\xib|^2/2}]\big{|}^2 \dxi \right)^{1/2}; U^c\right] \ . \label{eq:appliedbeurling}
\end{gather}
To evaluate the integrals on the RHS, we change the variables according to the isometry $i : \rthree\setminus\{\mathbf{0}\} \rightarrow (0, +\infty) \times S^2$ defined by $i : \xib \mapsto (|\xib|, \xib/|\xib|)$. In view of Theorem 3.11, Example 3.23 and Lemma 3.27 in \cite{gri}, denoting the $d$-dimensional Lebesgue measure by $\mathscr{L}^d$, integrals with respect to $\mathscr{L}^3(\ud \xib)$ become integrals with respect to $4\pi \rho^2\mathscr{L}^1 \otimes u_{S^2}(\ud\rho \ud\ub)$ and the standard Laplacian $\Delta_{\xib}$ changes into $\Delta_{(\rho, \ub)} := \frac{\partial^2}{\partial \rho^2} + \frac{2}{\rho}\frac{\partial}{\partial \rho} + \frac{1}{\rho^2}\Delta_{S^2}$, where $\Delta_{S^2}$ stands for the Laplace-Beltrami operator on $S^2$. Now, from $|z_1 + z_2 + z_3|^2 \leq 3(|z_1|^2 + |z_2|^2 + |z_3|^2)$, we write
\begin{eqnarray}
\big{|}\Delta_{(\rho, \ub)}[\hat{\Mrand}(\rho \ub) - e^{-\rho^2/2}] \big{|}^2 &\leq& 3 \ \Big{|} \frac{\partial^2}{\partial \rho^2} [\hat{\Mrand}(\rho \ub) - e^{-\rho^2/2}] \Big{|}^2 \nonumber \\
&+& \frac{12}{\rho^2} \ \Big{|} \frac{\partial}{\partial \rho} [\hat{\Mrand}(\rho \ub) - e^{-\rho^2/2}] \Big{|}^2
+ \frac{3}{\rho^4} \ \big{|}\Delta_{S^2} \hat{\Mrand}(\rho \ub) \big{|}^2 \nonumber
\end{eqnarray}
and then we define the random functions
\begin{eqnarray}
\IIr &:=& \big{|} \hat{\Mrand}(\rho \ub) - e^{-\rho^2/2} \big{|}^2 + 3 \ \Big{|} \frac{\partial^2}{\partial \rho^2} [\hat{\Mrand}(\rho \ub) - e^{-\rho^2/2}] \Big{|}^2 \nonumber \\
&+& \frac{12}{\rho^2} \ \Big{|} \frac{\partial}{\partial \rho} [\hat{\Mrand}(\rho \ub) - e^{-\rho^2/2}] \Big{|}^2 \nonumber \\
\IIs &:=& \frac{3}{\rho^4} \ \big{|}\Delta_{S^2} \hat{\Mrand}(\rho \ub) \big{|}^2 \ . \nonumber
\end{eqnarray}
Hence, for the sum of the two integrals on the RHS of (\ref{eq:appliedbeurling}) we obtain
\begin{eqnarray}
&{}& \intethree \big{|}\hat{\mathcal{M}}(\xib) - e^{-|\xib|^2/2}\big{|}^2 \dxi + \intethree \big{|}\Delta_{\xib}[\hat{\mathcal{M}}(\xib) - e^{-|\xib|^2/2}]\big{|}^2 \dxi \nonumber \\
&\leq& 4\pi \int_{[0, \RR] \times S^2} \left(\IIr + \IIs\right) \rho^2\mathscr{L}^1 \otimes u_{S^2}(\ud\rho \ud\ub) \nonumber \\
&+& 4\pi \int_{(\RR, +\infty) \times S^2} \left(\IIr + \IIs\right) \rho^2\mathscr{L}^1 \otimes u_{S^2}(\ud\rho \ud\ub) \label{eq:splitintegralI}
\end{eqnarray}
where
\begin{equation} \label{eq:r}
\RR := \frac{1}{2} \left(\frac{1}{\mq \WW}\right)^{1/4} \ .
\end{equation}
In the following sub-subsections we analyze the integrals appearing in (\ref{eq:splitintegralI}), calling \emph{inner} (\emph{outer}, respectively) any integral on $[0, \RR] \times S^2$ ($(\RR, +\infty) \times S^2$, respectively).

\subsubsection{Outer integral of $\IIr$} \label{sect:outer}

An application of the inequality $|z_1 + z_2|^2 \leq 2|z_1|^2 + 2|z_2|^2$ yields
\begin{eqnarray}
\IIr &\leq& 2 |\hat{\Mrand}(\rho \ub)|^2 + 2e^{-\rho^2} + 6 \Big{|}\frac{\partial^2}{\partial \rho^2} \hat{\Mrand}(\rho \ub)\Big{|}^2 + 6 \Big{|}\frac{\ud^2}{\ud \rho^2} (e^{-\rho^2/2})\Big{|}^2 \nonumber \\
&+& \frac{24}{\rho^2} \Big{|}\frac{\partial}{\partial \rho} \hat{\Mrand}(\rho \ub)\Big{|}^2 +
\frac{24}{\rho^2} \Big{|}\frac{\ud}{\ud \rho} (e^{-\rho^2/2})\Big{|}^2 \label{eq:splitIext}
\end{eqnarray}
and a first proposition is given to analyze those summands which contain the Gaussian c.f..
\begin{prop} \label{prop:extgauss}
\emph{Let} $m, s, k$ \emph{be real numbers such that} $m \geq 0$, $s \geq 1$ \emph{and} $k$ \emph{in} $\mathbb{N}_0$. \emph{Then, there exists a positive constant} $c(m, s, k)$ \emph{such that}
$$
\int_{x}^{+\infty} \left(\frac{\ud^k}{\ud \rho^k} (e^{-\rho^2/2})\right)^2 \rho^m \ud \rho \leq c(m, s, k) x^{-s}
$$
\emph{holds for every} $x > 0$.
\end{prop}
See \ref{a:gauss} for the proof and an evaluation of $c(m, s, k)$. At this stage, applying successively the above statement with $(x, m, s, k) = (\RR, 2, 8, 0), (\RR, 0, 8, 1)$ and $(\RR, 2, 8, 2)$ gives
\begin{equation} \label{eq:extgaussian}
\int_{\RR}^{+\infty} \Big{\{} 2e^{-\rho^2} +  6 \Big{|}\frac{\ud^2}{\ud \rho^2} (e^{-\rho^2/2})\Big{|}^2
+ \frac{24}{\rho^2} \Big{|}\frac{\ud}{\ud \rho} (e^{-\rho^2/2})\Big{|}^2 \Big{\}} \rho^2 \ud\rho \leq\ \overline{C}_1 \RR^{-8}
\end{equation}
with $\overline{C}_1 := [2c(2, 8, 0) + 6c(2, 8, 2) + 24c(0, 8, 1)]$.

Then we study those terms on the RHS of (\ref{eq:splitIext}) which depend on $\Mrand$ making use of the next proposition, whose statement involves the random function
\begin{equation}\label{eq:psigrande}
\Psi(\rho) := \prod_{j = 1}^{\nu} \left(\frac{\lambda^2}{\lambda^2 + \rho^2 \pi_{j, \nu}^2}\right)^q
\end{equation}
with $\lambda$ and $q$ as in Proposition \ref{prop:tails}.
\begin{prop} \label{prop:boundderivativesMhat}
\emph{If} (\ref{eq:fourthmom})-(\ref{eq:tailsCDGR}) \emph{and} (\ref{eq:normalizations})-(\ref{eq:covariancematrix}) \emph{are in force}, \emph{then}
\begin{equation} \label{eq:boundPsiN}
\sup_{\ub \in S^2} \big{|}\hat{\Nrand}(\rho; \ub)\big{|} \ \leq\ \Psi(\rho)
\end{equation}
\emph{and}
\begin{equation} \label{eq:boundPsi}
\sup_{\ub \in S^2} \big{|}\hat{\Mrand}(\rho \ub)\big{|}\ \leq\ \Psi(\rho)
\end{equation}
\emph{hold for every} $\rho$ \emph{in} $[0, \RR]$, \emph{with the exception of a set of} $\pt$-\emph{probability zero. Moreover, there are two non-random polynomials} $\wp_1$ \emph{and} $\wp_2$ \emph{of degree} 2 \emph{and} 4 \emph{respectively, with positive coefficients depending only on} $\mu_0$, \emph{such that}
\begin{equation} \label{eq:boundderinternalN}
\sup_{\ub \in S^2} \Big{|}\frac{\partial^k}{\partial \rho^k}\hat{\Nrand}(\rho; \ub)\Big{|}\ \leq\ \wp_k(\rho) \Psi(\rho)
\end{equation}
\emph{and}
\begin{equation} \label{eq:boundderinternal}
\sup_{\ub \in S^2} \Big{|}\frac{\partial^k}{\partial \rho^k}\hat{\Mrand}(\rho \ub)\Big{|}\ \leq\ \wp_k(\rho) \Psi(\rho)
\end{equation}
\emph{hold for every} $\rho$ \emph{in} $[0, \RR]$, \emph{with the exception of a set of} $\pt$-\emph{probability zero.}
\end{prop}
A complete characterization of $\wp_1$ and $\wp_2$ is given in the course of the proof of this proposition, in \ref{a:boundderivativesMhat}. For the sake of completeness, we observe that (\ref{eq:boundPsiN}) and (\ref{eq:boundderinternalN}) hold true for any choice of $\mathrm{B}$ in (\ref{eq:psijn}).

One of the advantages of the splitting (\ref{eq:split}) consists in the fact that all the realizations of $\Psi$ on $U^c$ share a property of uniform integrability, as shown in the following
\begin{prop} \label{prop:newtontrick}
\emph{Over} $U^c$, \emph{the inequality}
\begin{equation} \label{eq:newtontrick}
\prod_{j=1}^{\nu} \left(1 + \pi_{j, \nu}^{2} x^2\right) \geq\ \epsilon x^{2r}
\end{equation}
\emph{is valid for every} $x > 0$, \emph{with} $\epsilon := (2 r!)^{-1}$ \emph{and} $r$ \emph{given by} (\ref{eq:ra}). \emph{Therefore,}
\begin{equation} \label{eq:intenewtonm}
\sup_{\omega \in U^c} \int_{x}^{+\infty} \Psi^s(\rho) \rho^m \ud \rho \leq\ \frac{1}{2rqs - m - 1}\left(\frac{\lambda^{2r}}{\epsilon}\right)^{qs} x^{-2rqs + m + 1}
\end{equation}
\emph{holds true for every} $x > 0$, $s > 0$ \emph{and} $m < (2rqs - 1)$.
\end{prop}
See \ref{a:newton} for the proof. We are now in a position to complete the study of the outer integral of $\IIr$. First, taking into account that $2rq = 11$, combination of (\ref{eq:boundPsi}) with (\ref{eq:intenewtonm}) yields
\begin{equation} \label{eq:intextM}
\int_{\RR}^{+\infty} |\hat{\Mrand}(\rho \ub)|^2 \rho^2 \ud \rho \leq \int_{\RR}^{+\infty} \Psi^2(\rho) \rho^2 \ud \rho \leq\ \frac{1}{19}\left(\frac{\lambda^{2r}}{\epsilon}\right)^{2q} \RR^{-19} \ .
\end{equation}
The applicability of (\ref{eq:intenewtonm}) is guaranteed by the fact that, when $s = 2$ and $m = 2$, one has $m < 4rq - 1 = 21$. Second, since (\ref{eq:boundderivativesMhat}), (\ref{eq:boundPsi}) and (\ref{eq:newtontrick}) entail
$$
\lim_{y \rightarrow +\infty} \hat{\Mrand}(-y \ub) \left[\frac{\partial}{\partial \rho} \hat{\Mrand}(\rho \ub)\right]_{\rho = y} = 0
$$
on $U^c$, after integrating by parts we get
$$
\int_{\RR}^{+\infty} \Big{|}\frac{\partial}{\partial \rho} \hat{\Mrand}(\rho \ub)\Big{|}^2 \ud \rho \leq g_1 \Psi(\RR) + g_2 \int_{\RR}^{+\infty} \Psi(\rho) \ud \rho \ .
$$
Thus, (\ref{eq:newtontrick})-(\ref{eq:intenewtonm}) with $s=1$ and $m=0$ lead to
\begin{equation} \label{eq:intextM1}
\int_{\RR}^{+\infty} \Big{|}\frac{\partial}{\partial \rho} \hat{\Mrand}(\rho \ub)\Big{|}^2 \ud \rho \leq
\left(\frac{\lambda^{2r}}{\epsilon}\right)^q \cdot \left(g_1 \RR^{-11} + \frac{g_2}{10} \RR^{-10}\right) \ .
\end{equation}
To study the last integral, we recall that $\prod_{j=1}^{\nu} \pi_{j, \nu} \neq 0$ on $U^c$ and then combine (\ref{eq:boundPsi}) with (\ref{eq:boundderinternal})-(\ref{eq:newtontrick}) to prove that
\begin{eqnarray}
\lim_{y \rightarrow +\infty} y^2 \left[\frac{\partial^2}{\partial \rho^2} \hat{\Mrand}(\rho \ub) \cdot \left( \frac{\partial}{\partial \rho} \hat{\Mrand}(-\rho \ub)\right)\right]_{\rho = y} &=& 0 \nonumber \\
\lim_{y \rightarrow +\infty} y^2 \hat{\Mrand}(-y \ub) \left[\frac{\partial^3}{\partial \rho^3} \hat{\Mrand}(\rho \ub)\right]_{\rho = y} &=& 0 \nonumber \\
\lim_{y \rightarrow +\infty} y \hat{\Mrand}(-y \ub) \left[\frac{\partial^2}{\partial \rho^2} \hat{\Mrand}(\rho \ub)\right]_{\rho = y} &=& 0 \ . \nonumber
\end{eqnarray}
At this stage, after two integrations by parts, we have
\begin{eqnarray}
&{}& \int_{\RR}^{+\infty} \Big{|}\frac{\partial^2}{\partial \rho^2} \hat{\Mrand}(\rho \ub)\Big{|}^2 \rho^2\ud \rho \leq \RR^2 \wp_1(\RR)\wp_2(\RR)\Psi^2(\RR) + (g_3\RR^2 + 2g_2\RR) \Psi(\RR) \nonumber \\
&+& \int_{\RR}^{+\infty} (g_4 \rho^2 + 4g_3 \rho + 2g_2) \Psi(\rho) \ud \rho \nonumber
\end{eqnarray}
and, in view of Proposition \ref{prop:newtontrick}, the above RHS is bounded by
\begin{eqnarray}
&{}& \left(\frac{\lambda^{2r}}{\epsilon}\right)^{2q} \RR^{-20}\wp_1(\RR)\wp_2(\RR) + \big{(} g_3\RR^2 + 2g_2\RR \big{)} \cdot \left(\frac{\lambda^{2r}}{\epsilon}\right)^q \RR^{-11} \nonumber \\
&+& \left(\frac{\lambda^{2r}}{\epsilon}\right)^q \cdot \Big{(} \frac{g_4}{8} \RR^{-8} + \frac{4g_3}{9} \RR^{-9} + \frac{g_2}{5} \RR^{-10}\Big{)} \ . \label{eq:intextM2}
\end{eqnarray}

The final bound can be obtained, via the Tonelli theorem, by starting from (\ref{eq:splitIext}) and collecting the upper bounds in (\ref{eq:extgaussian}) and (\ref{eq:intextM})-(\ref{eq:intextM2}). Indeed, these last upper bounds are independent of $\ub$ and are expressed as sums of powers of $\RR$, of order less than or equal to $-8$. Therefore, recalling (\ref{eq:r}) and the inequality $\WW \leq 1$, we obtain
\begin{equation} \label{eq:finalextM}
4\pi \int_{(\RR, +\infty) \times S^2}  \IIr \rho^2\mathscr{L}^1 \otimes u_{S^2}(\ud\rho \ud\ub) \leq C_{\ast, r} \WW^2
\end{equation}
with
\begin{eqnarray}
C_{\ast, r} &:=& 4\pi \Big{\{} 2^8 \overline{C}_1 \mq^2 + \left(\frac{\lambda^{2r}}{\epsilon}\right)^{2q} \cdot \Big{(}
\frac{1}{19} 2^{20} \mq^{19/4} \nonumber \\
&+& 6 \cdot 2^{20} \mq^5 \wp_1(2^{-1}\mq^{-1/4}) \wp_2(2^{-1}\mq^{-1/4}) \Big{)} + \left(\frac{\lambda^{2r}}{\epsilon}\right)^q \cdot \Big{(}6 \cdot 2^5 \mq^2g_4
\nonumber \\
&+& \frac{26}{3} 2^9 \mq^{9/4}g_3 + \frac{78}{5} 2^{10} \mq^{5/2}g_2 + 24 \cdot 2^{11} \mq^{11/4}g_1 \Big{)}\Big{\}} \nonumber \ .
\end{eqnarray}

\subsubsection{Outer integral of $\IIs$} \label{sect:outerL}

As first step, we use the Tonelli theorem to write the outer integral of $\IIs$ as
\begin{equation}\label{eq:TonelliL}
\lim_{y \rightarrow +\infty} \int_{\RR}^{y} \frac{3}{\rho^2} \left(\int_{S^2} \big{|}\Delta_{S^2} \hat{\Mrand}(\rho \ub)\big{|}^2 \unifSu \right) \ud \rho \ .
\end{equation}
Then, we apply Theorem 3.16 in \cite{gri} to obtain
$$
\int_{S^2} \big{|}\Delta_{S^2} \hat{\Mrand}(\rho \ub)\big{|}^2 \unifSu = \int_{S^2} \hat{\Mrand}(-\rho \ub)
\Delta_{S^2}^2 \hat{\Mrand}(\rho \ub) \unifSu
$$
which, by virtue of (\ref{eq:boundPsi}), yields
\begin{equation}\label{eq:intpartsS2}
\int_{S^2} \big{|}\Delta_{S^2} \hat{\Mrand}(\rho \ub)\big{|}^2 \unifSu \leq\ \Psi(\rho) \sup_{\ub \in S^2}\big{|}\Delta_{S^2}^2 \hat{\Mrand}(\rho \ub)\big{|}\ .
\end{equation}
At this stage, to handle the computations involving the Laplace-Beltrami operator, we define the following plane domains
\begin{eqnarray}
D_1 = D_3 &:=& \{(u, v) \in \rone^2 \ | \ (u - \pi/2)^2/(5\pi)^2 + (v - \pi)^2/(11\pi)^2 < (1/12)^2\} \nonumber \\
D_2 = D_4 &:=& \{(u, v) \in \rone^2 \ | \ (u - \pi/2)^2/(5\pi)^2 + v^2/(11\pi)^2 < (1/12)^2\} \nonumber
\end{eqnarray}
along with the parametrizations
\begin{equation} \label{eq:parametrizationS2}
\begin{array}{lll}
\hb_k : D_k \ni (u, v) &\mapsto (\cos v \sin u, \sin v \sin u, \cos u) \in \rthree & k = 1, 2 \\
\hb_k : D_k \ni (u, v) &\mapsto (\cos u, \cos v \sin u, \sin v \sin u) \in \rthree & k = 3, 4
\end{array}
\end{equation}
to form the atlas $\mathcal{A}$ on $S^2$ composed by the charts $\Omega_k := \hb_k(D_k) \subset S^2$ for $k = 1, \dots, 4$.
Then, $\Delta_{S^2}^2$ can be expressed in local coordinates as
\begin{eqnarray}
\Delta_{(u, v)}^{2} &=& \partial_{u u u u} + 2\cot u \partial_{u u u} - \sin^{-2}u \partial_{u u} + \sin^{-2}u\cot u \partial_u + \sin^{-4}u \partial_{v v v v} \nonumber \\
&-& 2\sin^{-4}u (2 - \sin^{2}u) \partial_{v v} + 6 \sin^{-2}u\cot u \partial_{u v v} + 2\sin^{-2}u \partial_{u u v v} \nonumber
\end{eqnarray}
by virtue of (3.84) in \cite{gri}, and hence
\begin{gather}
\sup_{\ub \in S^2} \big{|} \Delta_{S^2}^{2}\hat{\mathcal{M}}(\rho\ub) \big{|} = \sup_{k \in \{1, \dots, 4\}} \sup_{(u, v) \in D_k} \big{|} \Delta_{(u, v)}^{2} \hat{\Mrand}(\rho \hb_k(u, v)) \big{|} \nonumber \\
\leq \overline{\Delta} \sum_{1 \leq |\boldsymbol{\alpha}| \leq 4} \sup_{k \in \{1, \dots, 4\}} \sup_{(u, v) \in D_k} \big{|} \partial_{\boldsymbol{\alpha}} \hat{\Mrand}(\rho \hb_k(u, v)) \big{|} \label{eq:coordlaplaceM}
\end{gather}
where $\boldsymbol{\alpha}$ indicates the multi-index $(\alpha_1, \alpha_2)$, $\partial_{\boldsymbol{\alpha}}$ stands for the partial derivative $\frac{\partial^{\alpha_1 + \alpha_2}}{\partial u^{\alpha_1} \partial v^{\alpha_2}}$, and $\overline{\Delta} = 4 (2 + \sqrt{3})^2 (6 + \sqrt{3})$ is the maximum absolute value of the coefficients of $\Delta_{(u, v)}^{2}$. To study $\partial_{\boldsymbol{\alpha}} \ \hat{\Mrand}(\rho \hb_k(u, v))$ we resort to the \emph{multi-dimensional Fa\`a di Bruno formula} stated and proved in \cite{sav}. Therefore, taking into account that $\big{|} \partial_{\boldsymbol{\alpha}} \ \hb_k(u, v) \big{|} \leq 1$ for every multi-index $\boldsymbol{\alpha}$, we have
\begin{equation} \label{eq:multiderC}
\big{|} \partial_{\boldsymbol{\alpha}} \ \hat{\Mrand}(\rho \hb_k(u, v)) \big{|} \leq
\sum_{h = 1}^{|\boldsymbol{\alpha}|}\sum_{l=1}^{|\boldsymbol{\alpha}|} a_{h,l}(\boldsymbol{\alpha}) \mathfrak{M}_l \rho^h
\end{equation}
where the $a_{h,l}$'s are constants specified in \cite{sav}, and $\mathfrak{M}_l := \intethree |\vb|^l \Mrand(\ud\vb)$. At this stage, (\ref{eq:TonelliL})-(\ref{eq:intpartsS2}) and (\ref{eq:coordlaplaceM})-(\ref{eq:multiderC}) yield
\begin{eqnarray}
&{}& \int_{(\RR, +\infty) \times S^2} \IIs \rho^2\mathscr{L}^1 \otimes u_{S^2}(\ud\rho \ud\ub) \nonumber \\
&\leq& 3\overline{\Delta} \sum_{1 \leq |\boldsymbol{\alpha}| \leq 4}\sum_{h = 1}^{|\boldsymbol{\alpha}|} \sum_{l=1}^{|\boldsymbol{\alpha}|} a_{h,l}(\boldsymbol{\alpha}) \mathfrak{M}_l \int_{\RR}^{+\infty} \Psi(\rho) \rho^{h-2} \ud \rho \ . \nonumber
\end{eqnarray}
Moreover, the Lyapunov inequality gives $\mathfrak{M}_l \leq \mathfrak{M}_{4}^{l/4}$ for $l$ in $[0, 4]$ and then, from (\ref{eq:boundderivativesMhat}), we get $\mathfrak{M}_4 \leq 3\sum_{i = 1}^{3} \left(\lim_{\rho \rightarrow 0}\frac{\partial^4}{\partial \rho^4} \hat{\Mrand}(\rho\mathbf{e}_i)\right) \leq\ 9g_4$. Now, an application of (\ref{eq:intenewtonm}) with $s = 1$ and $m = h - 2$, combined with (\ref{eq:r}), leads to
\begin{equation}\label{eq:outerLaplaceM}
4\pi\int_{(\RR, +\infty) \times S^2} \IIs \rho^2\mathscr{L}^1 \otimes u_{S^2}(\ud\rho \ud\ub) \leq C_{\ast, s} \WW^2
\end{equation}
with
$$
C_{\ast, s} := 12\pi \overline{\Delta} \left(\frac{\lambda^{2r}}{\epsilon}\right)^{q} \sum_{1 \leq |\boldsymbol{\alpha}| \leq 4}\sum_{h = 1}^{|\boldsymbol{\alpha}|} \sum_{l=1}^{|\boldsymbol{\alpha}|} a_{h,l}(\boldsymbol{\alpha}) (9g_4)^{l/4} \frac{1}{12 - h}(2\mq^{1/4})^{12 - h} \ .
$$

\subsubsection{Inner integral of $\IIr$} \label{sect:inner}

The analysis is essentially based on certain new Berry-Esseen-type inequalities presented in \cite{doBE}, after observing the analogy between $\rho \mapsto \hat{\Mrand}(\rho \ub)$ and the c.f. $\varphi_n(t)$ therein. Indeed, for any $\ub$ in $S^2$ and for every choice of $\mathrm{B}$ in (\ref{eq:psijn}), each realization of $\hat{\Nrand}(\rho; \ub)$, as a function of $\rho$, coincides with the c.f. of a weighted sum of independent random numbers, according to (\ref{eq:N}). Moreover, the definition of $\RR$ in (\ref{eq:r}) corresponds to the upper bound $\tau$ appearing in the Berry-Esseen-type inequalities proved in \cite{doBE}. To implement the aforesaid inequalities within the present framework, it is worth introducing the following entities
\begin{eqnarray}
T(\ub) &:=& \Big{\{}\sum_{j = 1}^{\nu} \pi_{j, \nu}^2 \psib_{j, \nu}^{t}(\ub) V[\mu_0] \psib_{j, \nu}(\ub) \leq 1/3\Big{\}} \label{eq:Vu} \\
\MM_{j, n}^{(m)}(\ub) &:=& \et\left[\left(\mathbf{V}_j \cdot \psib_{j, n}(\ub)\right)^m\ \big{|}\ \mathscr{G} \right] \label{eq:Mjnm} \\
\XX(\ub) &:=& \sum_{j = 1}^{\nu} \pi_{j, \nu}^2 |\MM_{j, \nu}^{(2)}(\ub) - 1| \label{eq:XXu} \\
\YY(\ub) &:=& \sum_{j = 1}^{\nu} \big{|}\pi_{j, \nu}^3 \MM_{j, \nu}^{(3)}(\ub)\big{|} \label{eq:YYu} \\
\ZZ(\ub) &:=& \et\left[\Big{[}\sum_{j = 1}^{\nu} \pi_{j, \nu}^2 \psib_{j, \nu}^{t}(\ub) V[\mu_0] \psib_{j, \nu}(\ub) - 1\Big{]}^2\ \big{|}\ \mathscr{G} \right] \label{eq:ZZu}
\end{eqnarray}
where $T(\ub)$ belongs to $\mathscr{H}$. With this new notation at hand, the Berry-Esseen-type inequality can be re-written as
\begin{eqnarray}
\Big{|}\frac{\partial^l}{\partial \rho^l} \Big{[} \hat{\mathcal{M}}(\rho \ub) - e^{-\rho^2/2}\Big{]}\Big{|} &\leq& \et\left[\Big{|}\frac{\partial^l}{\partial \rho^l} \hat{\Nrand}(\rho; \ub) \Big{|} \ \ind_{T(\ub)} \ | \ \mathscr{G}\right] + u_{2,l}(\rho) \XX(\ub) \nonumber \\
&+& u_{3,l}(\rho) \YY(\ub) + u_{4,l}(\rho) \mq\WW + v_{l}(\rho) \ZZ(\ub) \nonumber
\end{eqnarray}
for $l = 0, 1, 2$, $\rho$ in $[0, \RR]$ and $\ub$ in $S^2$, $u_{2,l}$, $u_{3,l}$, $u_{4,l}$, $v_l$ being non-random rapidly decreasing continuous functions depending only on $\mu_0$. See \cite{doBE} for their definition. The above inequality yields
\begin{gather}
\int_{0}^{\RR} \Big{|} \frac{\partial^l}{\partial \rho^l} \Big{[} \hat{\mathcal{M}}(\rho \ub) - e^{-\rho^2/2}\Big{]} \Big{|}^2 \rho^m \ud \rho \nonumber \\
\leq 5 \int_{0}^{\RR} \left(\et\left[\Big{|}\frac{\partial^l}{\partial \rho^l} \hat{\Nrand}(\rho; \ub) \Big{|} \ \ind_{T(\ub)} \ | \ \mathscr{G}\right]\right)^2 \rho^m \ud \rho + 5 \XX^2(\ub) \int_{0}^{+\infty} u_{2,l}^{2}(\rho) \rho^m \ud \rho \nonumber \\
+ 5 \YY^2(\ub) \int_{0}^{+\infty} u_{3,l}^{2}(\rho) \rho^m \ud \rho + 5 \mq^2\WW^2 \int_{0}^{+\infty} u_{4,l}^2(\rho) \rho^m \ud \rho + 5 \ZZ^2(\ub) \int_{0}^{+\infty} v_{l}^{2}(\rho) \rho^m \ud \rho \label{eq:BEapplied2}
\end{gather}
for $l = 0, 1, 2$, $m \geq 0$ and $\ub$ in $S^2$. The integrals $\overline{u}_{h, l, m} := \int_{0}^{+\infty} u_{h,l}^{2}(\rho) \rho^m \ud \rho$ and $\overline{v}_{l, m} := \int_{0}^{+\infty} v_{l}^{2}(\rho) \rho^m \ud \rho$ are finite and depend only on $\mu_0$ for $h = 2, 3, 4$, $l = 0, 1, 2$ and $m \geq 0$. As to the above conditional expectation, we have
\begin{gather}
\et[\ind_{T(\ub)} \ | \ \mathscr{G}] = \pt[T(\ub) \ | \ \mathscr{G}] \nonumber \\
\leq \pt\left[\Big{\{}\Big{|}\sum_{j = 1}^{\nu} \pi_{j, \nu}^2 \psib_{j, \nu}^{t}(\ub) V[\mu_0] \psib_{j, \nu}(\ub) - 1\Big{|} \geq 1/3\Big{\}} \ | \ \mathscr{G}\right]  \leq 9 \ZZ(\ub) \label{eq:Markov}
\end{gather}
the latter inequality following from the conditional Markov inequality. Now, we apply (\ref{eq:boundPsiN}) and (\ref{eq:boundderinternalN}) and, after observing that the upper bounds provided therein are $\mathscr{G}$-measurable, we
obtain
\begin{eqnarray}
\int_{0}^{\RR} \left(\et\left[\Big{|}\hat{\Nrand}(\rho; \ub) \Big{|} \ \ind_{T(\ub)} \ | \ \mathscr{G}\right]\right)^2 \rho^m \ud\rho &\leq& 81\ZZ^2(\ub)\int_{0}^{+\infty} \Psi^2(\rho) \rho^m \ud \rho \label{eq:1301a} \\
\int_{0}^{\RR} \left(\et\left[\Big{|}\frac{\partial^l}{\partial \rho^l} \hat{\Nrand}(\rho; \ub) \Big{|} \ \ind_{T(\ub)} \ | \ \mathscr{G}\right]\right)^2 \rho^m \ud\rho &\leq& 81\ZZ^2(\ub)\int_{0}^{+\infty} \wp_{l}^{2}(\rho) \Psi^2(\rho) \rho^m \ud \rho \nonumber \\
\label{eq:1301b}
\end{eqnarray}
for $l = 1, 2$ and any $m$ in $[0, 13)$. In addition, by virtue of Proposition \ref{prop:newtontrick}, the integrals $\overline{z}_m := \int_{0}^{+\infty} \Psi^2(\rho) \rho^m \ud \rho$ and $\overline{w}_{l, m} := \int_{0}^{+\infty} \wp_{l}^{2}(\rho) \Psi^2(\rho) \rho^m \ud \rho$ are finite and depend only on $\mu_0$ when $\omega$ varies in $U^c$. Coming back to the integral of interest, the Tonelli theorem can be applied to write
$$
\int_{(0, \RR) \times S^2} \IIr \rho^2\mathscr{L}^1 \otimes u_{S^2}(\ud\rho \ud\ub) = \int_{S^2} \left(\int_{0}^{\RR} \IIr \rho^2 \ud \rho\right) \unifSu \ .
$$
Since the inner integral on the RHS has already been studied, it remains to explain how it depends on $\ub$. For this, a fundamental role is played by $\mathrm{B}$, which appears in the RHS of (\ref{eq:BEapplied2}) through the random variables $\XX$, $\YY$ and $\ZZ$. Apropos of this, it should be recalled that the so-called \emph{hairy ball theorem} -- see, e.g., Chapter 5 of \cite{hir} -- asserts that a function $\mathrm{B}$, meeting the properties specified to write (\ref{eq:psijn}),
cannot be continuous everywhere. Nevertheless, we know that the definition of $\Mrand$ is independent of the choice of $\mathrm{B}$. We take advantage of this fact to overcome the aforesaid drawback by splitting $S^2$ into the charts $\Omega_k$ introduced in the previous subsection and by choosing for each $\Omega_k$ a specific $\mathrm{B}$, say $\mathrm{B}_k$, smooth on $\overline{\Omega}_k$. This possibility is guaranteed by the fact that $S^2\setminus \overline{\Omega}_k$ contains at least two antipodal points. We now have, by (\ref{eq:BEapplied2}) and (\ref{eq:1301a})-(\ref{eq:1301b}),
\begin{eqnarray}
&{}& 4\pi \int_{(0, \RR) \times S^2} \IIr \rho^2\mathscr{L}^1 \otimes u_{S^2}(\ud\rho \ud\ub) \nonumber \\
&\leq& \overline{B}_2\sum_{k = 1}^{4} \int_{\Omega_k} \XX^{2}_k(\ub)\unifSu + \overline{B}_3\sum_{k = 1}^{4} \int_{\Omega_k} \YY^{2}_{k}(\ub)\unifSu \nonumber \\
&+& \overline{B}_4\mq^2 \WW^2 + \overline{B}_5 \sum_{k = 1}^{4} \int_{\Omega_k} \ZZ^{2}_{k}(\ub)\unifSu \label{eq:finalBEpolar}
\end{eqnarray}
where $\XX_k$, $\YY_k$, $\ZZ_k$ are the same as in (\ref{eq:XXu})-(\ref{eq:ZZu}) respectively, with $\mathrm{B} = \mathrm{B}_k$ and
\begin{eqnarray}
\overline{B}_2 &:=& 20\pi [\overline{u}_{2, 0, 2} + 12\overline{u}_{2, 1, 0} + 3\overline{u}_{2, 2, 2}] \nonumber \\
\overline{B}_3 &:=& 20\pi [\overline{u}_{3, 0, 2} + 12\overline{u}_{3, 1, 0} + 3\overline{u}_{3, 2, 2}] \nonumber \\
\overline{B}_4 &:=& 80\pi [\overline{u}_{4, 0, 2} + 12\overline{u}_{4, 1, 0} + 3\overline{u}_{4, 2, 2}] \nonumber \\
\overline{B}_5 &:=& 1620\pi [\overline{z}_2 + 12\overline{w}_{1, 0} + 3\overline{w}_{2, 2}] + 20\pi[\overline{v}_{0, 2} + 12\overline{v}_{1, 0} + 3\overline{v}_{2, 2}] \ . \nonumber
\end{eqnarray}

\subsubsection{Inner integral of $\IIs$} \label{sect:innerL}

With reference to (\ref{eq:splitintegralI}), the integral at issue is analyzed by splitting $S^2$ into the charts $\Omega_k$ defined in Sub-subsection \ref{sect:outerL}. On the basis of considerations made apropos of $\mathrm{B}$ at the end of the previous sub-subsection, here we choose the $\mathrm{B}_k$'s as follows:
\begin{equation} \label{eq:B12}
\mathrm{B}_k(\hb_k(u, v)) := \left( \begin{array}{ccc}
\sin v & \cos v \cos u & \cos v \sin u \\
-\cos v & \sin v \cos u & \sin v \sin u \\
0 & -\sin u & \cos u \\
\end{array} \right)
\end{equation}
for $k = 1, 2$ and
\begin{equation} \label{eq:B34}
\mathrm{B}_k(\hb_k(u, v)) := \left( \begin{array}{ccc}
0 & -\sin u & \cos u \\
\sin v & \cos v \cos u & \cos v \sin u \\
-\cos v & \sin v \cos u & \sin v \sin u \\
\end{array} \right)
\end{equation}
for $k = 3, 4$. Then, equality $\hat{\Mrand}(\rho \ub) = \et\left[\hat{\Nrand}(\rho; \ub)\ | \ \mathscr{G}\right]$, in combination with the definition of $T(\ub)$ in (\ref{eq:Vu}), produces the following upper bound for $\big{|} \Delta_{S^2} \hat{\Mrand}(\rho \ub) \big{|}^2$, namely
\begin{equation}\label{eq:809s1}
2\left(\et\left[\big{|}\Delta_{S^2}\hat{\Nrand}_k(\rho; \ub)\big{|}\ind_{T_k(\ub)} \ | \ \mathscr{G} \right]\right)^2 + 2
\Big{|} \et\left[\Delta_{S^2}\hat{\Nrand}_k(\rho; \ub)\ind_{T_k(\ub)^{c}} \ | \ \mathscr{G} \right]\Big{|}^2
\end{equation}
for every $\ub$ in $\Omega_k$, where $\Nrand_k$ and $T_k(\ub)$ are the same as $\Nrand$ and $T(\ub)$, respectively, with $\mathrm{B} = \mathrm{B}_k$. To bound the former summand we make use of the following
\begin{prop} \label{prop:7091}
\emph{Assume that the tail condition} (\ref{eq:tailsCDGR}) \emph{is in force together with the moment assumptions} (\ref{eq:fourthmom}) \emph{and} (\ref{eq:normalizations})-(\ref{eq:covariancematrix}). \emph{Then, there exists a non-random polynomial} $\wp_L$ \emph{of degree 6, with positive coefficients which depend only on} $\mu_0$, \emph{such that}
\begin{equation} \label{eq:7091}
\sup_{k \in \{1, \dots, 4\}}\sup_{\ub \in \Omega_k} \big{|} \Delta_{S^2}\hat{\Nrand}_k(\rho; \ub)\big{|} \leq \rho^2\wp_L(\rho)\Psi(\rho)
\end{equation}
\emph{holds for every} $\rho$ \emph{in} $[0, \RR]$, \emph{with the exception of a set of} $\pt$-\emph{probability zero}.
\end{prop}
The proof is deferred to \ref{a:boundderivativesMhat}, where $\wp_L$ is given explicitly. At this stage, we note that the upper bound in (\ref{eq:7091}) is $\mathscr{G}$-measurable and, afterwards, we apply (\ref{eq:Markov}) to obtain
\begin{eqnarray}
&{}& \int_{(0, \RR) \times \Omega_k} \frac{1}{\rho^2} \left(\et\left[\big{|}\Delta_{S^2}\hat{\Nrand}_k(\rho; \ub)\big{|}\ind_{T_k(\ub)} \ | \ \mathscr{G} \right]\right)^2\mathscr{L}^1 \otimes u_{S^2}(\ud\rho \ud\ub) \nonumber \\
&\leq& \int_{[0, \RR] \times \Omega_k} \frac{1}{\rho^2} \rho^4\wp_{L}^{2}(\rho)\Psi^2(\rho) \pt[T(\ub) \ | \ \mathscr{G}]^2
\mathscr{L}^1 \otimes u_{S^2}(\ud\rho \ud\ub) \nonumber \\
&\leq& 81 \int_{0}^{+\infty} \rho^2\wp_{L}^{2}(\rho)\Psi^2(\rho) \ud\rho \cdot \int_{\Omega_k}\ZZ^{2}_{k}(\ub)\unifSu \ .
\label{eq:badlaplace}
\end{eqnarray}
If we consider the random variable $\int_{0}^{+\infty} \rho^2\wp_{L}^{2}(\rho)\Psi^2(\rho) \ud\rho$ on $U^c$, then Proposition \ref{prop:newtontrick} can be used to conclude that this random variable is bounded by the constant
$J_L := \int_{0}^{1} \rho^2\wp_{L}^{2}(\rho) \ud\rho + \left(\frac{\lambda^{2r}}{\epsilon}\right)^{2q} \int_{1}^{+\infty}
\rho^{-20} \wp_{L}^{2}(\rho)\ud\rho$.

In the final part of this sub-subsection we provide an upper bound for the latter summand in the RHS of (\ref{eq:809s1}), by
means of the following statement which involves new random quantities such as
\begin{eqnarray}
\XX_L(\ub) &:=& \sum_{j = 1}^{\nu} \pi_{j, \nu}^2 |\Delta_{S^2}\MM_{j, \nu}^{(2)}(\ub)| \label{eq:XXuL} \\
\YY_L(\ub) &:=& \sum_{j = 1}^{\nu} \big{|}\pi_{j, \nu}^3 \Delta_{S^2}\MM_{j, \nu}^{(3)}(\ub)\big{|} \label{eq:YYuL} \\
\ZZ_G(\ub) &:=& \et\left[\Blnorm \sum_{j = 1}^{\nu}\pi_{j, \nu}^{2}\gradient \Big{(}\psib_{j, \nu}^{t}(\ub) V[\mu_0] \psib_{j, \nu}(\ub)\Big{)} \Brnormg^2 \ | \ \mathscr{G} \right] \label{eq:ZG} \\
\ZZ_L(\ub) &:=& \et\left[\Big{[}\sum_{j = 1}^{\nu}\pi_{j, \nu}^{2}\Delta_{S^2}\Big{(}\psib_{j, \nu}^{t}(\ub) V[\mu_0] \psib_{j, \nu}(\ub)\Big{)}\Big{]}^2 \ | \ \mathscr{G} \right] \label{eq:ZL}
\end{eqnarray}
where the $\MM_{j, n}^{(m)}(\ub)$'s are the same as in (\ref{eq:Mjnm}), $\gradient$ is the Riemannian gradient on $S^2$ and $\lnorm \cdot \rnorm_{S^2}$ the Riemannian length.
\begin{prop} \label{prop:BElaplace}
\emph{Let the moment assumptions} (\ref{eq:fourthmom}) \emph{and} (\ref{eq:normalizations})-(\ref{eq:covariancematrix}) \emph{be in force. Then, for every} $k = 1, \dots, 4$, \emph{there exist (non-random) rapidly decreasing continuous functions} $z_1, \dots, z_6$, \emph{depending only on} $\mu_0$, \emph{such that}
\begin{eqnarray}
&& \Big{|}\et\left[\big{(} \Delta_{S^2}\hat{\Nrand}_k(\rho; \ub) \big{)} \ind_{T_k(\ub)^{c}} \ | \ \mathscr{G} \right]\Big{|} \nonumber \\
&\leq& \rho^2 \Big{[} z_1(\rho) \WW + z_2(\rho) \XX_{L, k}(\ub) + z_3(\rho) \YY_{L, k}(\ub) \nonumber \\
&+& z_4(\rho) \ZZ_k(\ub) + z_5(\rho) \ZZ_{G, k}(\ub) + z_6(\rho) \ZZ_{L, k}(\ub)\Big{]} \label{eq:BElaplace}
\end{eqnarray}
\emph{holds for every} $\ub$ \emph{in} $\Omega_k$ \emph{and} $\rho$ \emph{in} $[0, \RR]$, \emph{with the exception of a set of} $\pt$-\emph{probability zero}. $\XX_{L, k}$, $\YY_{L, k}$, $\ZZ_k$, $\ZZ_{G, k}$ \emph{and} $\ZZ_{L, k}$ \emph{are defined as in} (\ref{eq:XXuL})-(\ref{eq:ZL}) \emph{and} (\ref{eq:ZZu}) \emph{with} $\mathrm{B}_k$ \emph{in place of} $\mathrm{B}$.
\end{prop}
For the proof and the definition of the $z_i$'s see \ref{a:BElaplace}. Now, a straightforward application of the above proposition yields
\begin{eqnarray}
&{}& \int_{[0, \RR] \times \Omega_k} \frac{1}{\rho^2} \Big{|}\et\left[\big{(} \Delta_{S^2}\hat{\Nrand}_k(\rho; \ub) \big{)} \ind_{T_k(\ub)^c} \ | \ \mathscr{G} \right]\Big{|}^2\mathscr{L}^1 \otimes u_{S^2}(\ud\rho \ud\ub) \nonumber \\
&\leq& \overline{B}_{1, L} \WW^2 + \overline{B}_{2, L} \int_{\Omega_k} \XX_{L, k}^2(\ub) \unifSu + \overline{B}_{3, L}\int_{\Omega_k} \YY_{L, k}^2(\ub) \unifSu \nonumber \\
&+& \overline{B}_{4, L} \int_{\Omega_k}\ZZ_{k}^{2}(\ub)\unifSu + \overline{B}_{5, L} \int_{\Omega_k} \ZZ_{G, k}^2(\ub) \unifSu \nonumber \\
&+& \overline{B}_{6, L}\int_{\Omega_k} \ZZ_{L, k}^2(\ub) \unifSu \label{eq:integralBElaplace}
\end{eqnarray}
where $\overline{B}_{i, L} := 6 \int_{0}^{+\infty} z_{i}^{2}(\rho) \rho^2 \ud \rho$ for $i = 1, \dots, 6$.

The final bound is achieved by collecting inequalities (\ref{eq:809s1}), (\ref{eq:badlaplace}) and (\ref{eq:integralBElaplace}), according to
\begin{eqnarray}
&{}& 4\pi \int_{[0, \RR] \times S^2} \IIs \rho^2\mathscr{L}^1 \otimes u_{S^2}(\ud\rho \ud\ub)  \nonumber \\
&\leq& 96\pi \overline{B}_{1, L} \WW^2\ +\ 24\pi \sum_{k = 1}^{4}\Big{\{}\overline{B}_{2, L} \int_{\Omega_k} \XX_{L, k}^2(\ub) \unifSu \nonumber \\
&+& \overline{B}_{3, L} \int_{\Omega_k} \YY_{L, k}^2(\ub) \unifSu + (\overline{B}_{4, L} + 81J_L)  \int_{\Omega_k}\ZZ_{k}^{2}(\ub)\unifSu \nonumber \\
&+& \overline{B}_{5, L} \int_{\Omega_k} \ZZ_{G, k}^2(\ub) \unifSu + \overline{B}_{6, L} \int_{\Omega_k} \ZZ_{L, k}^2(\ub) \unifSu\Big{\}} \ . \label{eq:finalinnerintegral}
\end{eqnarray}

\subsubsection{The final step}

With a view to bounding the RHS of (\ref{eq:appliedbeurling}), we use the ultimate results of Sub-subsections \ref{sect:outer}-\ref{sect:innerL}, encapsulated in (\ref{eq:finalextM}), (\ref{eq:outerLaplaceM}), (\ref{eq:finalBEpolar}) and (\ref{eq:finalinnerintegral}) respectively, to write
\begin{eqnarray}
&{}& \left(\intethree \big{|}\hat{\Mrand}(\xib) - e^{-|\xib|^2/2}\big{|}^2 \dxi + \intethree \big{|}\Delta_{\xib}[\hat{\Mrand}(\xib) - e^{-|\xib|^2/2}]\big{|}^2 \dxi \right)^{1/2} \ind_{U^c}\nonumber \\
&\leq& (C_{\ast, r} + C_{\ast, s} + \overline{B}_4\mq^2 + 96\pi \overline{B}_{1, L})^{1/2} \WW\ +\ \sum_{k = 1}^{4}
\Big{\{} \nonumber \\
&+& \overline{B}_{2}^{1/2} \Big{(}\int_{\Omega_k} \XX^{2}_k(\ub)\unifSu\Big{)}^{1/2}
+ \sqrt{24\pi} \overline{B}_{2, L}^{1/2} \Big{(}\int_{\Omega_k} \XX_{L, k}^2(\ub) \unifSu\Big{)}^{1/2} \nonumber \\
&+& \overline{B}_{3}^{1/2} \Big{(}\int_{\Omega_k} \YY^{2}_{k}(\ub)\unifSu\Big{)}^{1/2} + \sqrt{24\pi} \overline{B}_{3, L}^{1/2} \Big{(}\int_{\Omega_k} \YY_{L, k}^2(\ub) \unifSu\Big{)}^{1/2} \nonumber \\
&+& [\overline{B}_5 + 24\pi(\overline{B}_{4, L} + 81J_L)]^{1/2} \Big{(}\int_{\Omega_k} \ZZ^{2}_k(\ub)\unifSu\Big{)}^{1/2} \nonumber \\
&+& \sqrt{24\pi}\overline{B}_{5, L}^{1/2} \Big{(}\int_{\Omega_k} \ZZ_{G, k}^2(\ub) \unifSu\Big{)}^{1/2} \nonumber \\
&+& \sqrt{24\pi} \overline{B}_{6, L}^{1/2} \Big{(}\int_{\Omega_k} \ZZ_{L, k}^2(\ub) \unifSu\Big{)}^{1/2} \Big{\}} \ .
\label{eq:vie}
\end{eqnarray}
Then, we proceed by taking expectation of both sides of (\ref{eq:vie}). Apropos of this computation it is worth noting that, if $\mu_0$ meets the additional conditions
$$
\begin{array}{lll}
& \sigma_1 = \sigma_2 = \sigma_3 = 1 & {} \\
& \intethree \xb^{\boldsymbol{\alpha}} \mu_0(\ud \xb) = 0 &\ \ \ \ \text{for\ every\ multi-index\ } \boldsymbol{\alpha}\ \text{with}\ |\boldsymbol{\alpha}| = 3 \ ,
\end{array}
$$
then $\MM_{j, n}^{(2)} \equiv 1$ and $\MM_{j}^{(3)} \equiv 0$, implying that all random variables in the RHS of (\ref{eq:vie}) vanish, except for $\WW$. Since $\et[\WW] = e^{\Lambda_b t}$ in view of (\ref{eq:gare})-(\ref{eq:l4spectral}), the proof of Theorem \ref{thm:CDGR} would be complete. Let us carry on with the computation of the aforesaid expectations to show they all admit an upper bound like $C e^{\Lambda_b t}$, even under the original more general conditions.

As for the random variables $\XX_k$ and $\XX_{L, k}$, a key role is played by the identity
\begin{equation} \label{eq:MMjn2}
\MM_{j, n}^{(2)}(\ub) - 1 = \left(\sum_{s = 1}^{3} \sigma_{s}^{2} (u_{s}^{2} - 1/3)\right) \cdot \zeta_{j, n}
\end{equation}
valid for $j = 1, \dots, n$, $n$ in $\mathbb{N}$ and $\ub$ in $S^2$, independently of the choice of $\mathrm{B}$ in (\ref{eq:psijn}). The $\zeta_{j, n}$'s are given by
\begin{equation}
\zeta_{j, n} := \zeta_{j, n}^{\ast}(\tau_n, (\phi_1, \dots, \phi_{n-1}))
\end{equation}
and the $\zeta_{j, n}^{\ast}$'s are defined on $\mathbb{T}(n) \times [0, \pi]^{n-1}$ as follows. Put $\zeta_{1, 1}^{\ast} \equiv 1$ and, for $n \geq 2$,
\begin{equation} \label{eq:zetajnast}
\zeta_{j, n}^{\ast}(\treen, \boldsymbol{\varphi}) := \left\{ \begin{array}{ll}
\zeta_{j, n_l}^{\ast}(\mathfrak{t}_{n}^{l}, \boldsymbol{\varphi}^l) \cdot (\frac{3}{2}\cos^2\varphi_{n-1} - \frac{1}{2})& \text{for} \ j = 1, \dots, n_l \\
\zeta_{j - n_l, n_r}^{\ast}(\mathfrak{t}_{n}^{r}, \boldsymbol{\varphi}^r) \cdot (\frac{3}{2}\sin^2\varphi_{n-1} - \frac{1}{2}) & \text{for} \ j = n_l + 1, \dots, n
\end{array} \right.
\end{equation}
for every $\boldsymbol{\varphi}$ in $[0, \pi]^{n-1}$. The reader is referred to \ref{a:momentirandom} for the proof of (\ref{eq:MMjn2}). Combination of (\ref{eq:MMjn2}) with (\ref{eq:XXu}) and (\ref{eq:XXuL}) yields
$\XX_{k}(\ub) = \big{|}\sum_{s = 1}^{3} \sigma_{s}^{2} (u_{s}^{2} - 1/3)\big{|} \cdot \sum_{j = 1}^{\nu} \pi_{j, \nu}^{2} |\zeta_{j, \nu}|$ and $\XX_{L, k}(\ub) = \big{|}\sum_{s = 1}^{3} \sigma_{s}^{2} \Delta_{S^2}(u_{s}^{2})\big{|} \cdot \sum_{j = 1}^{\nu} \pi_{j, \nu}^{2} |\zeta_{j, \nu}|$ for $k = 1, \dots, 4$. Whence,
\begin{eqnarray}
\Big{(}\int_{\Omega_k} \XX_{k}^{2}(\ub) \unifSu\Big{)}^{1/2} &=& \overline{\XX}_k \sum_{j = 1}^{\nu} \pi_{j, \nu}^{2} |\zeta_{j, \nu}| \nonumber \\
\Big{(}\int_{\Omega_k} \XX_{L, k}^{2}(\ub) \unifSu\Big{)}^{1/2} &=& \overline{\XX}_{L, k} \sum_{j = 1}^{\nu} \pi_{j, \nu}^{2} |\zeta_{j, \nu}| \nonumber
\end{eqnarray}
where
\begin{eqnarray}
\overline{\XX}_k &:=& \Big{(}\int_{\Omega_k}\big{[}\sum_{s = 1}^{3} \sigma_{s}^{2} (u_{s}^{2} - 1/3)\big{]}^2\unifSu\Big{)}^{1/2}  \nonumber \\
\overline{\XX}_{L, k} &:=& \Big{(}\int_{\Omega_k}\big{[}\sum_{s = 1}^{3} \sigma_{s}^{2}\Delta_{S^2}(u_{s}^{2}) \big{]}^2\unifSu\Big{)}^{1/2} \nonumber
\end{eqnarray}
are constants. At this stage, it is worth noticing that
\begin{equation}\label{eq:garezeta}
\et\Big{(}\sum_{j = 1}^{\nu} \pi_{j, \nu}^{2} |\zeta_{j, \nu}|\Big{)} = e^{-(1 - f(b))t}
\end{equation}
holds for every $t \geq 0$, with $f(b) := \int_{0}^{\pi} \sin^2\varphi \ \big{|} \frac{3}{2}\sin^2\varphi - \frac{1}{2} \big{|} \ \beta(\ud \varphi)$. See \ref{a:gare}. Then, we combine the inequality $\sin^2\varphi \ \big{|} \frac{3}{2}\sin^2\varphi - \frac{1}{2} \big{|} + \cos^2\varphi \ \big{|} \frac{3}{2}\cos^2\varphi - \frac{1}{2} \big{|} \ \leq \sin^4\varphi + \cos^4\varphi$ with (\ref{eq:bsymm}) to show that $\Lambda_b \geq -(1 - f(b))$, i.e. the RHS in (\ref{eq:garezeta}) approaches zero faster than $e^{\Lambda_b t}$ as $t$ goes to infinity. Therefore, we can conclude that
\begin{eqnarray}
\et\Big{[}\Big{(}\int_{\Omega_k} \XX_{k}^{2}(\ub) \unifSu\Big{)}^{1/2}\Big{]} &=& \overline{\XX}_k e^{-(1 - f(b))t} \label{eq:finalXXk} \\
\et\Big{[}\Big{(}\int_{\Omega_k} \XX_{L, k}^{2}(\ub) \unifSu\Big{)}^{1/2}\Big{]} &=& \overline{\XX}_{L, k} e^{-(1 - f(b))t}
\label{eq:finalXXkL} \ .
\end{eqnarray}

As for the random variables $\YY_k$ and $\YY_{L, k}$ are concerned, we write
\begin{equation} \label{eq:MMjn3+-}
\MM_{j, n}^{(3)}(\ub) = \et\left[\left(\mathbf{V}_j \cdot \psib_{j, n}(\ub)\right)^3 - \frac{3}{5} \Mtre \cdot \psib_{j, n}(\ub)\ \big{|}\ \mathscr{G} \right] + \frac{3}{5} \Mtre \cdot
\et\left[\psib_{j, n}(\ub)\ \big{|}\ \mathscr{G} \right]
\end{equation}
with $\Mtre := \intethree |\vb|^2 \vb \mu_0(\ud\vb)$. Now, the analog of (\ref{eq:MMjn2}) is given by the couple of identities
\begin{eqnarray}
\et\left[\left(\mathbf{V}_j \cdot \psib_{j, n}(\ub)\right)^3 - \frac{3}{5} \Mtre \cdot \psib_{j, n}(\ub)\ \big{|}\ \mathscr{G} \right] &=& l_3(\ub) \eta_{j, n} \label{eq:MMjn3} \\
\et\left[\psib_{j, n}(\ub)\ \big{|}\ \mathscr{G} \right] &=& \ub  \pi_{j,n} \label{eq:psijnpijn}
\end{eqnarray}
valid for $j = 1, \dots, n$, $n$ in $\mathbb{N}$ and $\ub$ in $S^2$, independently of the choice of $\mathrm{B}$ in (\ref{eq:psijn}), and for $l_3(\ub) := \intethree[(\ub \cdot \vb)^3 - \frac{3}{5}|\vb|^2 (\ub \cdot \vb)]\mu_0(\ud \vb)$. The $\eta_{j, n}$'s are given by
\begin{equation}
\eta_{j, n} := \eta_{j, n}^{\ast}(\tau_n, (\phi_1, \dots, \phi_{n-1}))
\end{equation}
while the $\eta_{j, n}^{\ast}$'s are defined on $\mathbb{T}(n) \times [0, \pi]^{n-1}$ as follows. Put $\eta_{1, 1}^{\ast} \equiv 1$ and, for $n \geq 2$,
\begin{eqnarray} \label{eq:etajnast}
&{}& \eta_{j, n}^{\ast}(\treen, \boldsymbol{\varphi}) \nonumber \\
&:=& \left\{ \begin{array}{ll}
\eta_{j, n_l}^{\ast}(\mathfrak{t}_{n}^{l}, \boldsymbol{\varphi}^l) \cdot (\frac{5}{2}\cos^2\varphi_{n-1} - \frac{3}{2})\cos\varphi_{n-1} & \text{for} \ j = 1, \dots, n_l \\
\eta_{j - n_l, n_r}^{\ast}(\mathfrak{t}_{n}^{r}, \boldsymbol{\varphi}^r) \cdot (\frac{5}{2}\sin^2\varphi_{n-1} - \frac{3}{2} )\sin\varphi_{n-1}  & \text{for} \ j = n_l + 1, \dots, n
\end{array} \right. \nonumber \\
\end{eqnarray}
for every $\boldsymbol{\varphi}$ in $[0, \pi]^{n-1}$. The reader is referred to \ref{a:momentirandom} for the proof of (\ref{eq:MMjn3})-(\ref{eq:psijnpijn}). Combination of (\ref{eq:MMjn3+-})-(\ref{eq:psijnpijn}) with (\ref{eq:YYu}) and (\ref{eq:YYuL}) entails $\YY_{k}(\ub) \leq |l_3(\ub)| \cdot \sum_{j = 1}^{\nu} |\pi_{j, \nu}^{3} \eta_{j, \nu}| + \frac{3}{5}|\Mtre \cdot \ub| \WW$ and $\YY_{L, k}(\ub) \leq |\Delta_{S^2}l_3(\ub)| \cdot \sum_{j = 1}^{\nu}|\pi_{j, \nu}^{3} \eta_{j, \nu}| + \frac{3}{5}|\Delta_{S^2}(\Mtre \cdot \ub)| \WW$ for $k = 1, \dots, 4$. By elementary inequalities we obtain \begin{eqnarray}
\Big{(}\int_{\Omega_k} \YY_{k}^{2}(\ub) \unifSu\Big{)}^{1/2} &\leq& \overline{\YY}_{k}^{(1)}\sum_{j = 1}^{\nu} |\pi_{j, \nu}^{3} \eta_{j, \nu}| + \overline{\YY}_{k}^{(2)} \WW \nonumber \\
\Big{(}\int_{\Omega_k} \YY_{L, k}^{2}(\ub) \unifSu\Big{)}^{1/2} &\leq& \overline{\YY}_{L, k}^{(1)}\sum_{j = 1}^{\nu} |\pi_{j, \nu}^{3} \eta_{j, \nu}| + \overline{\YY}_{L, k}^{(2)} \WW \nonumber
\end{eqnarray}
where
\begin{eqnarray}
\overline{\YY}_{k}^{(1)} &:=& \Big{(}2\int_{\Omega_k} |l_3(\ub)|^2 \unifSu\Big{)}^{1/2} \nonumber \\
\overline{\YY}_{k}^{(2)} &:=& \Big{(}\frac{18}{25} \int_{\Omega_k} |\Mtre \cdot \ub|^2 \unifSu\Big{)}^{1/2} \nonumber \\
\overline{\YY}_{L, k}^{(1)} &:=& \Big{(}2\int_{\Omega_k} |\Delta_{S^2}l_3(\ub)|^2 \unifSu\Big{)}^{1/2} \nonumber \\
\overline{\YY}_{L, k}^{(2)} &:=& \Big{(}\frac{18}{25}\int_{\Omega_k} |\Delta_{S^2}(\Mtre \cdot \ub)|^2 \unifSu\Big{)}^{1/2} \nonumber
\end{eqnarray}
are constants. At this stage, to compute the expectation in the above inequalities, it is worth highlighting that the identity
\begin{equation}\label{eq:gareeta}
\et\Big{(}\sum_{j = 1}^{\nu}|\pi_{j, \nu}^{3} \eta_{j, \nu}| \Big{)} = e^{-(1 - g(b))t}
\end{equation}
holds for every $t \geq 0$, with $g(b) := \int_{0}^{\pi} \sin^4\varphi \ \big{|} \frac{5}{2}\sin^2\varphi - \frac{3}{2} \big{|} \ \beta(\ud \varphi)$. See \ref{a:gare}. Now, we combine the inequality $\sin^4\varphi \ \big{|} \frac{5}{2}\sin^2\varphi - \frac{3}{2} \big{|} + \cos^4\varphi \ \big{|} \frac{5}{2}\cos^2\varphi - \frac{3}{2} \big{|} \ \leq \sin^4\varphi + \cos^4\varphi$ with (\ref{eq:bsymm}) to show that $\Lambda_b \geq -(1 - g(b))$, which says that the RHS in (\ref{eq:gareeta}) approaches zero faster than $e^{\Lambda_b t}$ as $t$ goes to infinity. Relations (\ref{eq:MMjn3+-})-(\ref{eq:gareeta}) lead to
\begin{eqnarray}
\et\Big{[}\Big{(}\int_{\Omega_k} \YY_{k}^{2}(\ub) \unifSu\Big{)}^{1/2}\Big{]} &\leq& \overline{\YY}_{k}^{(1)}e^{-(1 - g(b))t} + \overline{\YY}_{k}^{(2)}e^{\Lambda_b t} \label{eq:finalYYk} \\
\et\Big{[}\Big{(}\int_{\Omega_k} \YY_{L, k}^{2}(\ub) \unifSu\Big{)}^{1/2}\Big{]} &\leq& \overline{\YY}_{L, k}^{(1)}e^{-(1 - g(b))t} + \overline{\YY}_{L, k}^{(2)}e^{\Lambda_b t} \label{eq:finalYYkL} \ .
\end{eqnarray}

It remains only to deal with the expectations involving $\ZZ$, $\ZZ_G$ and $\ZZ_L$. Unfortunately, unlike the $\XX$'s and the $\YY$'s, it is not possible to write the random variables $\ZZ$, $\ZZ_G$ and $\ZZ_L$ as product of a given function of $\ub$ by some other random variable independent of $\ub$ and ``contracting'' in some sense. Nevertheless, such a contraction property can be found on the integrals of the $\ZZ$'s over $\Omega_k$. Accordingly, we show that the expectations of the last three random variables in (\ref{eq:vie}) admit bounds like $C e^{\Lambda_b t}$ with $C$ depending only on $\mu_0$. To prove this, we apply the Jensen inequality and exploit (\ref{eq:covariancematrix}) to get
\begin{eqnarray}
\Big{|}\sum_{j = 1}^{\nu} \pi_{j, \nu}^2 \psib_{j, \nu; k}^{t}(\ub) V[\mu_0] \psib_{j, \nu; k}(\ub) - 1\Big{|}^2 &\leq& \sum_{s = 1}^{3} \frac{\sigma_{s}^{2}}{3}\SSS_{k, s}^2 \\ \label{eq:jensenpsi}
\Blnorm \sum_{j = 1}^{\nu}\pi_{j, \nu}^{2}\gradient \Big{(}\psib_{j, \nu; k}^{t}(\ub) V[\mu_0] \psib_{j, \nu; k}(\ub)\Big{)} \Brnormg^2 &\leq& \sum_{s = 1}^{3} \frac{\sigma_{s}^{2}}{3}\Blnorm \gradient \SSS_{k, s} \Brnormg^2  \\ \label{eq:jensenpsiG}
\Big{|}\sum_{j = 1}^{\nu}\pi_{j, \nu}^{2}\Delta_{S^2}\Big{(}\psib_{j, \nu; k}^{t}(\ub) V[\mu_0] \psib_{j, \nu; k}(\ub)\Big{)}\Big{|}^2 &\leq& \sum_{s = 1}^{3} \frac{\sigma_{s}^{2}}{3}\Big{|}\Delta_{S^2} \SSS_{k, s}\Big{|}^2
\label{eq:jensenpsiL}
\end{eqnarray}
where $\psib_{j, n; k}$ is the analog of (\ref{eq:psijn}) when $\mathrm{B}$ is replaced by $\mathrm{B}_k$, $\psi_{j, n; k, s}$ denotes its $s$-th component and $\SSS_{k, s} := \sum_{j = 1}^{\nu} \pi_{j, \nu}^{2}\big{(} 3 \psi_{j, \nu; k, s}^{2} - 1 \big{)}$. Whence, by a further application of Jensen's inequality and of an obvious inequality concerning the square root of a sum,
\begin{eqnarray}
\Big{(}\int_{\Omega_k} \ZZ^{2}_k(\ub)\ud u_{S^2}\Big{)}^{1/2} &\leq& \frac{\sqrt{3}}{3} \sum_{s = 1}^{3} \sigma_s \Big{(}\int_{\Omega_k} \Big{\{} \et\Big{[}\SSS_{k, s}^2 \ | \ \mathscr{G}\Big{]} \Big{\}}^2 \ud u_{S^2}\Big{)}^{1/2}  \nonumber \\
\Big{(}\int_{\Omega_k} \ZZ_{G, k}^2(\ub) \ud u_{S^2}\Big{)}^{1/2} &\leq& \frac{\sqrt{3}}{3} \sum_{s = 1}^{3} \sigma_s \Big{(} \int_{\Omega_k}\Big{\{} \et\Big{[}\Blnorm \gradient \SSS_{k, s}\Brnormg^2\ | \ \mathscr{G}\Big{]} \Big{\}}^2 \ud u_{S^2}\Big{)}^{1/2}  \nonumber \\
\Big{(}\int_{\Omega_k} \ZZ_{L, k}^2(\ub) \ud u_{S^2} \Big{)}^{1/2} &\leq& \frac{\sqrt{3}}{3} \sum_{s = 1}^{3} \sigma_s
\Big{(}\int_{\Omega_k}\Big{\{} \et\Big{[}\Big{|}\Delta_{S^2} \SSS_{k, s}\Big{|}^2 \ | \ \mathscr{G}\Big{]} \Big{\}}^2 \ud u_{S^2} \Big{)}^{1/2}\ .  \nonumber
\end{eqnarray}
Both the square roots and the squares after the brackets constitute an obstacle for the interchange of the integral with the expectation $\et$ and for the consequent application of useful properties of conditional expectation. To overcome this difficulty, we resort to the imbedding of the Sobolev space $\mathrm{W}^{1, 1}(\Omega_k)$ into $\mathrm{L}^2(\Omega_k)$. See, e.g., Chapter 2 of \cite{aub}. Taking the same constants $A_1(0)$ and $K(2, 1)$ as in Theorem 2.28 therein, we write
\begin{gather}
\et\Big{[}\Big{(}\int_{\Omega_k}\Big{\{} \et\Big{[}\big{|}\diff \SSS_{k, s}\big{|}^2 \ | \ \mathscr{G}\Big{]} \Big{\}}^2 \unifSu \Big{)}^{1/2}\Big{]} \leq A_1(0) \et\Big{[}\int_{\Omega_k} \big{|}\diff \SSS_{k, s}\big{|}^2\unifSu\Big{]} \nonumber \\
+ K(2, 1)\et\Big{[}\int_{\Omega_k} \Blnorm \gradient \et\Big{[}\big{|}\diff \SSS_{k, s}\big{|}^2 \ | \ \mathscr{G}\Big{]}\Brnormg \unifSu\Big{]} \label{eq:sobolev}
\end{gather}
where $\diff$ can be $\mathrm{Id}$, $\gradient$, $\Delta_{S^2}$, and $\big{|}\diff \SSS_{k, s}\big{|}$ is to be interpreted in accordance with the meaning of $\diff$. To work out the last term in the above inequality, we use (5.1.25) in \cite{stro} to say that
\begin{equation} \label{eq:jensenS2}
\Blnorm \gradient \et\Big{[}\big{|}\diff \SSS_{k, s}\big{|}^2 \ | \ \mathscr{G}\Big{]}\Brnormg\ \leq \et\Big{[}\Blnorm \gradient \big{(}\big{|}\diff \SSS_{k, s}\big{|}^2\big{)}\Brnormg \ | \ \mathscr{G}\Big{]}
\end{equation}
holds true $\pt$-almost surely. Moreover, when $\diff$ is $\mathrm{Id}$ or $\Delta_{S^2}$, the Leibnitz rule for the gradient entails
\begin{equation} \label{eq:leibnitzG}
\Blnorm \gradient \big{(}\big{|}\diff \SSS_{k, s}\big{|}^2\big{)}\Brnormg\ \leq\ \big{|}\diff \SSS_{k, s}\big{|}^2 + \Blnorm \gradient \big{(}\diff \SSS_{k, s}\big{)}\Brnormg^2 \ .
\end{equation}
When $\diff$ is $\gradient$, the definition of the Hessian as symmetric bilinear form leads to
\begin{eqnarray}
\langle \gradient (\lnorm \gradient \SSS_{k, s}\rnorm_{S^2}^2), V\rangle &=& 2 \langle D_V \gradient \SSS_{k, s}, \gradient \SSS_{k, s}\rangle \nonumber \\
&=& 2 \hess[\SSS_{k, s}](\gradient \SSS_{k, s}, V) \nonumber
\end{eqnarray}
for every vector field $V$, $D$ standing for the Levi-Civita connection. See Exercise 11 in Chapter 6 of \cite{carmo}. Whence,
\begin{eqnarray}
\lnorm \gradient (\lnorm \gradient \SSS_{k, s}\rnorm_{S^2}^2) \rnorm_{S^2} &\leq& 2\lnorm \hess[\SSS_{k, s}]\rnorm_{\ast} \lnorm \gradient \SSS_{k, s}\rnorm_{S^2} \nonumber \\
&\leq& \lnorm \hess[\SSS_{k, s}]\rnorm_{\ast}^2 + \lnorm \gradient \SSS_{k, s}\rnorm_{S^2}^2 \label{eq:leibnitzD}
\end{eqnarray}
where $\lnorm \cdot \rnorm_{\ast}$ denotes the $\mathrm{L}^2$-norm of the Hessian given by $\lnorm \hess[\SSS_{k, s}]\rnorm_{\ast}^2 := \sum_{ij} [\hess[\SSS_{k, s}](V_i, V_j)]^2$ for some orthonormal basis $\{V_1, V_2\}$ of vector fields.
At this stage, it comes in useful to emphasize the fact that, in view of (\ref{eq:jensenS2})-(\ref{eq:leibnitzD}), the latter summand in (\ref{eq:sobolev}) can be bounded by a sum of terms sharing the same structure of the former summand. Then, to provide an effective bound for the RHS of (\ref{eq:sobolev}) it is enough to prove that
\begin{equation} \label{eq:laastbound}
\et\Big{[}\int_{\Omega_k} \big{|}\diff^{'} \SSS_{k, s}\big{|}^2\unifSu\Big{]} \leq c_k(\diff^{'}) e^{\Lambda_b t}
\end{equation}
holds for some suitable constant $c(\diff^{'})$, $\diff^{'}$ being one of the following operators: $\mathrm{Id}$, $\gradient$, $\Delta_{S^2}$, $\gradient \Delta_{S^2}$, $\hess$. For the proof of (\ref{eq:laastbound}), cf. \ref{a:laastbound}. Now, we are in a position to write explicit bounds for the last three terms in (\ref{eq:vie}), which read
\begin{eqnarray}
\et\Big{[}\Big{(}\int_{\Omega_k} \ZZ_{k}^{2}(\ub) \unifSu\Big{)}^{1/2}\Big{]} &\leq& \overline{Z}_k e^{\Lambda_b t} \label{eq:expQ} \\
\et\Big{[}\Big{(}\int_{\Omega_k} \ZZ_{G, k}^{2}(\ub) \unifSu\Big{)}^{1/2}\Big{]} &\leq& \overline{Z}_{G, k} e^{\Lambda_b t} \label{eq:expQG} \\
\et\Big{[}\Big{(}\int_{\Omega_k} \ZZ_{L, k}^2(\ub) \unifSu\Big{)}^{1/2}\Big{]} &\leq& \overline{Z}_{L, k} e^{\Lambda_b t} \label{eq:expQL}
\end{eqnarray}
with
\begin{eqnarray}
\overline{Z}_k &=& \frac{\sqrt{3}}{3} \left(\sum_{s = 1}^{3} \sigma_s\right) [A_1(0)c_k(\mathrm{Id}) + K(2,1) (c_k(\mathrm{Id}) + c_k(\gradient))] \nonumber \\
\overline{Z}_{G, k} &=& \frac{\sqrt{3}}{3} \left(\sum_{s = 1}^{3} \sigma_s\right) [A_1(0)c_k(\gradient) + K(2,1) (c_k(\hess) + c_k(\gradient))] \nonumber \\
\overline{Z}_{L, k} &=& \frac{\sqrt{3}}{3} \left(\sum_{s = 1}^{3} \sigma_s\right) [A_1(0)c_k(\Delta_{S^2}) + K(2,1) (c_k(\Delta_{S^2}) + c_k(\gradient\Delta_{S^2}))] \ . \nonumber
\end{eqnarray}

To conclude, we gather (\ref{eq:gare})-(\ref{eq:l4spectral}), (\ref{eq:finalXXk})-(\ref{eq:finalXXkL}), (\ref{eq:finalYYk})-(\ref{eq:finalYYkL}), (\ref{eq:expQ})-(\ref{eq:expQL}) and we resort to (\ref{eq:appliedbeurling}) and (\ref{eq:vie}) to obtain
$$
\et[\vartot(\Mrand, \gamma); U^c] \leq 2^{-5/4} \pi^{-1/2} C(U^c) e^{\Lambda_b t}
$$
with
\begin{eqnarray}
C(U^c) &:=& (C_{\ast, r} + C_{\ast, s} + \overline{B}_4\mq^2 + 96\pi \overline{B}_{1, L})^{1/2} \nonumber \\
&+& \sum_{k=1}^{4} \Big{\{} \overline{B}_{2}^{1/2} \overline{\XX}_k + \sqrt{24\pi} \overline{B}_{2, L}^{1/2}\overline{\XX}_{L, k} + \overline{B}_{3}^{1/2}(\overline{\YY}_k^{(1)} + \overline{\YY}_k^{(2)}) \nonumber \\
&+& \sqrt{24\pi} \overline{B}_{3, L}^{1/2}(\overline{\YY}_{L, k}^{(1)} + \overline{\YY}_{L, k}^{(2)}) + [\overline{B}_5 + 24\pi(\overline{B}_{4, L} + 81J_L)]^{1/2} \overline{Z}_k \nonumber \\
&+& \sqrt{24\pi}\overline{B}_{5, L}^{1/2} \overline{Z}_{G, k} + \sqrt{24\pi} \overline{B}_{6, L}^{1/2}\overline{Z}_{L, k} \Big{\}} \ . \nonumber
\end{eqnarray}
Finally, we recall (\ref{eq:split}) and combine the last inequality with (\ref{eq:bad}).

\subsection{Proof of Theorem \ref{thm:CLT}} \label{sect:CLT}

Without any loss of generality, we prove the sufficiency of (\ref{eq:secondmoment}) for the weak convergence to the Maxwellian distribution, under extra-conditions (\ref{eq:normalizations})-(\ref{eq:covariancematrix}) and
\begin{equation} \label{eq:intornolombardo}
\max_{i = 1, 2, 3} |\sigma_{i}^{2} - 1| \leq \frac{\sqrt{42 + \delta^2}}{21 + \delta^2} \delta
\end{equation}
with $\delta := -\Lambda_b/16$. This last assumption is not restrictive since the Cauchy problem associated with (\ref{eq:boltzmann}) is autonomous and $\max_{i = 1, 2, 3} \big{|}\intethree v_{i}^{2}\mu(\ud\vb, t) - 1\big{|}$ approaches zero as $t$ goes to infinity. See \cite{Dbasic, iktr}. The argument proceeds, as in Section 9.1 of \cite{chte}, on the basis of the L\'{e}vy continuity theorem. Therefore, fix $\xib \neq \mathbf{0}$ and write
\begin{equation}\label{eq:suff1}
\big{|}\hat{\mu}(\xib, t) - \hat{\gamma}(\xib)\big{|}\ \leq\ \et \Big{|}\hat{\Nrand}(\rho; \ub) - e^{-T^2 \rho^2/2}\Big{|}\ +\ \Big{|}\et\left[e^{-T^2 \rho^2/2}\right] - e^{-\rho^2/2}\Big{|}
\end{equation}
where $\rho = |\xib|$, $\ub = \xib/|\xib|$ and $T^2 := \sum_{j = 1}^{\nu} \pi_{j, \nu}^2 \psib_{j, \nu}^{t}(\ub) V[\mu_0] \psib_{j, \nu}(\ub)$. As to the first summand in (\ref{eq:suff1}), use (6)-(7) in Section 9.1 of \cite{chte} to obtain
\begin{eqnarray}
\Big{|}\hat{\Nrand}(\rho; \ub) - e^{-T^2 \rho^2/2}\Big{|} &\leq& \rho^2 \sum_{j = 1}^{\nu} \pi_{j, \nu}^2 \int_{A_j(\varepsilon)} (\psib_{j, \nu}(\ub) \cdot \vb)^2 \mu_0(\ud \vb)
+ \varepsilon |T|^3 \rho^3 \nonumber \\
&+& \frac{1}{8} T^4 \rho^4 \frac{\max_{1 \leq j \leq \nu} \pi_{j, \nu}^2 \psib_{j, \nu}^{t}(\ub) V[\mu_0] \psib_{j, \nu}(\ub)}{T^2} \label{eq:suff5}
\end{eqnarray}
with $\varepsilon > 0$ and $A_j(\varepsilon) := \{\vb \in \rthree \ \big{|} \ |\pi_{j, \nu} (\psib_{j, \nu}(\ub) \cdot \vb)| \geq \varepsilon |T|\}$ for $j = 1, \dots, \nu$. Then, one has $\sigma_{\ast}^2 := \min\{\sigma_{1}^2, \sigma_{2}^2, \sigma_{3}^2\} \leq T^2 \leq 3$ and
\begin{equation} \label{eq:suff3}
\frac{1}{8} T^2 \rho^4 \max_{1 \leq j \leq \nu} \pi_{j, \nu}^2 \psib_{j, \nu}^{t}(\ub) V[\mu_0] \psib_{j, \nu}(\ub) \leq \frac{9}{8} \rho^4 \pi_{o}^2
\end{equation}
with $\pi_o := \max_{1 \leq j \leq \nu} |\pi_{j, \nu}|$. Put $M(y) := \int_{\{|\vb| \geq 1/y\}} |\vb|^2 \mu_0(\ud \vb)$ for $y > 0$ and note that $M$ is a monotonically increasing bounded function satisfying $\lim_{y \downarrow 0} M(y) = 0$. Moreover, from
$$
A_j(\varepsilon) \subset \{\vb \in \rthree\ \big{|} \ \pi_o |\psib_{j, \nu}(\ub) \cdot \vb| \geq \varepsilon \sigma_{\ast}\} \subset \{\vb \in \rthree\ \big{|} \ \pi_o \cdot |\vb| \geq \varepsilon \sigma_{\ast}\}
$$
one can conclude that
\begin{equation} \label{eq:suff4}
\sum_{j = 1}^{\nu} \pi_{j, \nu}^2 \int_{A_j(\varepsilon)} (\psib_{j, \nu}(\ub) \cdot \vb)^2 \mu_0(\ud \vb) \leq M\left(\frac{\pi_o}{\varepsilon \sigma_{\ast}}\right)
\end{equation}
holds true for every strictly positive $\varepsilon$. At this stage, take $\varepsilon = \sqrt{\pi_o}$ and combine (\ref{eq:suff5})-(\ref{eq:suff4}) to get
\begin{equation} \label{eq:suff2}
\Big{|}\hat{\Nrand}(\rho; \ub) - e^{-T^2 \rho^2/2}\Big{|} \leq M\left(\frac{\sqrt{\pi_o}}{\sigma_{\ast}}\right)\rho^2  + \sqrt{27 \pi_o} \rho^3 + \frac{9}{8} \pi_{o}^2 \rho^4 \ .
\end{equation}
To complete the analysis of the first summand in the RHS of (\ref{eq:suff1}), one shows that the expectation of the RHS of (\ref{eq:suff2}) approaches zero as $t$ goes to infinity, for every $\rho$ in $[0, +\infty)$. Indeed, for any monotonically increasing bounded function $g : (0, \infty) \rightarrow (0, \infty)$ satisfying $\lim_{x \downarrow 0} g(x) = 0$, one has
\begin{eqnarray}
\et[g(\pi_{o})] &=& \et[g(\pi_{o})\ind\{\pi_{o} \leq e^{-z t}\}] + \et[g(\pi_{o})\ind\{\pi_{o} > e^{-z t}\}] \nonumber \\
&\leq& g(e^{-z t}) + \sup_{x \in (0, \infty)} g(x) \cdot \et[\pi_{o}^4] e^{4 z t} \nonumber
\end{eqnarray}
for every $z$ in $(0, \infty)$. By virtue of (\ref{eq:gare})-(\ref{eq:l4spectral}), $\et[\pi_{o}^4] \leq e^{\Lambda_b t}$
and, after choosing $z = -\Lambda_b/8$, one obtains $\lim_{t \rightarrow +\infty} \et[g(\pi_{o})] = 0$. This argument, applied with $g(x) = M\left(\frac{\sqrt{x}}{\sigma_{\ast}}\right)\rho^2  + \sqrt{27 x} \rho^3 + \frac{9}{8} x^2 \rho^4$, leads to the desired result. As far as the latter summand in (\ref{eq:suff1}) is concerned, a plain application of (\ref{eq:main}) implies that $\et[e^{- T^2 \rho^2/2}]$ can be thought of as the Fourier transform of the solution of (\ref{eq:boltzmann}) when the initial datum coincides with $\prod_{i = 1}^{3} \frac{1}{\sigma_i \sqrt{2\pi}} \exp\{-\frac{v_{i}^{2}}{2 \sigma_{i}^{2}}\} \ud v_i$, where the $\sigma_i$'s have been fixed initially. Now, in view of (\ref{eq:intornolombardo}), this initial datum belongs to a convenient neighborhood of the equilibrium $\gamma$ -- according to Theorem 1.1 in \cite{doLom} -- so that
$$
\sup_{\rho \in \rone} \big{|}\et\exp\{-T^2 \rho^2/2\} - \exp\{-\rho^2/2\}\big{|} \leq C_{\ast} e^{\frac{1}{2}\Lambda_b t}
$$
holds true for every $t \geq 0$ with the same $C_{\ast}$ as in the above-quoted theorem. \\

As to the necessity of (\ref{eq:secondmoment}), suppose that $\mu(\cdot, t)$ converges weakly to some limit as $t$ goes to infinity. Following a technique developed in \cite{flr}, the argument starts with the introduction of the random vector
$$
W = \big{(}\nu, \{\tau_n\}_{n \geq 1}, \{\phi_n\}_{n \geq 1}, \{\vartheta_n\}_{n \geq 1}, \boldsymbol{\lambda}, \Lambda, \mathbf{U}\big{)}
$$
defined on $(\Omega, \mathscr{F})$. To explain the three right-most symbols above, one fixes an arbitrary point $\ub_0$ in $S^2$ and defines:
\begin{enumerate}
\item[i)] $\boldsymbol{\lambda} := \{\lambda_1(\cdot), \dots, \lambda_{\nu}(\cdot), \delta_0(\cdot), \delta_0(\cdot), \dots\}$ to be the sequence of random p.d.'s on $(\rone, \mathscr{B}(\rone))$ such that $\hat{\lambda}_j(\xi) := \hat{\mu}_0(\xi \pi_{j, \nu} \psib_{j, \nu}(\ub_0))$, for $j = 1, \dots, \nu$ and $\xi$ in $\rone$.
\item[ii)] $\Lambda$ to be the random p.d. on $(\rone, \mathscr{B}(\rone))$ obtained as convolution of all elements of $\boldsymbol{\lambda}$, i.e. $\Lambda = \lambda_1 \ast \dots \ast \lambda_{\nu}$.
\item[iii)] $\mathbf{U} := \{U_1, U_2, \dots\}$ to be the sequence of random numbers defined by $U_k := \max_{1 \leq j \leq \nu} \lambda_j\left(\left[-\frac{1}{k}, \frac{1}{k}\right]^c\right)$ for every $k$ in $\mathbb{N}$.
\end{enumerate}
To grasp the usefulness of $W$, one can note that its components are the essential ingredients of the central limit problem for \emph{independent uniformly asymptotically negligible} summands. See Sections 16.6-9 of \cite{frgr}. Apropos of the negligibility condition, it is easy to prove that $\lim_{t \rightarrow +\infty} \pt[U_k > \alpha] = 0$ holds for every $k$ in $\mathbb{N}$ and for every $\alpha$ in $(0, +\infty)$. In fact, the inclusion $\{\vb \in \rthree \ \big{|} \ |\pi_{j, \nu} \psib_{j, \nu} \cdot \vb| \geq 1/k\} \subset \{\vb \in \rthree \ \big{|} \ |\pi_{j, \nu} \vb| \geq 1/k\}$ entails
$$
\{U_k > \alpha\} \subset \left\{\left[\max_{1 \leq j \leq \nu} \mu_0\{\vb \in \rthree \ \big{|} \ |\pi_{j, \nu} \vb| \geq 1/k\}\right] \geq \alpha\right\} \ .
$$
To conclude, apply the argument used to prove Lemma 2 in \cite{gr8}. Now, think of the range of $W$ as a subset of
$$
\mathbb{S} := \overline{\mathbb{N}} \times \overline{\mathbb{T}} \times [0, \pi]^{\infty} \times [0, 2\pi]^{\infty} \times (\mathcal{P}(\overline{\mathbb{R}}))^{\infty} \times \mathcal{P}(\overline{\mathbb{R}}) \times [0, 1]^{\infty}
$$
where: $\overline{\mathbb{N}} := \{1, 2, \dots, +\infty\}$ and $\overline{\mathbb{T}}$ are the one-point compactifications of $\mathbb{N}$ and $\mathbb{T}$, respectively; $\overline{\mathbb{R}} := [-\infty, +\infty]$; $\mathcal{P}(X)$ is the space of all p.d.'s on $X$. Here, $\mathcal{P}(\overline{\mathbb{R}})$ is metrized, consistently with the topology of weak convergence, in such a way that it turns out to be a separable, compact and complete metric space. Cf. Section 6.II of \cite{parth}. Then, $\mathbb{S}$ is a separable, compact and complete metric space w.r.t. the product topology and so the family of probability distributions $\{\pt \circ W^{-1}\}_{t \geq 0}$ is tight. This implies that any sequence $\{\textsf{P}_{t_m} \circ W^{-1}\}_{m \geq 1}$, when $t_m$ strictly increases to infinity, contains a subsequence $\{\textsf{Q}_l\}_{l \geq 1}$, with $\textsf{Q}_l := \textsf{P}_{t_{m_l}} \circ W^{-1}$, which converges weakly to a p.d. $\textsf{Q}$. It is worth noting that, thanks to the weak convergence of $\mu(\cdot, t)$, $\textsf{Q}$ is supported by
$$
\{+\infty\} \times \overline{\mathbb{T}} \times [0, \pi]^{\infty} \times [0, 2\pi]^{\infty} \times \{\delta_0\}^{\infty} \times \mathcal{P}(\mathbb{R}) \times \{0\}^{\infty} \ .
$$
This claim can be verified by recalling Lemma 3 in \cite{gr8}. Since $\mathbb{S}$ is Polish, one can now invoke the \emph{Skorokhod representation theorem} (see Theorem 4.30 in \cite{ka}). Therefore, there are a probability space $(\tilde{\Omega}, \tilde{\mathscr{F}}, \tilde{\pp})$ and $\mathbb{S}$-valued random elements on it, say $\tilde{W}_l = \big{(}\tilde{\nu}_l, \{\tilde{\tau}_{n, l}\}_{n \geq 1}, \{\tilde{\phi}_{n, l}\}_{n \geq 1}, \{\tilde{\vartheta}_{n, l}\}_{n \geq 1}, \tilde{\boldsymbol{\lambda}}_l, \tilde{\Lambda}_l, \tilde{\mathbf{U}}_l\big{)}$ and $\tilde{W}_{\infty}$, which have respective p.d.'s $\textsf{Q}_l$ and $\textsf{Q}$, for every $l$ in $\mathbb{N}$. Moreover, for every $\tilde{\omega}$ in $\tilde{\Omega}$, one has $\tilde{W}_l(\tilde{\omega}) \rightarrow \tilde{W}_{\infty}(\tilde{\omega})$ (in the metric of $\mathbb{S}$) as $l$ goes to infinity, which entails
\begin{equation} \label{eq:convergenzeflr}
\begin{array}{lll}
\tilde{\nu}_l \rightarrow +\infty, & \ \ \ \tilde{\mathbf{U}}_l \rightarrow \{0, 0, \dots\} \\
\tilde{\boldsymbol{\lambda}}_l \Rightarrow \{\delta_0, \delta_0, \dots\}, & \ \ \ \tilde{\Lambda}_l \Rightarrow \tilde{\Lambda}_{\infty}
\end{array}
\end{equation}
$\tilde{\Lambda}_{\infty}$ being an element of $\mathcal{P}(\mathbb{R})$. The distributional properties of $\tilde{W}_l$ imply that $\tilde{\Lambda}_l$ is the convolution of the elements of $\tilde{\boldsymbol{\lambda}}_l$, and that $\tilde{U}_{k, l}$
coincides with $\max_{1 \leq j \leq \tilde{\nu}_l} \tilde{\lambda}_{j, l}\left(\left[-\frac{1}{k}, \frac{1}{k}\right]^c\right)$ for every $k$ in $\mathbb{N}$, $\tilde{\pp}$-almost surely. For convenience, denote with $q^{(s)}$ the symmetrized form of the p.d. $q$, i.e. $\hat{q^{(s)}(\cdot)} := |\hat{q}(\cdot)|^2$. Now, (\ref{eq:convergenzeflr}) entails $\tilde{\Lambda}_{l}^{(s)} \Rightarrow \tilde{\Lambda}_{\infty}^{(s)}$ for every $\tilde{\omega}$ in $\tilde{\Omega}$ and the combination of this fact with Theorem 24 in Chapter 16 of \cite{frgr} yields
\begin{equation} \label{eq:frgr}
+\infty > \sigma^2(\tilde{\omega}) := \lim_{\varepsilon \downarrow 0} \overline{\underline{\lim}}_{l \rightarrow \infty} \sum_{j = 1}^{\tilde{\nu}_l(\tilde{\omega})} \int_{[-\varepsilon, \varepsilon]} x^2 \tilde{\lambda}_{j}^{(s)}(\ud x; \tilde{\omega})
\end{equation}
with the exception of a set of points $\tilde{\omega}$ of $\tilde{\pp}$-probability 0. The final argument is split into the following steps. First,
\begin{eqnarray}
\sum_{j = 1}^{\tilde{\nu}_l} \int_{[-\varepsilon, \varepsilon]} x^2 \tilde{\lambda}_{j}^{(s)}(\ud x) &=& \sum_{j = 1}^{\tilde{\nu}_l} \tilde{\pi}_{j, \tilde{\nu}_l}^{2} \intethree (\tilde{\psib}_{j, \tilde{\nu}_l} \cdot \vb)^2 \ind\{|\tilde{\pi}_{j, \tilde{\nu}_l} \tilde{\psib}_{j, \tilde{\nu}_l} \cdot \vb| \leq \varepsilon\} \mu_{0}^{(s)}(\ud \vb) \nonumber \\
&\geq& \sum_{j = 1}^{\tilde{\nu}_l} \tilde{\pi}_{j, \tilde{\nu}_l}^{2} \sum_{i = 1}^{3} \tilde{\psi}_{j, \tilde{\nu}_l; i}^{2} \int_{\{\tilde{\pi}_{l, o} |\vb| \leq \varepsilon\}} v_{i}^{2} \mu_{0}^{(s)}(\ud \vb)
\end{eqnarray}
where the $\tilde{\pi}$'s and $\tilde{\psib}$'s denote the counterparts, in the Skorokhod representation, of the $\pi$'s and $\psib(\ub_0)$'s, $\tilde{\pi}_{l, o} := \max_{1 \leq j \leq \tilde{\nu}_l} |\tilde{\pi}_{j, \tilde{\nu}_l}|$ and the inequality is a consequence of the inclusion $\{\vb \in \rthree \ \big{|} \ |\tilde{\pi}_{j, \tilde{\nu}_l} \tilde{\psib}_{j, \tilde{\nu}_l} \cdot \vb| \leq \varepsilon\} \supset \{\vb \in \rthree \ \big{|} \ \tilde{\pi}_{l, o} |\vb| \leq \varepsilon\} $. Second, define $d = d(\tilde{\omega}; j, l)$ to be an element of $\{1, 2, 3\}$ for which $\tilde{\psi}_{j, \tilde{\nu}_l; d}^2 = \max_{1 \leq i \leq 3} \tilde{\psi}_{j, \tilde{\nu}_l; i}^2$. Note that $\tilde{\psi}_{j, \tilde{\nu}_l; d}^2$ must be greater than $1/3$ since $\tilde{\psib}_{j, \tilde{\nu}_l}$ belongs to $S^2$, for every $\tilde{\omega}$ in $\tilde{\Omega}$, $l$ in $\mathbb{N}$ and $j = 1, \dots, \tilde{\nu}_l$. Then,
\begin{eqnarray}
\sum_{j = 1}^{\tilde{\nu}_l} \tilde{\pi}_{j, \tilde{\nu}_l}^{2} \sum_{i = 1}^{3} \tilde{\psi}_{j, \tilde{\nu}_l; i}^2 \int_{\{\tilde{\pi}_{l, o} |\vb| \leq \varepsilon\}} v_{i}^{2} \mu_{0}^{(s)}(\ud \vb) &\geq& \sum_{j = 1}^{\tilde{\nu}_l} \tilde{\pi}_{j, \tilde{\nu}_l}^{2} \tilde{\psi}_{j, \tilde{\nu}_l; d}^2 \int_{\{\tilde{\pi}_{l, o} |\vb| \leq \varepsilon\}} v_{d}^{2} \mu_{0}^{(s)}(\ud \vb) \nonumber \\
&\geq& \frac{1}{3} \sum_{j = 1}^{\tilde{\nu}_l} \tilde{\pi}_{j, \tilde{\nu}_l}^{2} \int_{\{\tilde{\pi}_{l, o} |\vb| \leq \varepsilon\}} v_{d}^{2} \mu_{0}^{(s)}(\ud \vb) \nonumber \\
&=& \frac{1}{3} \sum_{h = 1}^{3} \tilde{s}_{h, l} \int_{\{\tilde{\pi}_{l, o} |\vb| \leq \varepsilon\}} v_{h}^{2} \mu_{0}^{(s)}(\ud \vb) \nonumber
\end{eqnarray}
where $\tilde{s}_{h, l}$ denotes the sum of those $\tilde{\pi}_{j, \tilde{\nu}_l}^{2}$ for which $d(\tilde{\omega}; j, l) = h$. At this stage, observe that $\tilde{\pi}_{l, o}$ goes to zero with probability one as $l$ goes to infinity, in view of Lemma 1 in \cite{gr8}. Since $\sum_{h = 1}^{3} \tilde{s}_{h, l} = 1$ with probability one, there are some $\tilde{\omega}$ and $h$, say $\tilde{\omega}_{\ast}$ and $h_{\ast}$, such that $\tilde{\pi}_{l, o}(\tilde{\omega}_{\ast}) \rightarrow 0$ and $\overline{\lim}_l \tilde{s}_{h_{\ast}, l}(\tilde{\omega}_{\ast})$ is strictly positive. Then,
\begin{eqnarray}
\sigma^2(\tilde{\omega}_{\ast}) &\geq& \lim_{\varepsilon \downarrow 0} \overline{\lim}_{l \rightarrow \infty} \frac{1}{3} \sum_{h = 1}^{3} \tilde{s}_{h, l}(\tilde{\omega}_{\ast}) \int_{\{\tilde{\pi}_{l, o}(\tilde{\omega}_{\ast}) |\vb| \leq \varepsilon\}} v_{h}^{2} \mu_{0}^{(s)}(\ud \vb) \nonumber \\
&\geq& \frac{1}{3} \overline{\lim}_{r \rightarrow \infty} \int_{\{|\vb| \leq r\}} v_{h_{\ast}}^{2} \mu_{0}^{(s)}(\ud \vb) \cdot \overline{\lim}_{l \rightarrow \infty} \tilde{s}_{h_{\ast}, l}(\tilde{\omega}_{\ast}) \nonumber
\end{eqnarray}
which shows that the $h_{\ast}$-th marginal of $\mu_{0}^{(s)}$ -- and hence also the $h_{\ast}$-th marginal of $\mu_0$ -- has finite second moment. To complete the proof, observe that $h_{\ast}$ can be determined independently of $\mu_0$ and that weak convergence of $\mu(\cdot, t)$ entails weak convergence of $\mu(\cdot, t) \circ f_{Q}^{-1}$, $f_{Q}$ being the map $\vb \mapsto Q\vb$ and $Q$ an orthogonal matrix. Hence, since $\mu(\cdot, t) \circ f_{Q}^{-1}$ turns out to be the solution of (\ref{eq:boltzmann}) with initial datum $\mu_0 \circ f_{Q}^{-1}$ (cf. \cite{Dbasic}), the above argument can be used to prove that $\intethree v_{h_{\ast}}^{2} \mu_0 \circ f_{Q}^{-1}(\ud \vb)$ is finite, where $h_{\ast}$ is invariant w.r.t. $Q$ and $\mu_0$. At the end, choose $f_Q$ firstly equal to $(v_1, v_2, v_3) \mapsto (v_2, v_3, v_1)$ and, then, equal to $(v_1, v_2, v_3) \mapsto (v_3, v_1, v_2)$ to complete the proof.

\appendix
\section{Appendix}

Gathered here are the proofs of unproved propositions and formulas scattered throughout Sections \ref{sect:intro} and \ref{sect:proof2}.

\subsection{Proof of (\ref{eq:gare}), (\ref{eq:garezeta}) and (\ref{eq:gareeta})} \label{a:gare}

Fix $s > 0$ and define
\begin{eqnarray}
\mathrm{A}_{1}^{(s)}(\nu, \tau_{\nu}) &:=& \et\Big{(}\sum_{j = 1}^{\nu} |\pi_{j, \nu}|^s\ \ \big{|} \ \nu, \tau_{\nu} \Big{)}
\nonumber \\
\mathrm{A}_2(\nu, \tau_{\nu}) &:=& \et\Big{(}\sum_{j = 1}^{\nu} \pi_{j, \nu}^{2} |\zeta_{j, \nu}|\ \ \big{|} \ \nu, \tau_{\nu} \Big{)} \nonumber \\
\mathrm{A}_3(\nu, \tau_{\nu}) &:=& \et\Big{(}\sum_{j = 1}^{\nu}|\pi_{j, \nu}^{3} \eta_{j, \nu}|\ \ \big{|} \ \nu, \tau_{\nu} \Big{)}\ . \nonumber
\end{eqnarray}
These functions satisfy the relations
\begin{equation} \label{eq:recurrenceLR}
\begin{array}{lll}
\mathrm{A}(1, \tree_1) &= 1 & {} \\
\mathrm{A}(n, \treen) &= \alpha[\mathrm{A}(n_l, \treen^l) + \mathrm{A}(n_r, \treen^r)] & \text{if}\ n \geq 2
\end{array}
\end{equation}
for every $n$ in $\mathbb{N}$, $\treen$ in $\mathbb{T}(n)$ and for some suitable constant $\alpha$. This claim is checked for each of them, following a common scheme of reasoning. First,
$\mathrm{A}_{1}^{(s)}(1, \tree_1) = \mathrm{A}_2(1, \tree_1) = \mathrm{A}_3(1, \tree_1) = 1$ holds by definition. Then, to obtain
the latter identity in (\ref{eq:recurrenceLR}) as regards $\mathrm{A}_{1}^{(s)}$, utilize (\ref{eq:pijnast}) in the equality
$$
\mathrm{A}_{1}^{(s)}(n, \treen) = \int_{[0, \pi]^{n-1}} \sum_{j = 1}^{n} |\pi_{j, n}^{\ast}(\treen, \boldsymbol{\varphi})|^s \beta^{\otimes_{n-1}}(\ud \boldsymbol{\varphi}) \ .
$$
Thus, $\alpha = \int_{0}^{\pi} |\cos \varphi|^s \beta(\ud \varphi) = \int_{0}^{\pi} |\sin \varphi|^s \beta(\ud \varphi) = l_s(b)$, where the validity of the exchange of $\cos$ with $\sin$ is a consequence of (\ref{eq:bsymm}). As to $\mathrm{A}_2$, use (\ref{eq:pijnast}) and (\ref{eq:zetajnast}) in
$$
\mathrm{A}_2(n, \treen) = \int_{[0, \pi]^{n-1}} \sum_{j = 1}^{n} |\pi_{j, n}^{\ast}(\treen, \boldsymbol{\varphi})|^2 |\zeta_{j, n}^{\ast}(\treen, \boldsymbol{\varphi})| \beta^{\otimes_{n-1}}(\ud \boldsymbol{\varphi})
$$
to show that $\mathrm{A}_2$ satisfies the latter identity in (\ref{eq:recurrenceLR}) with $\alpha = f(b)$. Passing to $\mathrm{A}_3$, consider (\ref{eq:pijnast}) and (\ref{eq:etajnast}) in conjunction with
$$
\mathrm{A}_3(n, \treen) = \int_{[0, \pi]^{n-1}} \sum_{j = 1}^{n} |\pi_{j, n}^{\ast}(\treen, \boldsymbol{\varphi})|^3 |\eta_{j, n}^{\ast}(\treen, \boldsymbol{\varphi})| \beta^{\otimes_{n-1}}(\ud \boldsymbol{\varphi})
$$
to verify that $\mathrm{A}_3$ meets the latter identity in (\ref{eq:recurrenceLR}) with $\alpha = g(b)$.

At this stage, since $\delta_j(\treen^l) + 1 = \delta_j(\treen)$ for $j = 1, \dots, n_l$ and $\delta_j(\treen^r) + 1 = \delta_{j+n_l}(\treen)$ for $j = 1, \dots, n_r$, an induction argument yields $\mathrm{A}(n, \treen) = \sum_{j = 1}^{n} \alpha^{\delta_j(\treen)}$, where $\delta_j$ is the depth defined in Subsection \ref{sect:representation}. By the concept of germination explained in Subsection \ref{sect:representation}, $\delta_j(\treenk) = \delta_j(\treen) + \delta_{j, k} + \delta_{j, k+1}$ for $j = 1, \dots, k+1$, with $\delta_{r, s}$ standing for the Kronecker delta, and $\delta_j(\treenk) = \delta_{j+1}(\treen)$ for $j = k+2, \dots, n$. Then, the specific form of $\mathrm{A}(n, \treen)$ shows that
\begin{equation} \label{eq:bassettirecurrence}
\frac{1}{n} \sum_{k = 1}^{n} \mathrm{A}(n+1, \treenk) = \left(1 + \frac{2\alpha - 1}{n}\right) \mathrm{A}(n, \treen)
\end{equation}
holds for every $n$ in $\mathbb{N}$ and $\treen$ in $\mathbb{T}(n)$. Now, since $\et[\mathrm{A}(n+1, \tau_{n+1})\ | \ \tau_n = \treen] = \sum_{k = 1}^{n}\mathrm{A}(n+1, \treenk)\pt[\tau_{n+1} = \treenk\ | \ \tau_n = \treen]$, (\ref{eq:markov}) and $(\ref{eq:bassettirecurrence})$ imply that $a_n := \et[\mathrm{A}(n, \tau_n)]$ satisfies $a_1 = 1$ and $a_{n+1} = \left(1 + \frac{2\alpha - 1}{n}\right) a_n$ for every $n$ in $\mathbb{N}$. Hence, if $(1 - 2\alpha)$ does not belong to $\mathbb{N}$,  $a_n = \frac{\Gamma(n + 2\alpha - 1)}{\Gamma(n) \Gamma(2\alpha)}$ for every $n$ in $\mathbb{N}$. Otherwise, if $(1 - 2\alpha) = m$, then $a_n = (-1)^{n+1} \binom{m-1}{n-1}$ for $n = 1, \dots, m$ and $a_n = 0$ for $n > m$. Finally, note that the expectations in (\ref{eq:gare}), (\ref{eq:garezeta}) and (\ref{eq:gareeta}) coincide with $\et[\mathrm{A}_{1}^{(s)}]$, $\et[\mathrm{A}_2]$ and $\et[\mathrm{A}_3]$ respectively, and that $\et[\mathrm{A}(\nu, \tau_{\nu})\ | \ \nu] = a_{\nu}$, in view of the stochastic independence of $\nu$ and $\{\tau_n\}_{n \geq 1}$. Therefore, conclude by observing that $\et[a_{\nu}] = \sum_{n = 1}^{\infty} a_n e^{-t} (1 - e^{-t})^{n-1} = e^{-(1 - 2\alpha)t}$.
%\begin{eqnarray}
%\sum_{j = 1}^{n} |\pi_{j, n}^{\ast}(\treen, \boldsymbol{\varphi})|^s &=& |\cos \varphi_{n-1}|^s \sum_{j = 1}^{n_l}
%|\pi_{j, n_l}^{\ast}(\treen^l, \boldsymbol{\varphi}^l)|^s \nonumber \\
%&+& |\sin \varphi_{n-1}|^s \sum_{j = 1}^{n_r} |\pi_{j, n_r}^{\ast}(\treen^r, \boldsymbol{\varphi}^r)|^s \nonumber
%\end{eqnarray}
%\begin{eqnarray}
%&{}& \sum_{j = 1}^{n} |\pi_{j, n}^{\ast}(\treen, \boldsymbol{\varphi})|^2 |\zeta_{j, n}^{\ast}(\treen , %\boldsymbol{\varphi})| \nonumber \\
%&=& \cos^2\varphi_{n-1} \ \big{|} \frac{3}{2} \sin^2\varphi_{n-1} - 1 \big{|} \sum_{j = 1}^{n_l} |\pi_{j, %n_l}^{\ast}(\treen^l, \boldsymbol{\varphi}^l)|^2 |\zeta_{j, n_l}^{\ast}(\treen^l, \boldsymbol{\varphi}^l)| \nonumber \\
%&+& \sin^2\varphi_{n-1} \ \big{|} \frac{3}{2} \cos^2\varphi_{n-1} - 1 \big{|}
%\cdot \sum_{j = 1}^{n_r} |\pi_{j, n_r}^{\ast}(\treen^r, \boldsymbol{\varphi}^r)|^2 |\zeta_{j, n_r}^{\ast}(\treen^r, %\boldsymbol{\varphi}^r)| \nonumber
%\end{eqnarray}
%\begin{eqnarray}
%&{}& \sum_{j = 1}^{n} |\pi_{j, n}^{\ast}(\treen, \boldsymbol{\varphi})|^3 |\eta_{j, n}^{\ast}(\treen , \boldsymbol{\varphi})| %\nonumber \\
%&=& \cos^4\varphi_{n-1} \ \big{|} \frac{5}{2} \sin^2\varphi_{n-1} - 1 \big{|} \sum_{j = 1}^{n_l} |\pi_{j, %n_l}^{\ast}(\treen^l, \boldsymbol{\varphi}^l)|^3 |\eta_{j, n_l}^{\ast}(\treen^l, \boldsymbol{\varphi}^l)| \nonumber \\
%&+& \sin^4\varphi_{n-1} \ \big{|} \frac{5}{2} \cos^2\varphi_{n-1} - 1 \big{|}
%\cdot \sum_{j = 1}^{n_r} |\pi_{j, n_r}^{\ast}(\treen^r, \boldsymbol{\varphi}^r)|^3 |\eta_{j, n_r}^{\ast}(\treen^r, %\boldsymbol{\varphi}^r)| \nonumber
%\end{eqnarray}

\subsection{Probability law of $\{\tau_n\}_{n \geq 1}$} \label{a:pntn}

The aim is to show that the coefficient $p_n(\treen)$ in the Wild-McKean sum is equal to $\pt[\tau_n = \treen]$ for every $n$. Proceeding by mathematical induction, observe that the assertion is trivially true for $n = 1, 2$. To treat the case $n \geq 3$, introduce the symbol $\mathbb{P}(\treen)$ to denote the subset of $\mathbb{T}(n-1)$ of the trees which are able to produce $\treen$ by germination. Whence,
\begin{eqnarray}
\pt[\tau_n = \treen] &=& \sum_{\treesn \in \ptreen} \pt[\tau_n = \treen \ | \ \tau_{n-1} = \treesn] \ \pt[\tau_{n-1} = \treesn] \nonumber \\
&=& \frac{1}{n-1} \sum_{\treesn \in \ptreen} \pt[\tau_{n-1} = \treesn] = \frac{1}{n-1} \sum_{\treesn \in \ptreen} p_{n-1}(\treesn) \nonumber
\end{eqnarray}
the last equality being valid thanks to the inductive hypothesis. Now,
$$
\sum_{\treesn \in \ptreen} p_{n-1}(\treesn) = \sum_{\substack{\treesn \in \ptreen \\
\treesn^l = \treen^l}} p_{n-1}(\treesn) \ + \sum_{\substack{\treesn \in \ptreen \\
\treesn^r = \treen^r}} p_{n-1}(\treesn)
$$
and, by (\ref{eq:pntn}), the RHS turns out to be equal to
$$
\frac{1}{n-2} \Big{[} p_{n_l}(\treen^l) \sum_{\substack{\treesn \in \ptreen \\
\treesn^l = \treen^l}} p_{n_r - 1}(\treesn^r) \ + \ p_{n_r}(\treen^r) \sum_{\substack{\treesn \in \ptreen \\
\treesn^r = \treen^r}} p_{n_l - 1}(\treesn^l) \Big{]} \ . \nonumber
$$
At this stage, observe that
\begin{eqnarray}
&{}& \sum_{\substack{\treesn \in \ptreen \\
\treesn^l = \treen^l}}
p_{n_r - 1}(\treesn^r) = \sum_{\mathfrak{s}_{n_r - 1} \in \mathbb{P}(\treen^r)} p_{n_r - 1}(\mathfrak{s}_{n_r - 1}) \nonumber \\
&=& (n_r - 1) \sum_{\mathfrak{s}_{n_r - 1} \in \mathbb{P}(\treen^r)} \pt[\tau_{n_r - 1} = \mathfrak{s}_{n_r - 1}] \pt[\tau_{n_r} = \treen^r \ | \ \tau_{n_r - 1} = \mathfrak{s}_{n_r - 1}] \nonumber \\
&=& (n_r - 1) \pt[\tau_{n_r} = \treen^r] = (n_r - 1) p_{n_r}(\treen^r) \nonumber
\end{eqnarray}
and that the same procedure yields $\sum_{\substack{\treesn \in \ptreen \\ \treesn^r = \treen^r}} p_{n_l - 1}(\treesn^l) = (n_l - 1) p_{n_l}(\treen^l)$. To complete the proof it is enough to combine the previous equations and to recall that $n = n_l + n_r$.

\subsection{A few interesting characteristics of $\mathcal{C}[\zeta, \eta; \varphi]$} \label{a:C}

The first point concerns the invariance of (\ref{eq:internalintegralbobylev}) w.r.t. the choice of $\{\mathbf{a}(\ub), \mathbf{b}(\ub),\\ \ub \}$. Fix $\xib \neq \mathbf{0}$ and let $\{\mathbf{a}(\ub), \mathbf{b}(\ub), \ub \}$ and $\{\mathbf{a}^{'}(\ub), \mathbf{b}^{'}(\ub), \ub \}$ be distinct positive bases. Then, write $\boldsymbol{\psi}^l$ and $\boldsymbol{\psi}^r$ in (\ref{eq:psi}) with $\{\mathbf{a}^{'}(\ub), \mathbf{b}^{'}(\ub), \ub \}$ in the place of $\{\mathbf{a}(\ub), \mathbf{b}(\ub), \ub \}$. Since there exists some $\theta^{\ast}$ in $[0, 2\pi)$ such that $\mathbf{a}^{'} = \cos\theta^{\ast} \mathbf{a} - \sin\theta^{\ast} \mathbf{b}$ and $\mathbf{b}^{'} = \sin\theta^{\ast} \mathbf{a} + \cos\theta^{\ast} \mathbf{b}$, the change of basis gives
$$
\begin{array}{lll}
\boldsymbol{\psi}^l(\varphi, \theta, \ub) &= \cos(\theta - \theta^{\ast}) \sin\varphi \mathbf{a}(\ub) + \sin(\theta - \theta^{\ast}) \sin\varphi \mathbf{b}(\ub) + \cos\varphi \ub \\
\boldsymbol{\psi}^r(\varphi, \theta, \ub) &= -\cos(\theta - \theta^{\ast}) \cos\varphi \mathbf{a}(\ub) - \sin(\theta - \theta^{\ast}) \cos\varphi \mathbf{b}(\ub) + \sin\varphi \ub \ .
\end{array}
$$
After substituting these expressions in (\ref{eq:internalintegralbobylev}), the desired conclusion follows from an obvious change of variable.

To prove the measurability of $(\xib, \varphi) \mapsto I(\xib, \varphi)$, resort to Proposition 9 in Section 9.3 of \cite{frgr}, so that it is enough to verify the continuity of $\varphi \mapsto I(\xib, \varphi)$ for each fixed $\xib$ and the measurability of $\xib \mapsto I(\xib, \varphi)$ for each fixed $\varphi$. The former claim follows from the form of the dependence on $\varphi$ in (\ref{eq:psi})-(\ref{eq:internalintegralbobylev}). To verify the latter, one can show that also $\xib \mapsto I(\xib, \varphi)$ is continuous for each fixed $\varphi$. Continuity at $\xib = \mathbf{0}$ can be derived from the relation $|\boldsymbol{\psi}^l| = |\boldsymbol{\psi}^r| = 1$ and an ensuing application of the dominated convergence theorem. To check continuity at $\xib^{\ast} \neq \mathbf{0}$, take a sequence $\{\xib_n\}_{n \geq 1}$ converging to $\xib^{\ast}$ and observe that $|\xib_n| \rightarrow |\xib^{\ast}|$ and $\ub_n := \xib_n/|\xib_n| \rightarrow \ub^{\ast} := \xib^{\ast}/|\xib^{\ast}|$. Fix a small open neighborhood $\Omega(\ub^{\ast}) \subset S^2$ of $\ub^{\ast}$ in such a way that $S^2 \setminus \overline{\Omega(\ub^{\ast})}$ contains at least two antipodal points. In view of the first part of this appendix, choose a distinguished basis in such a way that the restrictions of $\ub \mapsto \mathbf{a}(\ub)$ and $\ub \mapsto \mathbf{b}(\ub)$ to $\Omega(\ub^{\ast})$ vary with continuity. As a consequence, $\boldsymbol{\psi}^l(\varphi, \theta, \ub_n)$ converges to $\boldsymbol{\psi}^l(\varphi, \theta, \ub^{\ast})$ and $\boldsymbol{\psi}^r(\varphi, \theta, \ub_n)$ converges to $\boldsymbol{\psi}^r(\varphi, \theta, \ub^{\ast})$ for every $\varphi$ in $[0, \pi]$ and $\theta$ in $(0, 2\pi)$, and the convergence of $I(\xib_n, \varphi)$ to $I(\xib^{\ast}, \varphi)$ follows again from the dominated convergence theorem. To show that $\xib \mapsto I(\xib, \varphi)$ is a c.f. for every $\varphi$ in $[0, \pi]$, resort to the multivariate version of the Bochner characterization. See Exercise 3.1.9 in \cite{stro}. The only point that requires some care is positivity. If this property were not in force, one could find a positive integer $N$, two $N$-vectors $(\omega_1, \dots, \omega_N)$ and $(\xib_1, \dots, \xib_N)$ in $\mathbb{C}^N$ and $(\rthree)^N$ respectively, and some $\varphi^{\ast}$ in $[0, \pi]$ in such a way that $\sum_{j = 1}^{N} \sum_{k = 1}^{N} \omega_j \overline{\omega}_k I(\xib_j - \xib_k, \varphi^{\ast}) < 0$. Note that the LHS of this inequality is a real number since $I(-\xib, \varphi) = \overline{I(\xib, \varphi)}$ for any $\xib$ and $\varphi$. Hence, by continuity of $\varphi \mapsto I(\xib, \varphi)$, there exists an open interval $J$ in $[0, \pi]$ containing $\varphi^{\ast}$ such that $\varphi \mapsto \sum_{j = 1}^{N} \sum_{k = 1}^{N} \omega_j \overline{\omega}_k I(\xib_j - \xib_k, \varphi)$ is strictly negative on $J$. Now, choose a specific $b_{\ast}$ for which the resulting p.m. in (\ref{eq:beta}), say $\beta_{\ast}$, is supported by $\overline{J}$. By construction, $L := \int_{0}^{\pi} \sum_{j = 1}^{N} \sum_{k = 1}^{N} \omega_j \overline{\omega}_k I(\xib_j - \xib_k, \varphi) \beta_{\ast}(\ud \varphi)$ is a strictly negative number, a fact which immediately leads to a contradiction. Indeed, denote by $\mathcal{Q}_{\ast}[\zeta, \eta]$ the RHS of (\ref{eq:extendedQ}) when $b_{\ast}$ replaces $b$ in the definition of $Q[p, q]$. Observe that $\mathcal{Q}_{\ast}[\zeta, \eta]$ is in any case a p.m. even if $b_{\ast}$ does not meet (\ref{eq:bsymm}). Now, $L$ must be equal to $\sum_{j = 1}^{N} \sum_{k = 1}^{N} \omega_j \overline{\omega}_k \hat{\mathcal{Q}}_{\ast}[\zeta, \eta](\xib_j - \xib_k)$, thanks to (\ref{eq:parametrizedbobylev}), and this quantity must be non-negative, from the Bochner criterion again.

To prove (\ref{eq:QCbis}), start by verifying that $\varphi \mapsto \mathcal{C}[\zeta, \eta; \varphi]$ is measurable, which is tantamount to checking that $\varphi \mapsto \mathcal{C}[\zeta, \eta; \varphi](K)$ is measurable for every $K = \textsf{X}_{\substack{i=1}}^{3} (-\infty, x_i]$, in view of Lemma 1.40 of \cite{ka}. To this aim, fix such a $K$ and use the Fubini theorem to show that
$$
(\mathbf{a}, \mathbf{b}, c, \varphi) \mapsto \left(\frac{1}{2\pi}\right)^3 \int_{-c}^{c} \int_{-c}^{c} \int_{-c}^{c} \left[\prod_{m=1}^{3} \frac{e^{-i\xi_m a_m} - e^{-i\xi_m b_m}}{i\xi_m} \right] \hat{\mathcal{C}}[\zeta, \eta; \varphi](\xib) \dxi
$$
is measurable, since $(\xib, \varphi) \mapsto \hat{\mathcal{C}}[\zeta, \eta; \varphi](\xib)$ does. To complete the argument, invoke the inversion formula and note that $\mathcal{C}[\zeta, \eta; \varphi](K)$ is equal to the limit of the above expression as $c \uparrow +\infty$, $a_m \downarrow -\infty$ and $b_m \downarrow x_m$ for $m = 1, 2, 3$. This paves the way for writing the integral in (\ref{eq:QCbis}), and the equality therein follows from (\ref{eq:parametrizedbobylev}) and (\ref{eq:internalintegralbobylev})
in view of the injectivity of the Fourier-transform operator.

\subsection{Proof of (\ref{eq:QCtnbis})} \label{a:QCtnbis}

The first step is to show that $\boldsymbol{\varphi} \mapsto \mathcal{C}_{\treen}[\mu_0; \boldsymbol{\varphi}]$ is measurable as a map from $[0, \pi]^{n-1}$ into $\prob$. Mimicking the argument in the last part of \ref{a:C}, it suffices to verify the measurability of $(\xib, \boldsymbol{\varphi}) \mapsto \hat{\mathcal{C}}_{\treen}[\mu_0; \boldsymbol{\varphi}](\xib)$ by means of
Proposition 9 in Section 9.3 of \cite{frgr}. On the one hand, the function $\xib \mapsto\hat{\mathcal{C}}_{\treen}[\mu_0; \boldsymbol{\varphi}](\xib)$ is continuous for every fixed $\boldsymbol{\varphi}$. On the other hand, measurability of $\boldsymbol{\varphi} \mapsto \hat{\mathcal{C}}_{\treen}[\mu_0; \boldsymbol{\varphi}](\xib)$, for every fixed $\xib$, can be proved by induction. When $n = 1$, $\hat{\mathcal{C}}_{\tree_1}[\mu_0; \emptyset](\xib)$ is independent of $\boldsymbol{\varphi}$ and the claim is obvious. When $n \geq 2$, it suffices to recall (\ref{eq:inductionCtn}) and to exploit the inductive hypothesis. To conclude, the equality
\begin{equation} \label{eq:QCtnbisfourier}
\hat{\mathcal{Q}}_{\treen}[\mu_0](\xib) = \int_{[0, \pi]^{n-1}} \hat{\mathcal{C}}_{\treen}[\mu_0; \boldsymbol{\varphi}](\xib) \beta^{\otimes_{n-1}}(\ud \boldsymbol{\varphi})
\end{equation}
for $n = 2, 3, \dots$ will be proved by mathematical induction. First, when $n = 2$, (\ref{eq:QCtnbisfourier}) is valid since it
coincides with (\ref{eq:parametrizedbobylev}). When $n \geq 3$, combine the definition of $\mathcal{Q}_{\treen}$ with (\ref{eq:parametrizedbobylev}) to obtain
$$
\hat{\mathcal{Q}}_{\treen}[\mu_0](\xib) = \int_{0}^{\pi}\int_{0}^{2\pi} \hat{\mathcal{Q}}_{\treen^l}[\mu_0](\rho \cos\varphi \boldsymbol{\psi}^l) \hat{\mathcal{Q}}_{\treen^r}[\mu_0](\rho \sin\varphi \boldsymbol{\psi}^r) u_{(0, 2\pi)}(\ud \theta) \beta(\ud \varphi)
$$
and the argument is completed by invoking the inductive hypothesis, the definition of $\mathcal{C}_{\treen}$ and (\ref{eq:parametrizedbobylev}). Therefore, (\ref{eq:QCtnbisfourier}) entails (\ref{eq:QCtnbis}) in view of the injectivity of the Fourier-transform operator.

\subsection{Proof of Proposition {\ref{prop:tails}}} \label{a:tails}

Put $k := \lceil 2/p \rceil$ with $p$ as in (\ref{eq:tailsCDGR}) and consider the random vector $\mathbf{S} = (S_1, S_2, S_3) := \sum_{j=1}^{2k} (-1)^{j} \mathbf{V}_j$, whose c.f. $\phi$ is given by $\phi(\xib) = |\hat{\mu}_{0}(\xib)|^{2k}$. The assumptions
(\ref{eq:normalizations})-(\ref{eq:covariancematrix}) plainly entail $\et\left[\mathbf{S}\right] = \mathbf{0}$, $\et\left[S_i S_j\right] = 0$ for $i \neq j$, and $\et\left[S_{i}^{2}\right] = 2k \sigma_{i}^{2}$ for $i = 1, 2, 3$. Note also that $\sigma_{i}^{2} > 0$ for $i = 1, 2, 3$ as a consequence of (\ref{eq:tailsCDGR}). Moreover, thanks to the Lyapunov inequality, $\et\left[|\mathbf{S}|^3\right] \leq (2k)^3 \mathfrak{m}_3$. Now, standard arguments explained, e.g., in Section 8.4 of \cite{chte} show that
$$
\phi(\xib) \leq 1 - k\sigma_{\ast}^{2} |\xib|^2 + \frac{(2k)^3 \mathfrak{m}_3}{6} |\xib|^3
$$
with $\sigma_{\ast}^{2} := \min\{\sigma_{1}^{2}, \sigma_{2}^{2}, \sigma_{3}^{2}\}$. Thus, $\phi(\xib) \leq 1 - \frac{k}{2} \sigma_{\ast}^{2} |\xib|^2$ whenever $|\xib| \leq (3 \sigma_{\ast}^{2})/(8 k^2 \mathfrak{m}_3)$, and elementary algebra entails $1 - \frac{k}{2} \sigma_{\ast}^{2} |\xib|^2 \leq \frac{\lambda^2}{\lambda^2 + |\xib|^2}$ for every $\xib$, provided that $\lambda^2 \geq 2/k\sigma_{\ast}^{2}$. Now, (\ref{eq:tailsCDGR}) gives $|\phi(\xib)| \leq L|\xib|^{-4}$ for every $\xib \neq \mathbf{0}$, with $L := (\sup_{\xib \in \rthree} |\xib|^p |\hat{\mu}_0(\xib)|)^{4/p}$, and again some algebra entails $L|\xib|^{-4} \leq \frac{\lambda^2}{\lambda^2 + |\xib|^2}$ if $|\xib|^2 \geq B(\lambda) := (L + \sqrt{L^2 + 4L\lambda^4})/(2\lambda^2)$.
Note that $B(\lambda) \leq 2\sqrt{L}$ holds true when $\lambda^2 \geq (2\sqrt{L})/3$. At this stage, choosing any $\lambda$ satisfying $\lambda^2 \geq \max\{2/k\sigma_{\ast}^{2}, (2\sqrt{L})/3\}$ yields
\begin{equation} \label{eq:globtailsphi}
|\phi(\xib)| \leq \frac{\lambda^2}{\lambda^2 + |\xib|^2}
\end{equation}
for every $\xib$ such that either $|\xib| \leq (3 \sigma_{\ast}^{2})/(8 k^2 \mathfrak{m}_3)$ or $|\xib|^2 \geq 2\sqrt{L}$. Therefore, the proof is completed if $(4L)^{1/4} \leq (3 \sigma_{\ast}^{2})/(8 k^2 \mathfrak{m}_3)$. Otherwise, if $(4L)^{1/4} > (3 \sigma_{\ast}^{2})/(8 k^2 \mathfrak{m}_3)$, define
$$
M := \sup_{\{(3 \sigma_{\ast}^{2})/(8 k^2 \mathfrak{m}_3) \leq |\xib| \leq (4L)^{1/4}\}} |\phi(\xib)|
$$
and resort to Corollary 2 in Section 8.4 of \cite{chte} to state that $M < 1$. Then, (\ref{eq:globtailsphi}) holds true also when $(3 \sigma_{\ast}^{2})/(8 k^2 \mathfrak{m}_3) \leq |\xib| \leq (4L)^{1/4}$ if $M \leq \inf_{(3 \sigma_{\ast}^{2})/(8 k^2 \mathfrak{m}_3) \leq |\xib| \leq (4L)^{1/4}}\left(\frac{\lambda^2}{\lambda^2 + |\xib|^2}\right)$, the last inequality being equivalent to $\lambda^2 \leq 2\sqrt{L} M/(1 - M)$. In conclusion, taking
$$
\lambda^2 := \max\{2/k\sigma_{\ast}^{2}, (2\sqrt{L})/3, 2\sqrt{L} M/(1 - M)\}
$$
leads to state that (\ref{eq:globtailsphi}) is valid for every $\xib$, and (\ref{eq:globtails}) follows.

\subsection{Proof of Proposition {\ref{prop:beurling}}} \label{a:beurling}

Initially, suppose that $\chi(\dx) = f(\xb) \dx$ for some $f$ in $\mathrm{L}^1(\rthree)$. Therefore, $\Delta \hat{\chi}(\xib) = \intethree |\xb|^2 f(\xb) e^{i \xb \cdot \xib} \dx$ and then, by the Plancherel identity,
$$
\intethree |f(\xb)|^2 (1 + |\xb|^4) \dx = \left(\frac{1}{2\pi}\right)^3 \intethree \left[|\hat{\chi}(\xib)|^2 + |\Delta \hat{\chi}(\xib)|^2\right] \dxi \ .
$$
Now, note that $|\chi|(\rthree) = \intethree |f(\xb)| \dx$ and apply the Cauchy-Schwartz inequality to get
$$
\intethree |f(\xb)| \dx\ \leq\ \Big{(}\intethree \frac{\dx}{1 + |\xb|^4} \Big{)}^{1/2} \cdot \Big{(}\intethree |f(\xb)|^2 (1 + |\xb|^4) \dx \Big{)}^{1/2}
$$
where $\intethree \frac{\dx}{1 + |\xb|^4} = \sqrt{2}\pi^2$. For a general $\chi$, consider the convolution $\chi_{\epsilon}$ of $\chi$ with the Gaussian distribution of zero mean and covariance matrix $\epsilon^2 \mathrm{I}$. Since $\chi_{\epsilon}$ is absolutely continuous, the first part of the proof gives
$$
|\chi_{\epsilon}|(\rthree)\ \leq 2^{-5/4}\pi^{-1/2} \Big{(}\intethree \left[|\hat{\chi}_{\epsilon}(\xib)|^2 + |\Delta \hat{\chi}_{\epsilon}(\xib)|^2\right] \dxi\Big{)}^{1/2}
$$
and thereby, taking account of $|\chi|(\rthree) \leq \liminf_{\epsilon \downarrow 0} |\chi_{\epsilon}|(\rthree)$ and letting $\epsilon \downarrow 0$,
$$
|\chi|(\rthree)\ \leq 2^{-5/4}\pi^{-1/2} \Big{(}\intethree \left[|\hat{\chi}(\xib)|^2 + |\Delta \hat{\chi}(\xib)|^2\right] \dxi\Big{)}^{1/2}\ .
$$
To complete the argument, observe that $\sup_{B \in \borelthree} |\chi(B)| \leq |\chi|(\rthree)$.

\subsection{Proof of Proposition \ref{prop:elmroth}} \label{a:elmroth}

Taking account of (\ref{eq:normalizations}), note that
$$
\et\left[(S(\ub))^2 \ | \ \mathscr{H} \right] = \sum_{j=1}^{\nu} \pi_{j, \nu}^2 \et\left[(\mathbf{V}_j \cdot \psib_{j, \nu}(\ub))^2 \ | \ \mathscr{H} \right] \leq \mathfrak{m}_2
$$
with $\mathfrak{m}_2 = 3$. The equality emanates by virtue of the stochastic independence of the $\mathbf{V}_j$'s while the inequality follows from the combination of the Cauchy-Schwartz inequality with (\ref{eq:sumpijn}) and the identity $|\psib_{j, n}(\ub)| = 1$. Thus, (\ref{eq:momlS}) holds true for $h = 2$ with $g_2 = \mathfrak{m}_2$. The case $h = 1$ can be derived from the case $h = 2$ thanks to the conditional Lyapunov inequality after putting $g_1 = \sqrt{g_2}$. When $h \geq 3$, an inequality due to Rosenthal (see Section 2.3 in \cite{pe1}) yields
\begin{eqnarray}
\et\left[|S(\ub)|^h \ | \ \mathscr{H} \right] &\leq& c(h) \Big{\{} \sum_{j = 1}^{\nu} \et\left[ |\pi_{j, \nu} \mathbf{V}_j \cdot \boldsymbol{\psi}_{j, \nu}(\ub)|^h \ | \ \mathscr{H} \right] \nonumber \\
&+& \Big{(} \sum_{j = 1}^{\nu} \et\left[|\pi_{j, \nu} \mathbf{V}_j \cdot
\boldsymbol{\psi}_{j, \nu}(\ub)|^2\ | \ \mathscr{H} \right] \Big{)}^{h/2} \Big{\}}  \nonumber
\end{eqnarray}
where $c(h)$ is a positive constant depending only on $h$. An additional application of the Cauchy-Schwartz inequality, combined with (\ref{eq:sumpijn}) and $|\psib_{j, n}(\ub)| = 1$, gives
$$
\et\left[|S(\ub)|^h \ | \ \mathscr{H} \right] \leq c(h) \Big{\{} \mathfrak{m}_h \sum_{j = 1}^{\nu} |\pi_{j, \nu}|^h + \Big{(}\mathfrak{m}_2 \sum_{j = 1}^{\nu} \pi_{j, \nu}^2 \Big{)}^{h/2}\Big{\}} \leq c(h) \{\mathfrak{m}_h  + \mathfrak{m}_{2}^{h/2}\}
$$
which entails (\ref{eq:momlS}) with $g_h = c(h) \{\mathfrak{m}_h  + \mathfrak{m}_{2}^{h/2}\}$. Now, $\frac{\partial^h}{\partial \rho^h}\Nrand(\rho; \ub)$ exists and is uniformly bounded by $g_h$, for $h = 1, \dots, 2k$. Then, since $\hat{\Mrand}(\rho\ub) = \et\left[\hat{\Nrand}(\rho; \ub) \ | \ \mathscr{G} \right]$ and the interchanging of the derivative with the expectation is here valid, one gets (\ref{eq:boundderivativesMhat}). Finally, taking $\ub = \mathbf{e}_i$ in (\ref{eq:boundderivativesMhat}) yields $\intethree v_{i}^{2k}\Mrand(\ud\vb) < +\infty$ for $i = 1, 2, 3$ which, in turn, entails (\ref{eq:finmom2M}).

\subsection{Proof of Proposition \ref{prop:extgauss}} \label{a:gauss}

The definition of the $k$-th Hermite polynomial shows that $\frac{\ud^k}{\ud \rho^k} e^{-\rho^2/2} = 2^{-k/2} H_k\left(\frac{\rho}{\sqrt{2}}\right) e^{-\rho^2/2}$ for every $k$ in $\mathbb{N}_0$ and $\rho$ in $\rone$. See, for example, (1) in Section 2.IV of \cite{san}. Moreover, according to ($9_2$) therein,
$$
H_k\left(\frac{\rho}{\sqrt{2}}\right) = k! \sum_{h =0}^{[k/2]} \frac{(-1)^{k+h}}{h! (k - 2h)!} (\sqrt{2} \rho)^{k - 2h}
$$
where $[n]$ stands for the integral part of $n$, and hence
$$
\int_{x}^{+\infty} \left( \frac{\ud^k}{\ud \rho^k} e^{-\rho^2/2} \right)^2 \rho^m \ud \rho
= \sum_{h =0}^{[k/2]} \sum_{l =0}^{[k/2]} \gamma_{k, h, l} \int_{x}^{+\infty} \rho^{m + 2(k - h -l)} e^{-\rho^2} \ud \rho
$$
with $\gamma_{k, h, l} := \frac{(-2)^{-h -l}(k!)^2}{h! l! (k - 2h)! (k - 2l)!}$. Now, take account of the following elementary inequalities: $\int_{x}^{+\infty} e^{- \rho^2/2} \ud \rho \leq \ \frac{1}{x} e^{- x^2/2}$ for
$x > 0$, and $\rho^t e^{-\rho^2/2} \leq \ (t/e)^{t/2}$ for $\rho \geq 0$ and $t \geq 0$, with the proviso that $0^0 := 1$ when $t = 0$. Whence,
\begin{eqnarray}
&& \int_{x}^{+\infty} \left( \frac{\ud^k}{\ud \rho^k} e^{-\rho^2/2} \right)^2 \rho^m \ud \rho \nonumber \\
&\leq& \sum_{h =0}^{[k/2]} \sum_{l =0}^{[k/2]} |\gamma_{k, h, l}| \left(\frac{m + 2(k - h -l)}{e}\right)^{m/2 + k - h -l} \frac{1}{x} e^{- x^2/2} \nonumber \\
&\leq& c(m, s, k) x^{-s} \nonumber
\end{eqnarray}
with $c(m, s, k) := \sum_{h =0}^{[k/2]} \sum_{l =0}^{[k/2]} |\gamma_{k, h, l}| \left(\frac{m + 2(k - h -l)}{e}\right)^{m/2 + k - h -l} \left(\frac{s - 1}{e}\right)^{(s - 1)/2}$.

\subsection{Proof of Propositions {\ref{prop:boundderivativesMhat}} and \ref{prop:7091}} \label{a:boundderivativesMhat}

The main task is to prove (\ref{eq:boundPsiN}), (\ref{eq:boundderinternalN}) and (\ref{eq:7091}). The remaining inequalities (\ref{eq:boundPsi}) and (\ref{eq:boundderinternal}) can be derived by interchanging derivative with expectation
in the equality $\hat{\Mrand}(\rho\ub) = \et\left[\hat{\Nrand}(\rho; \ub) \ | \ \mathscr{G} \right]$, since $\Psi(\rho)$ is a $\mathscr{G}$-measurable random variable for every fixed $\rho$. To start, (\ref{eq:boundPsiN}) follows from the combination of (\ref{eq:N}), (\ref{eq:globtails}) and (\ref{eq:psigrande}), upon recalling that $|\psib_{j, \nu}| = 1$. With a view to proving (\ref{eq:boundderinternalN}) and (\ref{eq:7091}), it is worth noting that $0 \leq \rho \leq \RR$ entails $\sup_{\ub \in S^2} \big{|}\hat{\mu}_0(\rho \pi_{j, \nu} \psib_{j, \nu}(\ub)) - 1\big{|} \leq 19/128$ for $j = 1, \dots, \nu$ and for every choice of $\mathrm{B}$ in (\ref{eq:psijn}), as shown in \cite{doBE}. This paves the way for considering the principal value of the logarithm and then for writing
\begin{equation} \label{eq:principallog}
\hat{\Nrand}(\rho; \ub) = \exp\Big{\{}\sum_{j=1}^{\nu} \Log[ \hat{\mu}_0(\rho \pi_{j, \nu} \psib_{j, \nu}(\ub))]\Big{\}} \ .
\end{equation}
The next step concerns the computation of certain derivatives of $\hat{\Nrand}(\rho; \ub)$ by means of the above identity. To this aim, the system of coordinates introduced in Sub-subsection \ref{sect:outerL} comes now in useful. Then, for $k = 1, \dots, 4$,
\begin{eqnarray}
\frac{\partial}{\partial x} \hat{\Nrand}(\rho; \hb_k(u, v)) &=& \hat{\Nrand}(\rho; \hb_k(u, v)) \sum_{j=1}^{\nu} \frac{\partial}{\partial x} \Log[ \hat{\mu}_0(\rho \pi_{j, \nu} \psib_{j, \nu}(\hb_k(u, v)))] \label{eq:dNdx} \\
\frac{\partial^2}{\partial x^2} \hat{\Nrand}(\rho; \hb_k(u, v)) &=& \hat{\Nrand}(\rho; \hb_k(u, v))\Big{\{}\Big{(}\sum_{j=1}^{\nu}
\frac{\partial}{\partial x} \Log[ \hat{\mu}_0(\rho \pi_{j, \nu} \psib_{j, \nu}(\hb_k(u, v)))]\Big{)}^2 \nonumber \\
&+& \sum_{j=1}^{\nu} \frac{\partial^2}{\partial x^2} \Log[ \hat{\mu}_0(\rho \pi_{j, \nu} \psib_{j, \nu}(\hb_k(u, v)))]\Big{\}} \label{eq:dN2dx2}
\end{eqnarray}
where $x$ can be $\rho$, $u$ or $v$. To bound each of these products, use (\ref{eq:boundPsiN}) as far as $\hat{\Nrand}(\rho; \hb_k)$ is concerned, and proceed with the detailed computation of bounds for the derivatives of the logarithms. As a starting point for all these calculations, consider the following equalities from \cite{doBE}:
\begin{eqnarray}
\hat{\mu}_0(\rho \pi_{j, \nu} \psib_{j, \nu}(\ub)) &=& 1 - \frac{1}{2} \rho^2\pi_{j, \nu}^2 \intethree [\psib_{j, \nu}(\ub) \cdot \vb]^2 \mu_0(\ud \vb) \nonumber \\
&-& \frac{i}{3!}\rho^3\pi_{j, \nu}^3 \intethree [\psib_{j, \nu}(\ub) \cdot \vb]^3 \mu_0(\ud \vb) +
R_j(\rho, \ub) \ \ \ \ \ \ \ \label{eq:expansionmuhat}
\end{eqnarray}
and
\begin{gather}
\Log[ \hat{\mu}_0(\rho \pi_{j, \nu} \psib_{j, \nu}(\ub))] = - \frac{1}{2} \rho^2\pi_{j, \nu}^2 \intethree [\psib_{j, \nu}(\ub) \cdot \vb]^2 \mu_0(\ud \vb) \nonumber \\
- \frac{i}{3!}\rho^3\pi_{j, \nu}^3 \intethree [\psib_{j, \nu}(\ub) \cdot \vb]^3 \mu_0(\ud \vb) +
R_j(\rho, \ub) - \Phi(w_j(\rho, \ub)) w_{j}^{2}(\rho, \ub)\ . \label{eq:expansionphi}
\end{gather}
Here, $w_j(\rho, \ub) := \hat{\mu}_0(\rho \pi_{j, \nu} \psib_{j, \nu}(\ub)) - 1$,
$$
\Phi(z) := \frac{z - \Log(1 + z)}{z^2} = \int_{0}^{+\infty}\Big{(}\int_{x}^{+\infty} \frac{s - x}{s} e^{-s} \ud s\Big{)} e^{-z x} \ud x \ \ \ \ \ (\Re z > -1)
$$
and the remainder $R_j(\rho, \ub)$ can assume one of the following forms:
\begin{eqnarray}
&{}& \frac{1}{3!} \rho^4 \pi_{j, \nu}^4 \intethree\int_{0}^{1} [\psib_{j, \nu}(\ub) \cdot \vb]^4 (1 - s)^3
e^{i \rho \pi_{j, \nu} [\psib_{j, \nu}(\ub) \cdot \vb] s} \ud s \mu_0(\ud \vb) \nonumber \\
&=& -\frac{i}{2} \rho^3 \pi_{j, \nu}^3 \intethree\int_{0}^{1} [\psib_{j, \nu}(\ub) \cdot \vb]^3 (1 - s)^2
(e^{i \rho \pi_{j, \nu} [\psib_{j, \nu}(\ub) \cdot \vb] s} - 1) \ud s \mu_0(\ud \vb) \nonumber \\
&=& -\rho^2 \pi_{j, \nu}^2 \intethree\int_{0}^{1} [\psib_{j, \nu}(\ub) \cdot \vb]^2 (1 - s) \times \nonumber \\
&\times& \Big{(} e^{i \rho \pi_{j, \nu} [\psib_{j, \nu}(\ub) \cdot \vb] s} - 1 - i \rho \pi_{j, \nu} [\psib_{j, \nu}(\ub) \cdot \vb] s \Big{)} \ud s \mu_0(\ud \vb) \ . \nonumber
\end{eqnarray}
The aim is now to show that $\sum_{j=1}^{\nu}\frac{\partial^l}{\partial x^l} \Log[ \hat{\mu}_0(\rho \pi_{j, \nu} \psib_{j, \nu})]$ admits, for $l = 1, 2$, an upper bound presentable as a non-random polynomial in $\rho$, independent of $\ub$.

As far as the derivatives w.r.t. $\rho$ are concerned, for the first two terms on the RHS of (\ref{eq:expansionphi}) one gets
\begin{eqnarray}
&& \sup_{\ub \in S^2} \Big{|} \frac{\partial^l}{\partial \rho^l}\Big{[}- \frac{1}{2} \rho^2\pi_{j, \nu}^2 \intethree [\psib_{j, \nu} \cdot \vb]^2 \ud\mu_0 - \frac{i}{3!}\rho^3\pi_{j, \nu}^3 \intethree [\psib_{j, \nu} \cdot \vb]^3 \ud\mu_0 \Big{]} \Big{|} \nonumber \\
&\leq& \frac{1}{(2-l)!} \mathfrak{m}_2 \pi_{j, \nu}^2 \rho^{2-l} + \frac{1}{(3-l)!} \mathfrak{m}_3|\pi_{j, \nu}|^3 \rho^{3-l} \label{eq:BEE1}
\end{eqnarray}
for $l = 0, 1, 2$, thanks to the fact that $|\psib_{j, \nu}| = 1$. Moreover, recall that $\mathfrak{m}_2 = 3$ in view of (\ref{eq:normalizations}). Standard manipulations of the above expressions of $R_j(\rho, \ub)$ lead to
\begin{equation} \label{eq:BEE2}
\sup_{\ub \in S^2} \Big{|} \frac{\partial^l}{\partial \rho^l} R_j(\rho, \ub)\Big{|} \leq c_l(R) \mq \pi_{j, \nu}^{4} \rho^{4-l}
\end{equation}
for $l = 0, 1, 2$, with $c_0(R) = 1/24$, $c_1(R) = 1/6$ and $c_2(R) = 1/3$. See \cite{doBE} for the details. After recalling (\ref{eq:r}), this last inequality plainly entails
\begin{equation} \label{eq:BEE3}
\sup_{\substack{\rho \in [0, \RR] \\ \ub \in S^2}} \sum_{j=1}^{\nu} \Big{|} \frac{\partial^l}{\partial \rho^l} R_j(\rho, \ub)\Big{|} \leq \left(\frac{1}{2}\right)^{4-l} c_l(R) \mq^{l/4}
\end{equation}
As to the last term in (\ref{eq:expansionphi}), one has
\begin{equation} \label{eq:bounddxPHI}
\Big{|} \frac{\partial}{\partial x} \left(\Phi(w_j) w_{j}^{2}\right) \Big{|} \leq |\Phi^{'}(w_j)| \cdot \Big{|}\frac{\partial}{\partial x} w_j \Big{|} \cdot |w_j|^2 + 2|\Phi(w_j)| \cdot \Big{|}\frac{\partial}{\partial x} w_j \Big{|} \cdot |w_j|
\end{equation}
and
\begin{eqnarray}
&& \Big{|} \frac{\partial^2}{\partial x^2} \left(\Phi(w_j) w_{j}^{2}\right) \Big{|} \leq |\Phi^{''}(w_j)| \cdot \Big{|}\frac{\partial}{\partial x} w_j \Big{|}^2 \cdot |w_j|^2 \nonumber \\
&+& |\Phi^{'}(w_j)| \left(4\Big{|}\frac{\partial}{\partial x} w_j \Big{|}^2 \cdot |w_j| + \Big{|}\frac{\partial^2}{\partial x^2} w_j \Big{|} \cdot |w_j|^2\right) \nonumber \\
&+& |\Phi(w_j)| \left(2 \Big{|}\frac{\partial^2}{\partial x^2} w_j \Big{|} \cdot |w_j| + 2\Big{|}\frac{\partial}{\partial x} w_j \Big{|}^2\right) \ . \label{eq:bounddx2PHI}
\end{eqnarray}
Since $\Phi$ is completely monotone, $|w_j| \leq \frac{19}{128}$ yields $|\Phi^{(l)}(w_j)| \leq |\Phi^{(l)}(-\frac{19}{128})|$ for every $l$ in $\mathbb{N}$. Then, combining (\ref{eq:expansionmuhat}) with (\ref{eq:BEE1})-(\ref{eq:BEE2}) gives
\begin{eqnarray}
\sup_{\ub \in S^2} \Big{|}\frac{\partial^l}{\partial \rho^l} w_j(\rho, \ub) \Big{|} &\leq& \frac{1}{(2-l)!}\mathfrak{m}_{2} \pi_{j, \nu}^2 \rho^{2-l} + \frac{1}{(3-l)!} \mathfrak{m}_{3}|\pi_{j, \nu}|^3 \rho^{3-l} \nonumber \\
&+& c_{l}(R) \mq \pi_{j, \nu}^{4} \rho^{4-l} \label{eq:BEE10} \\
\sup_{\ub \in S^2} \Big{|}\frac{\partial^l}{\partial \rho^l} w_j(\rho, \ub) \Big{|}^2 &\leq& \frac{3}{[(2-l)!]^2}\mathfrak{m}_{2}^{2} \pi_{j, \nu}^4 \rho^{4-2l} + \frac{3}{[(3-l)!]^2} \mathfrak{m}_{3}^{2}\pi_{j, \nu}^6 \rho^{6-2l} \nonumber \\
&+& 3c^{2}_{l}(R) \mq^2 \pi_{j, \nu}^{8} \rho^{8-2l} \label{eq:BEEE1}
\end{eqnarray}
for $l = 0, 1, 2$. By virtue of the Lyapunov inequality and Theorem 19 in \cite{hlp}, (\ref{eq:BEEE1}) entails
\begin{equation} \label{eq:BEE4}
\sup_{\substack{\rho \in [0, \RR] \\ \ub \in S^2}} \sum_{j=1}^{\nu} \Big{|}\frac{\partial^l}{\partial \rho^l} w_j(\rho, \ub) \Big{|}^2 \leq k_l(w) \mq^{l/2}
\end{equation}
for $l = 0, 1, 2$, with
$$
k_l(w) := 4^{l-2}\left[\frac{3}{[(2-l)!]^2} + \frac{3}{4[(3-l)!]^2} + \frac{3}{16} c^{2}_{l}(R)\right]\ .
$$
Thus, starting from (\ref{eq:dNdx})-(\ref{eq:dN2dx2}) and utilizing (\ref{eq:BEE1})-(\ref{eq:BEE3}), (\ref{eq:bounddxPHI})-(\ref{eq:bounddx2PHI}) and (\ref{eq:BEE4}) with $x = \rho$, one can define the $\wp_k$'s in (\ref{eq:boundderinternalN})-(\ref{eq:boundderinternal}) as follows:
\begin{eqnarray}
\wp_1(\rho) &=& \frac{1}{2} \mathfrak{m}_3 \rho^2 + \mathfrak{m}_2 \rho + \frac{1}{8} c_1(R) \mq^{1/4} +
\big{|}\Phi^{'}(-\frac{19}{128})\big{|} k_0(w) \sqrt{k_1(w)} \mq^{1/4} \nonumber \\
&+& \big{|}\Phi(-\frac{19}{128})\big{|} \cdot \left[k_0(w) + k_1(w)\mq^{1/2}\right]  \nonumber
\end{eqnarray}
and
\begin{eqnarray}
\wp_2(\rho) &=& \wp_{1}^{2}(\rho) + \mathfrak{m}_3 \rho + \mathfrak{m}_2 + \frac{1}{4}c_2(R) \mq^{1/2} +
\big{|}\Phi^{''}(-\frac{19}{128})\big{|} k_0(w) k_1(w) \mq^{1/2} \nonumber \\
&+& \big{|}\Phi^{'}(-\frac{19}{128})\big{|} \cdot \left[4 \sqrt{k_0(w)} k_1(w)\mq^{1/2} + k_0(w)\sqrt{k_2(w)} \mq^{1/2}\right] \nonumber \\
&+& \big{|}\Phi(-\frac{19}{128})\big{|} \cdot \left[k_0(w) + 2k_1(w)\mq^{1/2} + k_2(w)\mq\right] \ . \nonumber
\end{eqnarray}
This completes the proof of (\ref{eq:boundderinternalN}), showing also that the bound therein is independent of the choice of $\mathrm{B}$ in (\ref{eq:psijn}).

To prove (\ref{eq:7091}), one begins by considering $(u, v)$ in $D_k$ and taking $\mathrm{B}$ in (\ref{eq:psijn}) equal to $\mathrm{B}_k$ according to (\ref{eq:B12})-(\ref{eq:B34}). In this way, every map $\psib_{j, \nu; k} : (u, v) \mapsto \psib_{j, \nu}(\hb_k(u, v))$, and hence the map $\hat{\Nrand}_k : (u, v) \mapsto \hat{\Nrand}(\rho; \hb_k(u, v))$, turns out to belong to $\mathrm{C}^4(D_k)$ for $k = 1, \dots, 4$. Then, one resorts to (\ref{eq:dNdx})-(\ref{eq:dN2dx2}), with $x$ standing either for $u$ or $v$, and uses (\ref{eq:boundPsiN}) to bound the common factor $\hat{\Nrand}_k$. As to the derivatives w.r.t. $x$, one evaluates the expression of $\frac{\partial^l}{\partial x^l} [\psib_{j, \nu; k} \cdot \vb]^m$ for $l = 1, 2$, and applies the Cauchy-Schwartz inequality. Whence, after recalling (\ref{eq:psijn}) and introducing the $\mathrm{L}^2$ norm $\lnorm \cdot \rnormm$ of matrices, one gets
\begin{equation} \label{eq:dxRj}
\Big{|} \frac{\partial^l}{\partial x^l}[\psib_{j, \nu; k} \cdot \vb]^m \Big{|} \leq |\vb|^m \sum_{h = 1}^{l} \frac{m!}{(m-h)!} \Lnorm \frac{\partial^{l-h+1}}{\partial x^{l-h+1}} \mathrm{B}_k \Rnormm^h
\end{equation}
when $l = 1, 2$ and $m \geq l$. Since $\lnorm \frac{\partial^s}{\partial x^s} \mathrm{B}_k \rnormm\ \leq \sqrt{3}$ for every $s$ in $\mathbb{N}$, one has
\begin{eqnarray}
&& \sup_{(u, v) \in D_k} \Big{|} \frac{\partial^l}{\partial x^l}\Big{[}- \frac{1}{2} \rho^2\pi_{j, \nu}^2 \intethree [\psib_{j, \nu; k} \cdot \vb]^2 \ud\mu_0 - \frac{i}{3!}\rho^3\pi_{j, \nu}^3 \intethree [\psib_{j, \nu; k} \cdot \vb]^3 \ud\mu_0 \Big{]} \Big{|} \nonumber \\
&\leq& \left(\sum_{h = 1}^{l} 3^{h/2}\right) \mathfrak{m}_2\pi_{j, \nu}^2\rho^2 + \left(\sum_{h = 1}^{l} \frac{3^{h/2}}{(3-h)!} \right) \mathfrak{m}_3|\pi_{j, \nu}|^3 \rho^3 \label{eq:BEE5}
\end{eqnarray}
for $l = 1, 2$. Then, one proceeds with the study of the derivatives of the third term in the RHS of (\ref{eq:expansionphi}). As far as the first order derivative is concerned, one resorts to the second of the expressions of $R_j$, given in the first part of this appendix, to write
\begin{eqnarray}
&& \frac{\partial}{\partial x} R_j(\rho, \hb_k(u, v)) \nonumber \\
&=& -\frac{i}{2} \rho^3 \pi_{j, \nu}^3 \intethree\int_{0}^{1} (1 - s)^2 \Big{\{} (e^{i \rho \pi_{j, \nu} [\psib_{j, \nu; k} \cdot \vb] s} - 1)\Big{(}\frac{\partial}{\partial x} [\psib_{j, \nu; k} \cdot \vb]^3 \Big{)} \nonumber \\
&+& [\psib_{j, \nu; k} \cdot \vb]^3 i \rho \pi_{j, \nu} s \Big{(}\frac{\partial}{\partial x} [\psib_{j, \nu; k} \cdot \vb] \Big{)} e^{i \rho \pi_{j, \nu} [\psib_{j, \nu; k} \cdot \vb] s}\Big{\}} \ud s \mu_0(\ud \vb) \ . \nonumber
\end{eqnarray}
By virtue of (\ref{eq:dxRj}), one gets
\begin{equation} \label{eq:BEE6}
\sup_{(u, v) \in D_k} \Big{|} \frac{\partial}{\partial x} R_j(\rho, \hb_k(u, v)) \Big{|} \leq \frac{\sqrt{3}}{6} \mq \pi_{j, \nu}^4 \rho^4
\end{equation}
which, recalling (\ref{eq:r}), entails
\begin{equation} \label{eq:8096}
\sup_{\substack{\rho \in [0, \RR] \\ (u, v) \in D_k}} \sum_{j = 1}^{\nu} \Big{|}\frac{\partial}{\partial x} R_j(\rho, \hb_k(u, v)) \Big{|} \leq \frac{\sqrt{3}}{96} \ .
\end{equation}
To compute the second order derivatives of $R_j$ one employs the third of its expressions to write
\begin{eqnarray}
&& \frac{\partial^2}{\partial x^2} R_j(\rho, \hb_k(u, v)) \nonumber \\
&=&  -\pi_{j, \nu}^{2} \rho^2 \intethree\int_{0}^{1} (1 - s) \Big{\{} (e^{i \rho \pi_{j, \nu} [\psib_{j, \nu; k} \cdot \vb] s} - 1- i \rho \pi_{j, \nu} [\psib_{j, \nu; k} \cdot \vb] s) \times \nonumber \\
&\times& \Big{(}\frac{\partial^2}{\partial x^2} [\psib_{j, \nu; k} \cdot \vb]^2 \Big{)} \nonumber \\
&+& 2i \rho \pi_{j, \nu} s\ (e^{i \rho \pi_{j, \nu} [\psib_{j, \nu; k} \cdot \vb] s} - 1) \cdot \Big{(}\frac{\partial}{\partial x} [\psib_{j, \nu; k} \cdot \vb]^2 \Big{)} \cdot \Big{(}\frac{\partial}{\partial x} [\psib_{j, \nu; k} \cdot \vb] \Big{)} \nonumber \\
&+& i \rho \pi_{j, \nu} [\psib_{j, \nu; k} \cdot \vb]^2 s\ \Big{[} (e^{i \rho \pi_{j, \nu} [\psib_{j, \nu; k} \cdot \vb] s} - 1) \cdot \Big{(}\frac{\partial^2}{\partial x^2} [\psib_{j, \nu; k} \cdot \vb] \Big{)} \nonumber \\
&+& i \rho \pi_{j, \nu} s\ e^{i \rho \pi_{j, \nu} [\psib_{j, \nu; k} \cdot \vb] s}\ \Big{(}\frac{\partial}{\partial x} [\psib_{j, \nu; k} \cdot \vb] \Big{)}^2 \Big{]}\Big{\}}\ud s \mu_0(\ud \vb) \ . \nonumber
\end{eqnarray}
From (\ref{eq:dxRj}) and the inequality $|e^{i x} - \sum_{r = 1}^{N - 1} (i x)^r/r!| \leq |x|^N/N!$ one obtains the bound
\begin{equation} \label{eq:BEE9}
\sup_{(u, v) \in D_k} \Big{|} \frac{\partial^2}{\partial x^2} R_j(\rho, \hb_k(u, v)) \Big{|} \leq \frac{\sqrt{3} + 9}{6} \mq \pi_{j, \nu}^4 \rho^4
\end{equation}
which, taking account of (\ref{eq:r}), becomes
\begin{equation} \label{eq:80910}
\sup_{\substack{\rho \in [0, \RR] \\ (u, v) \in D_k}} \sum_{j = 1}^{\nu} \Big{|}\frac{\partial^2}{\partial x^2} R_j(\rho, \hb_k(u, v)) \Big{|} \leq \frac{\sqrt{3} + 9}{96} \ .
\end{equation}
Finally, as to the remaining term in the RHS of (\ref{eq:expansionphi}), one utilizes (\ref{eq:bounddxPHI})-(\ref{eq:bounddx2PHI}) with $x = u, v$. Then, combining (\ref{eq:expansionmuhat}) with (\ref{eq:BEE5}) and (\ref{eq:BEE6}) gives
\begin{eqnarray}
\sup_{(u, v) \in D_k} \Big{|}\frac{\partial}{\partial x} w_j \Big{|} &\leq& \sqrt{3} \mathfrak{m}_2 \pi_{j, \nu}^2 \rho^2 + \frac{\sqrt{3}}{2} \mathfrak{m}_3|\pi_{j, \nu}|^3 \rho^3 + \frac{\sqrt{3}}{6} \mq \pi_{j, \nu}^{4} \rho^4 \ \ \ \ \ \label{eq:BEE12} \\
\sup_{(u, v) \in D_k} \Big{|}\frac{\partial}{\partial x} w_j \Big{|}^2 &\leq& 9 \mathfrak{m}_{2}^{2} \pi_{j, \nu}^4 \rho^4 + \frac{9}{4} \mathfrak{m}_{3}^{2}\pi_{j, \nu}^6 \rho^6 + \frac{1}{4} \mq^2 \pi_{j, \nu}^{8} \rho^8 \label{eq:BEEE2} \ .
\end{eqnarray}
By virtue of the Lyapunov inequality and Theorem 19 in \cite{hlp}, (\ref{eq:BEEE2}) yields
\begin{equation} \label{eq:BEE7}
\sup_{\substack{\rho \in [0, \RR] \\ (u, v) \in D_k}} \sum_{j=1}^{\nu} \Big{|}\frac{\partial}{\partial x} w_j(\rho, \hb_k(u, v)) \Big{|}^2 \leq \frac{613}{1024} \ .
\end{equation}
As for the second order derivatives, from the combination of (\ref{eq:expansionmuhat}) with (\ref{eq:BEE5}) and (\ref{eq:BEE9}) one gets
\begin{eqnarray}
&& \sup_{(u, v) \in D_k}  \Big{|}\frac{\partial^2}{\partial x^2} w_j(\rho, \hb_k(u, v)) \Big{|}^2 \leq 3\left(\sum_{h = 1}^{2} 3^{h/2}\right)^2 \mathfrak{m}_{2}^{2} \pi_{j, \nu}^4 \rho^4 \nonumber \\
&+& 3\left(\sum_{h = 1}^{2} \frac{3^{h/2}}{(3-h)!} \right)^2 \mathfrak{m}_{3}^{2}\pi_{j, \nu}^6 \rho^6 +
3\left(\frac{\sqrt{3} + 9}{6}\right)^2 \mq^2 \pi_{j, \nu}^{8} \rho^8 \label{eq:luisamiller}
\end{eqnarray}
and hence
\begin{equation} \label{eq:BEE8}
\sup_{\substack{\rho \in [0, \RR] \\ (u, v) \in D_k}} \sum_{j=1}^{\nu} \Big{|}\frac{\partial^2}{\partial x^2} w_j(\rho, \hb_k(u, v)) \Big{|}^2 \leq W_{2}^{\ast}
\end{equation}
where
$$
W_{2}^{\ast} := \frac{3}{16}\left(\sum_{h = 1}^{2} 3^{h/2}\right)^2 + \frac{3}{64}\left(\sum_{h = 1}^{2} \frac{3^{h/2}}{(3-h)!} \right)^2 + \frac{3}{256}\left(\frac{\sqrt{3} + 9}{6}\right)^2\ .
$$
In view of (\ref{eq:bounddxPHI}), (\ref{eq:BEE10}), (\ref{eq:BEE4}), and (\ref{eq:BEE7}),
\begin{eqnarray}
&& \sup_{(u, v) \in D_k} \sum_{j=1}^{\nu} \Big{|} \frac{\partial}{\partial x} \left(\Phi(w_j) w_{j}^{2}\right) \Big{|} \leq \sqrt{\frac{613}{1024}} \left(\frac{1}{2}\mathfrak{m}_2 \rho^2 + \frac{1}{6} \mathfrak{m}_3 |\rho|^3 + \frac{1}{24} \mq \rho^4\right)\times \nonumber \\
&\times& \left(|\Phi^{'}(-\frac{19}{128})|\sqrt{k_0(w)} + 2|\Phi(-\frac{19}{128})|\right)  \label{eq:BEE11}
\end{eqnarray}
and, utilizing (\ref{eq:expansionphi}), (\ref{eq:BEE5}), (\ref{eq:BEE6}) and (\ref{eq:BEE11}), one concludes that
\begin{eqnarray}
&& \sup_{(u, v) \in D_k} \sum_{j=1}^{\nu} \Big{|} \frac{\partial}{\partial x} \Log[ \hat{\mu}_0(\rho \pi_{j, \nu} \psib_{j, \nu}(\hb_k(u, v)))] \Big{|} \leq
\sqrt{3}\mathfrak{m}_2\rho^2 + \frac{\sqrt{3}}{2}\mathfrak{m}_3\rho^3 \nonumber \\
&+& \frac{\sqrt{3}}{6} \mq \rho^4 + \sqrt{\frac{613}{1024}} \left(|\Phi^{'}(-\frac{19}{128})|\sqrt{k_0(w)} + 2|\Phi(-\frac{19}{128})|\right) \times \nonumber \\
&\times& \left(\frac{1}{2}\mathfrak{m}_2 \rho^2 + \frac{1}{6} \mathfrak{m}_3 \rho^3 + \frac{1}{24} \mq \rho^4\right) \label{eq:BEE13}
\end{eqnarray}
for $\rho$ in $[0, \RR]$. To obtain a bound of the same type for the second derivative, one can first combine (\ref{eq:bounddx2PHI}), (\ref{eq:BEE10}), (\ref{eq:BEE4}) and (\ref{eq:BEEE2})-(\ref{eq:BEE8}) to get
\begin{eqnarray}
&& \sup_{(u, v) \in D_k} \sum_{j=1}^{\nu} \Big{|} \frac{\partial^2}{\partial x^2} \left(\Phi(w_j) w_{j}^{2}\right) \Big{|} \nonumber \\
&\leq& \Big{[} |\Phi^{''}(-\frac{19}{128})\frac{613}{1024}\sqrt{k_0(w)} + |\Phi^{'}(-\frac{19}{128})|\left(\frac{613}{256} + \sqrt{W_{2}^{\ast}k_0(w)}\right) \nonumber \\
&+& 2|\Phi(-\frac{19}{128})|\sqrt{W_{2}^{\ast}}\Big{]} \cdot \left(\frac{1}{2}\mathfrak{m}_2 \rho^2 + \frac{1}{6} \mathfrak{m}_3 \rho^3 + \frac{1}{24} \mq \rho^4\right) \nonumber \\
&+& 2|\Phi(-\frac{19}{128})| \left(9 \mathfrak{m}_{2}^{2} \rho^4 + \frac{9}{4} \mathfrak{m}_{3}^{2}\rho^6 + \frac{1}{4} \mq^2 \rho^8\right) \label{eq:BEE14}
\end{eqnarray}
and, then, utilize (\ref{eq:expansionphi}), (\ref{eq:BEE5}), (\ref{eq:BEE9}) and (\ref{eq:BEE14}), to conclude that
\begin{eqnarray}
&& \sup_{(u, v) \in D_k} \sum_{j=1}^{\nu} \Big{|} \frac{\partial^2}{\partial x^2} \Log[\hat{\mu}_0(\rho \pi_{j, \nu} \psib_{j, \nu}(\hb_k(u, v)))] \Big{|} \nonumber \\
&\leq& \left(\sum_{h = 1}^{2} 3^{h/2}\right) \mathfrak{m}_2\rho^2 + \left(\sum_{h = 1}^{2} \frac{3^{h/2}}{(3-h)!} \right) \mathfrak{m}_3 \rho^3 + \frac{\sqrt{3} + 9}{6} \mq \rho^4 \nonumber \\
&+& \Big{[} |\Phi^{''}(-\frac{19}{128})\frac{613}{1024}\sqrt{k_0(w)} + |\Phi^{'}(-\frac{19}{128})|\left(\frac{613}{256} + \sqrt{W_{2}^{\ast}k_0(w)}\right) \nonumber \\
&+& 2|\Phi(-\frac{19}{128})|\sqrt{W_{2}^{\ast}}\Big{]} \cdot \left(\frac{1}{2}\mathfrak{m}_2 \rho^2 + \frac{1}{6} \mathfrak{m}_3 \rho^3 + \frac{1}{24} \mq \rho^4\right) \nonumber \\
&+& 2|\Phi(-\frac{19}{128})| \left(9 \mathfrak{m}_{2}^{2} \rho^4 + \frac{9}{4} \mathfrak{m}_{3}^{2}\rho^6 + \frac{1}{4} \mq^2 \rho^8\right) \ . \label{eq:BEE15}
\end{eqnarray}
At this stage, one observes that the RHSs of (\ref{eq:BEE13}) and (\ref{eq:BEE15}) can be written as $\rho^2 \wp_{L, 1}(\rho)$ and $\rho^2 \wp_{L, 2}(\rho)$ respectively, for specific non-random polynomials $\wp_{L, 1}$ and $\wp_{L, 2}$ with positive coefficients. As final step of the proof, expressing $\Delta_{S^2}$ in local coordinates leads to
\begin{gather}
\sup_{\ub \in \Omega_k} |\Delta_{S^2}\hat{\Nrand}_k(\rho; \ub)| \leq 4(2 + \sqrt{3}) \times \nonumber \\
\times \sup_{(u, v) \in D_k} \left(\Big{|}\frac{\partial^2}{\partial u^2}\hat{\Nrand}(\rho; \hb_k(u,v))\Big{|} +
\Big{|}\frac{\partial}{\partial u}\hat{\Nrand}(\rho; \hb_k(u, v))\Big{|} + \Big{|}\frac{\partial^2}{\partial v^2}\hat{\Nrand}(\rho; \hb_k(u, v))\Big{|}\right) \nonumber
\end{gather}
where $4(2 + \sqrt{3}) = \max_{u \in [\frac{1}{12}\pi, \frac{11}{12}\pi]} \max\{|\cot u|, 1/\sin^2u\}$ and hence
$$
\wp_L(\rho) = 4(2 + \sqrt{3}) [2\rho^2\wp_{L, 1}^2(\rho) + \wp_{L, 1}(\rho) + 2\wp_{L, 2}(\rho)]\ .
$$

\subsection{Proof of Proposition {\ref{prop:newtontrick}}} \label{a:newton}

Fix the sample point $\omega$ in $U^c$ and denote by $n$ the value of $\nu$ at $\omega$. Then,
designate the values of $\pi_{1, \nu}^{2}, \dots, \pi_{\nu, \nu}^{2}$ at $\omega$ by $a_1, \dots, a_n$ respectively, so that each $a_j$ belongs to $[0, 1]$ and $\sum_{j = 1}^{n} a_j = 1$ in view of (\ref{eq:sumpijn}). The argument continues by resorting to the following combinatorial tools:
\begin{enumerate}
\item[i)] The $k$-th elementary symmetric function $S_k(a_1, \dots, a_n)$ defined by
$$
S_k(a_1, \dots, a_n) := \sum_{1 \leq i_1 < \dots < i_k \leq n} a_{i_1} \dots a_{i_k}
$$
for $k$ in $\{1, \dots, n\}$.
\item[ii)] The $k$-th Newton symmetric function given by
$$
N_k(a_1, \dots, a_n) := \sum_{j = 1}^{n} a_{j}^{k} \ .
$$
\item[iii)] The group of relations, known as Newton's identities, which read
$$
k S_k = \sum_{j = 1}^{k} (-1)^{j + 1} N_j S_{k - j}
$$
for $k$ in $\{1, \dots, n\}$, with the proviso that $S_0(a_1, \dots, a_n) := 1$.
\end{enumerate}
See Section 1.9 of \cite{mer} for details. The way is now paved to prove that, if $\as \in (0, 1)$, $N_1 = 1$ and $N_2 \leq \as$, then
\begin{equation} \label{eq:newtoon}
S_k \geq 1/k! - 2^{k-1}\as
\end{equation}
holds for each $k$ in $\{1, \dots, n\}$. Proceeding by mathematical induction, when $k = 1$, one has $S_1 = N_1 = 1$ and (\ref{eq:newtoon}) follows. When $k \geq 2$, combine the Newton identities with the inductive hypothesis to get
$$
S_k \geq \frac{1}{k}S_{k-1} - \frac{1}{k} \sum_{j = 2}^{k} N_j S_{k - j} \geq \frac{1}{k!} - \frac{1}{k}\Big{(}2^{k-2}\as + \sum_{j = 2}^{k} N_j S_{k - j}\Big{)} \ .
$$
At this stage, note that $N_j \leq \as$ for each $j$ in $\{2, \dots, k\}$. Moreover, thanks to the multinomial identity (see, e.g., 1.7.2 in \cite{mer}), $N_1 = 1$ entails $S_m \leq 1/m! \leq 1$ for each $m$ in $\{0, \dots, n\}$. Hence,
$$
S_k \geq \frac{1}{k!} - \frac{1}{k}[2^{k-2} + (k-1)] \as
$$
which concludes the proof of (\ref{eq:newtoon}), after noting that $2^{k-2} + (k-1) \leq k2^{k-1}$ for every $k$ in $\mathbb{N}$. To complete the proof of (\ref{eq:newtontrick}), observe that $\omega \in U^c$ entails $n \geq r$ by virtue of (\ref{eq:ra}), so that
$$
\prod_{j = 1}^{n} (1 + a_j x^2) = \sum_{k = 0}^{n} S_k(a_1, \dots, a_n) x^{2k} \geq S_r(a_1, \dots, a_n) x^{2r} \ .
$$
Finally, recall the relation between $r$ and $\as$ given by (\ref{eq:ra}) and apply (\ref{eq:newtoon}) to obtain
$$
S_r \geq \frac{1}{r!} - 2^{r - 1}\as = \frac{1}{2 r!} = \epsilon \ .
$$
To prove (\ref{eq:intenewtonm}), an obvious change of variable entails
$$
\int_{x}^{+\infty} \Psi^s(\rho) \rho^m \ud \rho = \lambda^{m+1} \int_{x/\lambda}^{+\infty} \left[\frac{1}{\prod_{j=1}^{\nu} \left({1 + \pi_{j, \nu}^{2} y^2}\right)}\right]^{sq} y^m \ud y
$$
and conclude by using (\ref{eq:newtontrick}).

\subsection{Proof of Proposition \ref{prop:BElaplace}} \label{a:BElaplace}

The analysis to be developed is concerned with each of the charts $\Omega_1, \dots, \Omega_4$, but it is of the same kind for all of them. Therefore, even if the notation agrees with that introduced in \ref{a:boundderivativesMhat}, the subscript $k$ referring to the $k$-th chart will be dropped. The computation of the Laplacian in (\ref{eq:principallog}) yields
$$
\Delta_{S^2} \hat{\Nrand}(\rho; \ub) = \hat{\Nrand}(\rho; \ub) \Big{\{}\Delta_{S^2} \text{Log}\hat{\Nrand}(\rho; \ub)
+ \Blnorm \gradient\text{Log}\hat{\Nrand}(\rho; \ub)\Brnormg^2\Big{\}}
$$
for every fixed $\rho$. It is worth mentioning that here any differential operator $\diff$, applied to the complex-valued function $h = f + ig$, must be intended as $\diff h := \diff f + i \diff g$ and that the scalar product $\langle U_1 + i U_2, V_1 + i V_2\rangle$ is defined, by linearity, as $\langle U_1 , V_1\rangle - \langle U_2 , V_2\rangle + i \langle U_1 , V_2\rangle + i \langle U_2 , V_1\rangle$ for the vector fields $U_1, U_2, V_1, V_2$. Now, after observing that $\ind_{T^c} = 1 - \ind_{T}$, one gets
\begin{eqnarray}
&& \Big{|}\et\left[\big{(} \Delta_{S^2}\hat{\Nrand} \big{)} \ind_{T^{c}} \ | \ \mathscr{G} \right]\Big{|} \ \leq\ \et\left[ |\hat{\Nrand}| \cdot \Big{|}\ \Blnorm \gradient\text{Log}\hat{\Nrand}\Brnormg^2 \Big{|} \ind_{T^{c}} \ | \ \mathscr{G} \right] \nonumber \\
&+& e^{-\rho^2/2} \Big{|}\et\left[\Delta_{S^2} \text{Log}\hat{\Nrand}\ | \ \mathscr{G} \right]\Big{|} + e^{-\rho^2/2} \et\left[ |\Delta_{S^2} \text{Log}\hat{\Nrand}| \ind_T\ | \ \mathscr{G}\right] \nonumber \\
&+& \et\left[\big{|}\hat{\Nrand} - e^{-\rho^2/2}\big{|} \cdot \big{|} \Delta_{S^2} \text{Log}\hat{\Nrand}\big{|} \ind_{T^c}\ | \ \mathscr{G}\right] \label{eq:basisBElaplace}
\end{eqnarray}
which represents the starting point for the proof at issue. To bound the term $|\hat{\Nrand}|$, use (\ref{eq:principallog}), (\ref{eq:expansionphi}), (\ref{eq:Vu}), (\ref{eq:BEE3}) and (\ref{eq:BEE4}) to write
\begin{equation}\label{eq:BEl1}
\big{|}\hat{\Nrand}(\rho; \ub)\big{|} \ind_{T(\ub)^c} \leq c(N) e^{-\rho^2/6}
\end{equation}
for $\rho$ in $[0, \RR]$, with $c(N) := \exp\{\frac{1}{16}c_0(R) + |\Phi(-\frac{19}{128})| k_0(w)\}$. As to the term containing the gradient, an application of the triangular inequality in (\ref{eq:expansionphi}) shows that
\begin{eqnarray}
&& \et\left[\Big{|}\ \Blnorm \gradient\text{Log}\hat{\Nrand}(\rho; \ub)\Brnormg^2\Big{|}\ | \ \mathscr{G}\right] \leq \rho^4 \ZZ_G(\ub) + \frac{1}{9}\rho^6 \times \nonumber \\
&\times& \et\Big{[}\Blnorm \sum_{j = 1}^{\nu}\pi_{j, \nu}^{3} \gradient \intethree [\psib_{j, \nu}(\ub) \cdot \vb]^3
\mu_0(\ud \vb)\Brnormg^2\ | \ \mathscr{G}\Big{]} + 4E_1 + 4E_2 \ \ \ \ \ \label{eq:9091}
\end{eqnarray}
for $(\rho, \ub)$ in $[0, \RR] \times \Omega$, where $E_1$ and $E_2$ are conditional expectations, to be specified below, involving the remainders $R_j$ and $\Phi(w_j)w_{j}^2$ respectively. As to the second summand, the convexity of the square of the Riemannian length entails
\begin{eqnarray}
&& \Blnorm \sum_{j = 1}^{\nu}\pi_{j, \nu}^{2} \Big{(}\pi_{j, \nu} \gradient \intethree [\psib_{j, \nu} \cdot \vb]^3 \ud\mu_0\Big{)}\Brnormg^2 \nonumber \\
&\leq& \WW \sup_{j, \ub} \Blnorm \gradient\intethree [\psib_{j, \nu} \cdot \vb]^3 \ud\mu_0\Brnormg^2
\leq \WW \Big{(} \intethree \sup_{j, \ub} \lnorm \gradient [\psib_{j, \nu} \cdot \vb]^3\rnorm_{S^2} \ud\mu_0\Big{)}^2 \ . \nonumber
\end{eqnarray}
To evaluate the last integral, recall that $\psib_{j, \nu}$ is given by (\ref{eq:psijn}) with the proper choice of $\mathrm{B}$ as in (\ref{eq:B12})-(\ref{eq:B34}), which makes $\psib_{j, \nu} : \Omega \rightarrow S^2$ smooth. Writing the gradient in coordinates yields
\begin{eqnarray}
&& \sup_{j, \ub} \lnorm \gradient [\psib_{j, \nu}(\ub) \cdot \vb]^3\rnorm_{S^2} \nonumber \\
&=& \sup_{j, (u,v)} \left[(\partial_u [\psib_{j, \nu}(\hb(u, v)) \cdot \vb]^3)^2 + \frac{1}{\sin^2 u} (\partial_v [\psib_{j, \nu}(\hb(u, v)) \cdot \vb]^3)^2\right]^{1/2} \ . \nonumber
\end{eqnarray}
Since $1/\sin^2 u \leq 4(2 + \sqrt{3})$ for every $(u, v)$ in $D$, (\ref{eq:dxRj}) leads to
$$
\Big{(} \intethree \sup_{\substack{j = 1, \dots, \nu \\ \ub \in \Omega}} \lnorm \gradient [\psib_{j, \nu}(\ub) \cdot \vb]^3\rnorm_{S^2}
\mu_0(\ud \vb)\Big{)}^2 \leq 27(9 + 4\sqrt{3}) \mathfrak{m}_{3}^{2}\ .
$$
The terms $E_1$ and $E_2$ in (\ref{eq:9091}) can be derived as uniform bounds w.r.t. $\ub$ by writing $\Blnorm \gradient\text{Log}\hat{\Nrand}(\rho; \ub)\Brnormg^2$ in coordinates, according to
\begin{eqnarray}
E_1 &:=& \sup_{(u, v)} \et\Big{[}\Big{(}\sum_{j = 1}^{\nu} |\partial_u R_j(\rho, \hb(u, v))|\Big{)}^2 + \frac{1}{\sin^2 u}\Big{(}\sum_{j = 1}^{\nu} |\partial_v R_j(\rho, \hb(u, v))|\Big{)}^2\ | \ \mathscr{G}\Big{]} \nonumber \\
E_2 &:=& \sup_{(u, v)} \et\Big{[}\Big{(}\sum_{j = 1}^{\nu} |\partial_u \Phi(w_j(\rho, \hb(u, v))) w_{j}^{2}(\rho, \hb(u, v))|\Big{)}^2  \nonumber \\
&+& \frac{1}{\sin^2 u}\Big{(}\sum_{j = 1}^{\nu} |\partial_v\Phi(w_j(\rho, \hb(u, v))) w_{j}^{2}(\rho, \hb(u, v))|\Big{)}^2\ | \ \mathscr{G}\Big{]} \ . \nonumber
\end{eqnarray}
As for $E_1$, (\ref{eq:BEE6}) and (\ref{eq:8096}) give
\begin{equation} \label{eq:5giu1}
\sup_{(u, v)} \Big{[}\Big{(}\sum_{j = 1}^{\nu} |\partial_u R_j|\Big{)}^2 + \frac{1}{\sin^2 u}\Big{(}\sum_{j = 1}^{\nu} |\partial_v R_j|\Big{)}^2\Big{]} \leq \frac{9 + 4\sqrt{3}}{192} \mq \mathrm{W} \rho^4
\end{equation}
for every $\rho$ in $[0, \RR]$, the RHS being a $\mathscr{G}$-measurable function. Apropos of $E_2$, start from (\ref{eq:bounddxPHI}) and notice that, in view of (\ref{eq:BEE4}) and the complete monotonicity of $\Phi$,
$$
(|\Phi^{'}(w_j)| \cdot |w_j| + 2|\Phi(w_j)|) \leq \left(|\Phi^{'}(-\frac{19}{128})|\sqrt{k_0(w)} + 2|\Phi(-\frac{19}{128})|\right)
$$
holds for every $j = 1, \dots, \nu$, and $(\rho, \ub)$ in $[0, \RR] \times \Omega$. Then, by virtue of Lyapunov's inequality and Theorem 19 in \cite{hlp}, inequalities (\ref{eq:BEEE1}), with $l= 0$, (\ref{eq:BEEE2}) and (\ref{eq:luisamiller}) become
\begin{eqnarray}
\sup_{(u, v) \in D} \sum_{j = 1}^{\nu} |w_j(\rho, \hb(u, v))|^2 &\leq& 16k_0(w) \mq \WW \rho^4 \label{eq:boundwj2} \\
\sup_{(u, v) \in D} \sum_{j = 1}^{\nu} \Big{|}\frac{\partial}{\partial x} w_j(\rho, \hb(u, v)) \Big{|}^2 &\leq& \frac{613}{64} \mq \WW \rho^4 \label{eq:bounddxwj2} \\
\sup_{(u, v) \in D} \sum_{j = 1}^{\nu} \Big{|}\frac{\partial^2}{\partial x^2} w_j(\rho, \hb(u, v)) \Big{|}^2 &\leq& 16 W_{2}^{\ast} \mq \WW \rho^4 \label{eq:bounddx2wj2}
\end{eqnarray}
respectively. These last inequalities, in combination with (\ref{eq:BEE4}) and (\ref{eq:BEE7}), entail
\begin{equation} \label{eq:5giu2}
\sup_{(u, v)} \Big{[}\Big{(}\sum_{j = 1}^{\nu} |\partial_u \Phi(w_j) w_{j}^{2}|\Big{)}^2  + \frac{1}{\sin^2 u}\Big{(}\sum_{j = 1}^{\nu} |\partial_v \Phi(w_j) w_{j}^{2}|\Big{)}^2\Big{]} \leq \overline{E}_2 \mq \mathrm{W} \rho^4
\end{equation}
for every $\rho$ in $[0, \RR]$ with
$$
\overline{E}_2 := 4(9 + 4\sqrt{3}) \left(|\Phi^{'}(-\frac{19}{128})|\sqrt{k_0(w)} + 2|\Phi(-\frac{19}{128})|\right)^2 \left(k_0(w) + \frac{613}{64}\right)^2\ .
$$
This concludes the analysis of the first summand in the RHS of (\ref{eq:basisBElaplace}), after noting that the upper bound provided by (\ref{eq:5giu2}) is $\mathscr{G}$-measurable. To proceed, for notational simplicity put
\begin{eqnarray}
A &:=& -\frac{1}{2} \sum_{j = 1}^{\nu} \pi_{j, \nu}^2 \Big{(} \intethree [\psib_{j, \nu}(\ub) \cdot \vb]^2 \mu_0(\ud \vb) - 1 \Big{)}  \nonumber \\
B &:=& -\frac{1}{3!} \sum_{j = 1}^{\nu} \pi_{j, \nu}^3 \intethree [\psib_{j, \nu}(\ub) \cdot \vb]^3 \mu_0(\ud \vb) \nonumber \\
H &:=& \sum_{j = 1}^{\nu} R_j(\rho, \ub) + \sum_{j = 1}^{\nu} \Phi(w_j(\rho, \ub)) w_{j}^{2}(\rho, \ub) \nonumber
\end{eqnarray}
so that (\ref{eq:principallog}) can be re-written as $\hat{\Nrand} = \exp\{-\rho^2/2 + A\rho^2 + iB\rho^3 + H\}$. Observe that $\et[A^2\ | \ \mathscr{G}] = \frac{1}{4} \ZZ(\ub)$ and $\et[(\Delta_{S^2} A)^2\ | \ \mathscr{G}] = \frac{1}{4} \ZZ_L(\ub)$ hold by definition, and invoke (\ref{eq:XXuL})-(\ref{eq:YYuL}) to write
\begin{eqnarray}
|\et[\Delta_{S^2} A\ | \ \mathscr{G}]| &\leq& \frac{1}{2} \XX_L(\ub) \nonumber \\
|\et[\Delta_{S^2} B\ | \ \mathscr{G}]| &\leq&  \frac{1}{6} \YY_L(\ub) \nonumber
\end{eqnarray}
for every $\ub$ in $\Omega$. To deal with the Laplacian of $H$, start from the sum of the $R_j$'s. After writing the Laplacian in coordinates, the combination of (\ref{eq:BEE6}) with (\ref{eq:BEE9}) gives
\begin{equation} \label{eq:4giu1}
\sup_{\ub \in \Omega} \sum_{j = 1}^{\nu} |\Delta_{S^2} R_j(\rho, \ub)| \leq \frac{105 + 53\sqrt{3}}{6}\mq \WW \rho^4\ .
\end{equation}
Analogously, combining (\ref{eq:boundwj2})-(\ref{eq:bounddx2wj2}) with (\ref{eq:BEE4}), (\ref{eq:BEE7}) and (\ref{eq:BEE8}) yields
\begin{equation} \label{eq:4giu2}
\sup_{\ub} \sum_{j = 1}^{\nu} |\Delta_{S^2} \Phi(w_j(\rho, \ub)) w_{j}^{2}(\rho, \ub)| \leq \overline{\Phi}(\Delta) \mq \WW \rho^4
\end{equation}
with
\begin{eqnarray}
\overline{\Phi}(\Delta) &:=& 4(2 + \sqrt{3})\Big{[} 16\sqrt{\frac{613}{1024}}\ \big{|}\Phi^{'}(-\frac{19}{128})\big{|}\ k_0(w) + \frac{613}{1024}\ \big{|}\Phi(-\frac{19}{128})\big{|} \nonumber \\
&+& 16\big{|}\Phi(-\frac{19}{128})\big{|}\ k_0(w) \Big{]} + (9 + 4\sqrt{3}) \Big{[} \frac{613}{64}\ \big{|}\Phi^{''}(-\frac{19}{128})\big{|}\ k_0(w) \nonumber \\
&+& \big{|}\Phi^{'}(-\frac{19}{128})\big{|} \cdot \big{(} \frac{613}{16}\sqrt{k_0(w)} + 16 k_0(w) \sqrt{W_{2}^{\ast}}\big{)} \nonumber \\
&+& \big{|}\Phi(-\frac{19}{128})\big{|} \cdot \big{(} 16W_{2}^{\ast} + 16k_0(w) + \frac{613}{32}\big{)} \Big{]} \ . \nonumber
\end{eqnarray}
Hence, the second summand in the RHS of (\ref{eq:basisBElaplace}) admits the following bound:
\begin{eqnarray}
&& e^{-\rho^2/2} \Big{|}\et\left[\Delta_{S^2} \text{Log}\hat{\Nrand}(\rho; \ub)\ | \ \mathscr{G} \right]\Big{|} \leq \frac{1}{2} \rho^2 e^{-\rho^2/2} \XX_L(\ub) \nonumber \\
&+& \frac{1}{6} \rho^3 e^{-\rho^2/2} \YY_L(\ub) + \left(\frac{105 + 53\sqrt{3}}{6} + \overline{\Phi}(\Delta)\right)\rho^4e^{-\rho^2/2} \mq \WW \label{eq:5giu7}
\end{eqnarray}
for every $(\rho, \ub)$ in $[0, \RR] \times \Omega$. Apropos of the third summand in the RHS of (\ref{eq:basisBElaplace}), recall from \ref{a:boundderivativesMhat} that
$\sup_{\ub} |\Delta_{S^2} \text{Log}\hat{\Nrand}| \leq \rho^2 \wp_L(\rho)$ holds for every $\rho$ in $[0, \RR]$. Therefore, in view of (\ref{eq:Markov}), one has
\begin{equation} \label{eq:5giu3}
e^{-\rho^2/2} \et\left[ |\Delta_{S^2} \text{Log}\hat{\Nrand}(\rho, \ub)| \ind_T\ \ | \ \mathscr{G}\right] \leq 9 e^{-\rho^2/2} \rho^2 \wp_L(\rho) \ZZ(\ub) \ .
\end{equation}
To deal with the last summand in the RHS of (\ref{eq:basisBElaplace}), a bound for $\big{|}\hat{\Nrand} - e^{-\rho^2/2}\big{|}$ can be derived from the elementary inequalities $|e^{i x} - 1| \leq |x|$ and $|e^z - 1| \leq |z|e^{|z|}$, valid for every $x$ in $\rone$ and $z$ in $\mathbb{C}$, respectively. Whence, one gets
$$
\big{|}\hat{\Nrand} - e^{-\rho^2/2}\big{|}\ind_{T(\ub)^c} \leq e^{-\rho^2/6} \left(|H|e^{|H|} + |B|\rho^3 + |A|\rho^2\right)
$$
which, in turn, yields
\begin{gather}
\big{|}\hat{\Nrand} - e^{-\rho^2/2}\big{|} \cdot \big{|} \Delta_{S^2} \text{Log}\hat{\Nrand}\big{|} \ind_{T(\ub)^c} \leq
e^{-\rho^2/6} \Big{(}|H|e^{|H|}\rho^2 \wp_L(\rho) + |B \Delta_{S^2} A| \rho^5 \nonumber \\
+ |B \Delta_{S^2} B| \rho^6 + |B \Delta_{S^2} H| \rho^3 + |A \Delta_{S^2} A| \rho^4 + |A \Delta_{S^2} B| \rho^5 + |A \Delta_{S^2} H| \rho^2\Big{)} \label{eq:4giu3} \ .
\end{gather}
At this stage, note that $\sup_{\ub} |A| \leq \frac{1}{2}(1 + \mathfrak{m}_2)$, $\sup_{\ub} |B| \leq \frac{1}{6}\mathfrak{m}_3$
and $\sup_{\ub} e^{|H|} \leq c(N)$ for every $\rho$ in $[0, \RR]$, in view of (\ref{eq:BEE3}) and (\ref{eq:BEE4}). Thus, taking account of (\ref{eq:BEE2}), (\ref{eq:boundwj2}) and (\ref{eq:4giu1})-(\ref{eq:4giu2}) gives
\begin{equation} \label{eq:5giu4}
\left(|H|e^{|H|}\rho^2 \wp_L(\rho) + |B \Delta_{S^2} H| \rho^3 + |A \Delta_{S^2} H| \rho^2\right) \leq \rho^2 \wp_H(\rho) \mq \WW
\end{equation}
with
\begin{eqnarray}
\wp_H(\rho) &:=& c(N) \left(c_0(R) + 16 |\Phi(-\frac{19}{128})| k_0(w)\right) \rho^4 \wp_L(\rho) \nonumber \\
&+& \left(\frac{105 + 53\sqrt{3}}{6} + \overline{\Phi}(\Delta)\right) \cdot \left(\frac{1}{6}\mathfrak{m}_3\rho^5 + \frac{1}{2}(1 + \mathfrak{m}_2) \rho^4\right) \ . \nonumber
\end{eqnarray}
For the remaining terms in (\ref{eq:4giu3}), take the conditional expectation and write
\begin{gather}
\et\left[|B \Delta_{S^2} A| \rho^5 + |B \Delta_{S^2} B| \rho^6 + |A \Delta_{S^2} A| \rho^4 + |A \Delta_{S^2} B| \rho^5\ | \ \mathscr{G}\right] \nonumber \\
\leq \frac{1}{2} \Big{\{} \et[(B^2 + (\Delta_{S^2} B)^2)\ | \ \mathscr{G}] \cdot (\rho^5 + \rho^6) + \frac{1}{4} (\ZZ(\ub) + \ZZ_L(\ub)) \cdot (\rho^4 + \rho^5) \Big{\}} \ . \label{eq:5giu8}
\end{gather}
Then, an application of the Lyapunov inequality shows that
\begin{eqnarray}
B^2 &\leq& \frac{1}{36}\mathfrak{m}_{3}^{2} \WW \label{eq:5giu5} \\
(\Delta_{S^2} B)^2 &\leq& \frac{1}{36} \Big{(} \intethree \sup_{\substack{j = 1, \dots, \nu \\ \ub \in \Omega}} |\Delta_{S^2} [\psib_{j, \nu}(\ub) \cdot \vb]^3| \mu_0(\ud \vb)\Big{)}^2 \WW \label{eq:5giu6}
\end{eqnarray}
the two RHSs being $\mathscr{G}$-measurable. To evaluate the integral in (\ref{eq:5giu6}), it is enough to write the Laplacian w.r.t. the coordinates $(u, v)$ and  to recall that $\max\{1/\sin^2 u, 1/|\cot u|\} \leq 4(2 + \sqrt{3})$ for every $(u, v)$ in $D$, so that (\ref{eq:dxRj}) leads to
$$
\intethree \sup_{\substack{j = 1, \dots, \nu \\ \ub \in \Omega}} |\Delta_{S^2} [\psib_{j, \nu}(\ub) \cdot \vb]^3| \mu_0(\ud \vb)
\leq (234 + 123\sqrt{3}) \mathfrak{m}_3 \ .
$$
There are now all the elements to complete the proof of Proposition \ref{prop:BElaplace} by setting, in view of (\ref{eq:basisBElaplace})-(\ref{eq:5giu1}), (\ref{eq:5giu2}) and (\ref{eq:5giu7})-(\ref{eq:5giu6}),
\begin{eqnarray}
z_1(\rho) &:=& c(N)e^{-\rho^2/6}\Big{[}3(9 + 4\sqrt{3})\mathfrak{m}_{3}^{2}\rho^4 + \frac{9 + 4\sqrt{3}}{48}\mq \rho^2 + 4 \overline{E}_2 \mq \rho^2\Big{]} \nonumber \\
&+& \Big{(}\frac{105 + 53\sqrt{3}}{6} + \overline{\Phi}(\Delta)\Big{)} \mq \rho^2e^{-\rho^2/2} +  \mq \wp_H(\rho)e^{-\rho^2/6} \nonumber \\
&+& \frac{1 + (234 + 123\sqrt{3})^2}{72} \mathfrak{m}_{3}^{2} (\rho^3 + \rho^4) e^{-\rho^2/6} \nonumber
\end{eqnarray}
and
\begin{eqnarray}
z_2(\rho) &:=& \frac{1}{2}e^{-\rho^2/2} \nonumber \\
z_3(\rho) &:=& \frac{1}{6}\rho e^{-\rho^2/2} \nonumber \\
z_4(\rho) &:=& 9 \wp_L(\rho) e^{-\rho^2/2} + \frac{1}{8}(\rho^2 + \rho^3) e^{-\rho^2/6} \nonumber \\
z_5(\rho) &:=& c(N)\rho^2 e^{-\rho^2/6} \nonumber \\
z_6(\rho) &:=& \frac{1}{8}(\rho^2 + \rho^3) e^{-\rho^2/6} \nonumber \ .
\end{eqnarray}

\subsection{Proof of (\ref{eq:MMjn2}) and (\ref{eq:MMjn3})-(\ref{eq:psijnpijn})} \label{a:momentirandom}

The identities at issue are proved by induction on $n$. They hold true for $n = 1$ in view of the following remarks. As to (\ref{eq:MMjn2}), it suffices to observe that $\zeta_{1, 1} \equiv 1$, $\psib_{1, 1}(\ub) = \ub$ and to exploit (\ref{eq:normalizations})-(\ref{eq:covariancematrix}). Identity (\ref{eq:MMjn3}) holds thanks to $\eta_{1, 1} \equiv 1$, $\psib_{1, 1}(\ub) = \ub$ and the definition of $l_3(\ub)$. As far as (\ref{eq:psijnpijn}) is concerned, it is enough to observe that $\pi_{1, 1} \equiv 1$ and $\psib_{1, 1}(\ub) = \ub$. When $n \geq 2$, one has to verify the identities
$$
\int_{(0, 2\pi)^{n - 1}} \Big{[} \Big{(} \mathrm{B}(\ub) \mathrm{O}_{j, n}^{\ast}(\treen, \boldsymbol{\varphi}, \boldsymbol{\theta})\mathbf{e}_3 \cdot \mathbf{e}_s\Big{)}^2 - \frac{1}{3} \Big{]} u_{(0, 2\pi)}^{\otimes_{n - 1}}(\ud \boldsymbol{\theta}) = \Big{(}(\ub \cdot \mathbf{e}_s)^{2} - \frac{1}{3}\Big{)} \zeta_{j, n}^{\ast}(\treen, \boldsymbol{\varphi})\ ,
$$
\begin{gather}
\int_{(0, 2\pi)^{n - 1}} \intethree \Big{[}\Big{(} \mathrm{B}(\ub) \mathrm{O}_{j, n}^{\ast}(\treen, \boldsymbol{\varphi}, \boldsymbol{\theta})\mathbf{e}_3 \cdot \vb\Big{)}^3 - \frac{3}{5}|\vb|^2 \Big{(} \mathrm{B}(\ub) \mathrm{O}_{j, n}^{\ast}(\treen, \boldsymbol{\varphi}, \boldsymbol{\theta})\mathbf{e}_3 \cdot \vb\Big{)} \Big{]} \nonumber \\
u_{(0, 2\pi)}^{\otimes_{n - 1}}(\ud \boldsymbol{\theta}) \mu_0(\ud \vb) = l_3(\ub) \eta_{j, n}^{\ast}(\treen, \boldsymbol{\varphi})\ , \nonumber
\end{gather}
$$
\int_{(0, 2\pi)^{n - 1}} \Big{(} \mathrm{B}(\ub) \mathrm{O}_{j, n}^{\ast}(\treen, \boldsymbol{\varphi}, \boldsymbol{\theta})\mathbf{e}_3 \cdot \mathbf{e}_s\Big{)} u_{(0, 2\pi)}^{\otimes_{n - 1}}(\ud \boldsymbol{\theta}) =  (\ub \cdot \mathbf{e}_s) \pi_{j, n}^{\ast}(\treen, \boldsymbol{\varphi})
$$
for every $s = 1, 2, 3$, $\treen$ in $\mathbb{T}(n)$, $\boldsymbol{\varphi}$ in $[0, \pi]^{n-1}$, $\ub$ in $S^2$ and every choice of $\mathrm{B}$ as in (\ref{eq:psijn}). After recalling the definition of the $k$-th Legendre polynomial $P_k$, all the above equalities can be deduced from the common formula
\begin{equation}\label{eq:recursionlegendre}
\int_{(0, 2\pi)^{n - 1}} P_k\Big{(} \mathrm{B}(\ub) \mathrm{O}_{j, n}^{\ast}(\treen, \boldsymbol{\varphi}, \boldsymbol{\theta})\mathbf{e}_3 \cdot \xib\Big{)}u_{(0, 2\pi)}^{\otimes_{n - 1}}(\ud \boldsymbol{\theta}) = P_k(\ub \cdot \xib)
f_{j,n}^{(k)}(\treen, \boldsymbol{\varphi}) \ .
\end{equation}
Here, $\xib$ denotes any unit vector while, for any $k$ in $\mathbb{N}$, $f_{1,1}^{(k)}(\tree_1, \emptyset) \equiv 1$ and
$$
f_{j, n}^{(k)}(\treen, \boldsymbol{\varphi}) := \left\{ \begin{array}{ll}
f_{j, n_l}^{(k)}(\mathfrak{t}_{n}^{l}, \boldsymbol{\varphi}^l) P_k(\cos\varphi_{n-1}) & \text{for} \ j = 1, \dots, n_l \\
f_{j - n_l, n_r}^{(k)}(\mathfrak{t}_{n}^{r}, \boldsymbol{\varphi}^r) P_k(\sin\varphi_{n-1}) & \text{for} \ j = n_l + 1, \dots, n\ .
\end{array} \right.
$$
It is worth noting that $f_{j, n}^{(1)} = \pi_{j, n}^{\ast}$, $f_{j, n}^{(2)} = \zeta_{j, n}^{\ast}$ and $f_{j, n}^{(3)} = \eta_{j, n}^{\ast}$. Now, in view of the same argument used in Subsection \ref{sect:proofrepresentation} to verify that (\ref{eq:FourierCtnast}) and (\ref{eq:FourierCtn}) are equal, one gets
$$
\int_{(0, 2\pi)^{n - 1}} \left[\mathrm{B}(\ub) \mathrm{O}_{j, n}^{\ast}(\treen, \boldsymbol{\varphi}, \boldsymbol{\theta})\mathbf{e}_3 \cdot \xib\right]^m \ud \boldsymbol{\theta} = \int_{(0, 2\pi)^{n - 1}} \left[\qb_{j, n}(\treen, \boldsymbol{\varphi}, \boldsymbol{\theta}, \ub) \cdot \xib \right]^m \ud \boldsymbol{\theta}
$$
for any unit vector $\xib$ and $m$ in $\mathbb{N}$, which implies that the LHS of (\ref{eq:recursionlegendre}) can be written as
$\int_{(0, 2\pi)^{n - 1}} P_k(\qb_{j, n}(\treen, \boldsymbol{\varphi}, \boldsymbol{\theta}, \ub) \cdot \xib) u_{(0, 2\pi)}^{\otimes_{n - 1}}(\ud \boldsymbol{\theta})$. Taking $j$ in $\{1, \dots, n_l\}$, (\ref{eq:psijnabstract}) and the inductive hypothesis yield
\begin{eqnarray}
&& \int_{(0, 2\pi)^{n - 1}} P_k(\qb_{j, n}(\treen, \boldsymbol{\varphi}, \boldsymbol{\theta}, \ub) \cdot \xib) u_{(0, 2\pi)}^{\otimes_{n - 1}}(\ud \boldsymbol{\theta}) \nonumber \\
&=& \int_{(0, 2\pi)} \int_{(0, 2\pi)^{n_l - 1}} P_k(\qb_{j, n_l}(\treen^l, \boldsymbol{\varphi}^l, \boldsymbol{\theta}^l, \psib^l(\varphi_{n-1}, \theta_{n-1}, \ub))\cdot \xib) \nonumber \\
&& u_{(0, 2\pi)}^{\otimes_{n_l - 1}}(\ud \boldsymbol{\theta}^l) u_{(0, 2\pi)}(\ud \theta_{n-1})\nonumber \\
&=& f_{j, n_l}^{(k)}(\mathfrak{t}_{n}^{l}, \boldsymbol{\varphi}^l) \int_{(0, 2\pi)} P_k(\psib^l(\varphi_{n-1}, \theta_{n-1}, \ub) \cdot \xib) u_{(0, 2\pi)}(\ud \theta_{n-1}) \ . \nonumber
\end{eqnarray}
Then, one can write $\xib$ as $\cos \beta \sin \alpha \mathbf{a}(\ub) + \sin \beta \sin \alpha \mathbf{b}(\ub) + \cos \alpha \ub$ for a suitable $(\alpha, \beta)$ in $[0, \pi] \times [0, 2\pi)$, so that $\psib^l \cdot \xib = \sin \varphi \sin \alpha \cos (\theta - \beta) + \cos \varphi \cos \alpha$. Then, in view of the well-known addition theorem for the Legendre polynomials (see, e.g., (VII') on page 268 of \cite{san}),
$$
\int_{(0, 2\pi)} P_k(\psib^l(\varphi_{n-1}, \theta_{n-1}, \ub) \cdot \xib) u_{(0, 2\pi)}(\ud \theta_{n-1}) = P_k(\ub \cdot \xib)P_k(\cos \varphi_{n-1})
$$
which completes the proof of (\ref{eq:recursionlegendre}) for $j \leq n_l$, thanks to the definition of $f_{j, n}^{(k)}$. The proof is completed by applying, \emph{mutatis mutandis}, this very same argument to the case $j > n_l$.

\subsection{Proof of (\ref{eq:laastbound})} \label{a:laastbound}

The main aim is to find a recursive inequality -- reminiscent of (\ref{eq:bassettirecurrence}) -- for the conditional expectation
$$
\mathrm{A}_{\lambda}(\nu, \tau_{\nu}; k, s) := \et\Big{[}\Big{(}\int_{\Omega_k} \big{|}\diff^{'} \SSS_{k, s}\big{|}^2\unifSu - \lambda \sum_{j=1}^{\nu} \pi_{j, \nu}^4\Big{)}\ \big{|} \ \nu, \tau_{\nu}\Big{]}
$$
where $\lambda$ is a positive parameter. For the sake of notational simplicity, the following devices will be adopted: Omission of the asterisks appearing in (\ref{eq:pijnast}) and (\ref{eq:Ojnast}); removal of indices $(k, s)$ in $\mathrm{A}_{\lambda}(\nu, \tau_{\nu}; k, s)$ and of the subscript $k$ in $\Omega_k$ and $\mathrm{B}_k$; introduction of the symbols $\boldsymbol{\varphi}, \boldsymbol{\theta}, \overline{\boldsymbol{\varphi}}, \overline{\boldsymbol{\theta}}$ to indicate $(\varphi_1, \dots, \varphi_n), (\theta_1, \dots, \theta_n), (\varphi_1, \dots, \varphi_{n-1}), (\theta_1, \dots, \theta_{n-1})$ respectively. In this notation one can write
\begin{eqnarray}
&& \mathrm{A}_{\lambda}(n+1, \treenk) = \int_{[0, \pi]^n}\int_{(0, 2\pi)^n}\int_{\Omega} \Big{|}\diff^{'} \sum_{j = 1}^{n+1}
\pi_{j, n+1}^{2}(\treenk, \boldsymbol{\varphi}) \Big{[}3 \Big{(}\mathrm{B}(\ub) \times \nonumber \\
&\times& \mathrm{O}_{j, n+1}(\treenk, \boldsymbol{\varphi}, \boldsymbol{\theta})\mathbf{e}_3 \cdot \mathbf{e}_s\Big{)}^2 - 1 \Big{]} \Big{|}^2 \unifSu u_{(0, 2\pi)}^{\otimes_n}(\ud \boldsymbol{\theta}) \beta^{\otimes_n}(\ud \boldsymbol{\varphi}) \nonumber \\
&-& \lambda \int_{[0, \pi]^n} \left[\sum_{j = 1}^{n+1} \pi_{j, n+1}^{4}(\treenk, \boldsymbol{\varphi})\right] \beta^{\otimes_n}(\ud \boldsymbol{\varphi}) \label{eq:Fn1} \ .
\end{eqnarray}
The concept of germination explained in Subsection \ref{sect:representation} is used to express the $\pi$'s and the $\mathrm{O}$'s relative to $\treenk$ in terms of the $\pi$'s and the $\mathrm{O}$'s associated with $\treen$ according to
\begin{equation} \label{eq:bassettirecurrencepi}
\pi_{j, n+1}(\treenk, \boldsymbol{\varphi}) = \left\{ \begin{array}{ll}
\pi_{j, n}(\treen, \Sigma[\treen, k] \overline{\boldsymbol{\varphi}}) & \text{for} \ j < k \\
\pi_{k, n}(\treen, \Sigma[\treen, k] \overline{\boldsymbol{\varphi}})\cos\varphi_{h} & \text{for} \ j = k \\
\pi_{k, n}(\treen, \Sigma[\treen, k] \overline{\boldsymbol{\varphi}})\sin\varphi_{h} & \text{for} \ j = k + 1 \\
\pi_{j-1, n}(\treen, \Sigma[\treen, k] \overline{\boldsymbol{\varphi}}) & \text{for} \ j > k + 1
\end{array} \right.
\end{equation}
and
\begin{gather}
\mathrm{O}_{j, n+1}(\treenk, \boldsymbol{\varphi}, \boldsymbol{\theta}) \nonumber \\
= \left\{ \begin{array}{ll}
\mathrm{O}_{j, n}\big{(}\treen, \Sigma[\treen, k] \overline{\boldsymbol{\varphi}}, \Sigma[\treen, k] \overline{\boldsymbol{\theta}}\big{)} & \text{for} \ j < k \\
\mathrm{O}_{k, n}\big{(}\treen, \Sigma[\treen, k] \overline{\boldsymbol{\varphi}}, \Sigma[\treen, k] \overline{\boldsymbol{\theta}}\big{)} \ \mathrm{M}^l(\varphi_h, \theta_h) & \text{for} \ j = k \\
\mathrm{O}_{k, n}\big{(}\treen, \Sigma[\treen, k] \overline{\boldsymbol{\varphi}}, \Sigma[\treen, k] \overline{\boldsymbol{\theta}}\big{)} \ \mathrm{M}^r(\varphi_h, \theta_h) & \text{for} \ j = k + 1 \\
\mathrm{O}_{j-1, n}\big{(}\treen, \Sigma[\treen, k] \overline{\boldsymbol{\varphi}}, \Sigma[\treen, k] \overline{\boldsymbol{\theta}}\big{)} & \text{for} \ j > k + 1
\end{array} \right. \label{eq:bassettirecurrencepsi}
\end{gather}
where $\Sigma[\treen, k] : \{1, \dots, n-1\} \rightarrow \{1, \dots, n\}$ is an injection depending on $\treen$ and $k$, while
$h$ is the element of $\{1, \dots, n\}$ excluded from the range of $\Sigma[\treen, k]$. If $k = 1$ ($n$, respectively) the first line (the last line, respectively) in (\ref{eq:bassettirecurrencepi})-(\ref{eq:bassettirecurrencepsi}) must be omitted. Therefore, the terms in (\ref{eq:Fn1}) become
\begin{gather}
\sum_{j = 1}^{n+1} \pi_{j, n+1}^{2}(\treenk, \boldsymbol{\varphi}) \Big{[}3 \Big{(}\mathrm{B}(\ub)\mathrm{O}_{j, n+1}(\treenk, \boldsymbol{\varphi}, \boldsymbol{\theta})\mathbf{e}_3 \cdot \mathbf{e}_s\Big{)}^2 - 1 \Big{]} \nonumber \\
= \sum_{j = 1}^{n} \pi_{j, n}^{2}(\treen, \Sigma[\treen, k] \overline{\boldsymbol{\varphi}}) \Big{[}3 \Big{(}\mathrm{B}(\ub)\mathrm{O}_{j, n}(\treen, \Sigma[\treen, k]\overline{\boldsymbol{\varphi}}, \Sigma[\treen, k] \overline{\boldsymbol{\theta}})\mathbf{e}_3 \cdot \mathbf{e}_s\Big{)}^2 - 1 \Big{]} \nonumber \\
- 3\pi_{k, n}^{2}(\treen, \Sigma[\treen, k] \overline{\boldsymbol{\varphi}})\Big{(}\mathrm{B}(\ub)\mathrm{O}_{k, n}(\treen, \Sigma[\treen, k] \overline{\boldsymbol{\varphi}}, \Sigma[\treen, k] \overline{\boldsymbol{\theta}})\mathbf{e}_3 \cdot \mathbf{e}_s\Big{)}^2 \nonumber \\
+ 3\cos^2\varphi_h \pi_{k, n}^{2}(\treen, \Sigma[\treen, k] \overline{\boldsymbol{\varphi}})\Big{(}\mathrm{B}(\ub)\mathrm{O}_{k, n}(\treen, \Sigma[\treen, k]  \overline{\boldsymbol{\varphi}}, \Sigma[\treen, k]  \overline{\boldsymbol{\theta}})\mathrm{M}^l(\varphi_h, \theta_h) \mathbf{e}_3 \cdot \mathbf{e}_s\Big{)}^2 \nonumber \\
+ 3\sin^2\varphi_h \pi_{k, n}^{2}(\treen, \Sigma[\treen, k]  \overline{\boldsymbol{\varphi}}) \Big{(}\mathrm{B}(\ub)\mathrm{O}_{k, n}(\treen, \Sigma[\treen, k]  \overline{\boldsymbol{\varphi}}, \Sigma[\treen, k]  \overline{\boldsymbol{\theta}})\mathrm{M}^r(\varphi_h, \theta_h) \mathbf{e}_3 \cdot \mathbf{e}_s\Big{)}^2 \label{eq:2091}
\end{gather}
and
\begin{eqnarray}
&& \sum_{j = 1}^{n+1} \pi_{j, n+1}^{4}(\treenk, \boldsymbol{\varphi}) \nonumber \\
&=& \sum_{j = 1}^{n} \pi_{j, n}^{4}(\treen, \Sigma[\treen, k] \overline{\boldsymbol{\varphi}}) - 2\cos^2\varphi_h \sin^2\varphi_h \pi_{k, n}^{4}(\treen, \Sigma[\treen, k] \overline{\boldsymbol{\varphi}}) \ . \ \ \ \ \ \label{eq:2092}
\end{eqnarray}
At this stage, apply the operator $\diff^{'}$ to the RHS of (\ref{eq:2091}) and then consider the square of the corresponding norm. With a view to this application, it is useful to introduce the symbol $\bullet$ to indicate: The product when $\diff^{'}$ is either $\mathrm{Id}$ or $\Delta_{S^2}$; the scalar product when $\diff^{'}$ is either $\gradient$ or $\gradient\Delta_{S^2}$ and, when $\diff^{'}$ is the Hessian, for any pair of symmetric 2-forms $(\omega_1, \omega_2)$, $\omega_1 \bullet \omega_2$ stands for $\sum_{ij} \omega_1(V_i, V_j) \omega_2(V_i, V_j)$ where $\{V_1, V_2\}$ is any orthonormal basis of vector fields. This procedure leads to the sum of the following three terms
\begin{eqnarray}
T1 &:=& \Big{|}\sum_{j = 1}^{n} \pi_{j, n}^{2}(\treen, \Sigma[\treen, k] \overline{\boldsymbol{\varphi}}) \diff^{'}\Big{[}3 \Big{(}\mathrm{B}(\ub)\mathrm{O}_{j, n}(\treen, \Sigma[\treen, k]\overline{\boldsymbol{\varphi}}, \Sigma[\treen, k]\overline{\boldsymbol{\theta}})\mathbf{e}_3 \cdot \mathbf{e}_s\Big{)}^2 - 1 \Big{]}\Big{|}^2 \nonumber \\
T2 &:=& 6 \Big{\{}\sum_{j = 1}^{n} \pi_{j, n}^{2}(\treen, \Sigma[\treen, k] \overline{\boldsymbol{\varphi}}) \diff^{'}\Big{[}3 \Big{(}\mathrm{B}(\ub)\mathrm{O}_{j, n}(\treen, \Sigma[\treen, k]\overline{\boldsymbol{\varphi}}, \Sigma[\treen, k] \overline{\boldsymbol{\theta}})\mathbf{e}_3 \cdot \mathbf{e}_s\Big{)}^2 - 1 \Big{]}\Big{\}} \bullet \nonumber \\
&\bullet& \pi_{k, n}^{2}(\treen, \Sigma[\treen, k] \overline{\boldsymbol{\varphi}}) \cdot \Big{\{} - \diff^{'}\Big{(}\mathrm{B}(\ub)\mathrm{O}_{k, n}(\treen, \Sigma[\treen, k] \overline{\boldsymbol{\varphi}}, \Sigma[\treen, k] \overline{\boldsymbol{\theta}})\mathbf{e}_3 \cdot \mathbf{e}_s\Big{)}^2 \nonumber \\
&+& \cos^2\varphi_h \diff^{'}\Big{(}\mathrm{B}(\ub)\mathrm{O}_{k, n}(\treen, \Sigma[\treen, k] \overline{\boldsymbol{\varphi}}, \Sigma[\treen, k] \overline{\boldsymbol{\theta}})\mathrm{M}^l(\varphi_h, \theta_h) \mathbf{e}_3 \cdot \mathbf{e}_s\Big{)}^2 \nonumber \\
&+& \sin^2\varphi_h \diff^{'} \Big{(}\mathrm{B}(\ub)\mathrm{O}_{k, n}(\treen, \Sigma[\treen, k] \overline{\boldsymbol{\varphi}}, \Sigma[\treen, k] \overline{\boldsymbol{\theta}})\mathrm{M}^r(\varphi_h, \theta_h) \mathbf{e}_3 \cdot \mathbf{e}_s\Big{)}^2\Big{\}} \nonumber \\
T3 &:=& \pi_{k, n}^{4}(\treen, \Sigma[\treen, k] \overline{\boldsymbol{\varphi}}) \cdot \Big{|} - \diff^{'}\Big{(}\mathrm{B}(\ub)\mathrm{O}_{k, n}(\treen, \Sigma[\treen, k] \overline{\boldsymbol{\varphi}}, \Sigma[\treen, k] \overline{\boldsymbol{\theta}})\mathbf{e}_3 \cdot \mathbf{e}_s\Big{)}^2 \nonumber \\
&+& \cos^2\varphi_h \diff^{'}\Big{(}\mathrm{B}(\ub)\mathrm{O}_{k, n}(\treen, \Sigma[\treen, k] \overline{\boldsymbol{\varphi}}, \Sigma[\treen, k] \overline{\boldsymbol{\theta}})\mathrm{M}^l(\varphi_h, \theta_h) \mathbf{e}_3 \cdot \mathbf{e}_s\Big{)}^2 \nonumber \\
&+& \sin^2\varphi_h \diff^{'} \Big{(}\mathrm{B}(\ub)\mathrm{O}_{k, n}(\treen, \Sigma[\treen, k] \overline{\boldsymbol{\varphi}}, \Sigma[\treen, k] \overline{\boldsymbol{\theta}})\mathrm{M}^r(\varphi_h, \theta_h) \mathbf{e}_3 \cdot \mathbf{e}_s\Big{)}^2\Big{|}^2 \ . \nonumber
\end{eqnarray}
Following  (\ref{eq:Fn1}), one has to consider the integral $\int_{[0, \pi]^n} \int_{(0, 2\pi)^n} \int_{\Omega} $ of $T1$, $T2$ and $T3$ respectively, as well as the integral $\int_{[0, \pi]^n}$ of the RHS of (\ref{eq:2092}). Then, observe that
\begin{eqnarray}
&& \int_{[0, \pi]^n} \int_{(0, 2\pi)^n}\int_{\Omega} (T1) \unifSu u_{(0, 2\pi)}^{\otimes_n}(\ud \boldsymbol{\theta}) \beta^{\otimes_n}(\ud \boldsymbol{\varphi}) \nonumber \\
&-& \lambda \int_{[0, \pi]^n} \left[\sum_{j = 1}^{n} \pi_{j, n}^{4}(\treen, \Sigma[\treen, k] \overline{\boldsymbol{\varphi}})\right]  \beta^{\otimes_n}(\ud \boldsymbol{\varphi}) = \mathrm{A}_{\lambda}(n, \treen) \ \ \ \ \ \ \label{eq:intS1}
\end{eqnarray}
holds since the measures $u_{(0, 2\pi)}^{\otimes_n}$ and $\beta^{\otimes_n}$ are exchangeable, i.e. invariant under permutation of the coordinates. As to the integral of $T2$, it is worth remarking that $T2$ depends on $(\varphi_h, \theta_h)$ only through $\cos^2\varphi_h$, $\sin^2\varphi_h$, $\mathrm{M}^{l}(\varphi_h, \theta_h)$ and $\mathrm{M}^{r}(\varphi_h, \theta_h)$. Therefore,
since $\bullet$ behaves like the scalar product, one is led to consider the integral
$$
\int_{0}^{2\pi} \Big{(}\mathrm{B}(\ub)\mathrm{O}_{k, n}(\treen, \Sigma[\treen, k] \overline{\boldsymbol{\varphi}}, \Sigma[\treen, k] \overline{\boldsymbol{\theta}})\mathrm{M}^e(\varphi_h, \theta_h) \mathbf{e}_3 \cdot \mathbf{e}_s\Big{)}^2 u_{(0, 2\pi)}(\ud \theta_h)
$$
which, after putting $\xib := (\mathrm{B}\mathrm{O}_{k, n})^{t}\ \mathbf{e}_s$, becomes
$$
\int_{0}^{2\pi} \left(\mathrm{M}^l(\varphi_h, \theta_h) \ \mathbf{e}_3 \cdot \xib\right)^2 u_{(0, 2\pi)}(\ud \theta_h) = \frac{1}{2}\sin^2\varphi_h + (1 - \frac{3}{2}\sin^2\varphi_h) \xi_{3}^{2}
$$
when $e = l$ and
$$
\int_{0}^{2\pi} \left(\mathrm{M}^r(\varphi_h, \theta_h) \ \mathbf{e}_3 \cdot \xib\right)^2 u_{(0, 2\pi)}(\ud \theta_h)
= \frac{1}{2}\cos^2\varphi_h + (1 - \frac{3}{2}\cos^2\varphi_h) \xi_{3}^{2}
$$
when $e = r$. At this stage, the identities
\begin{eqnarray}
&& - \xi_{3}^{2} + \cos^2\varphi_h \sin^2\varphi_h + \cos^2\varphi_h (1 - \frac{3}{2}\sin^2\varphi_h) \xi_{3}^{2} \nonumber \\
&+& \sin^2\varphi_h (1 - \frac{3}{2}\cos^2\varphi_h) \xi_{3}^{2} = - \cos^2\varphi_h \sin^2\varphi_h (3\xi_{3}^{2} - 1) \nonumber
\end{eqnarray}
and $\int_{0}^{\pi}  (-6 \cos^2\varphi_h \sin^2\varphi_h) \beta(\ud \varphi_h) = 3 \Lambda_b$ show that
\begin{eqnarray}
&& \sum_{k=1}^{n} \int_{[0, \pi]^n}\int_{(0, 2\pi)^n}\int_{\Omega} (T2) \unifSu u_{(0, 2\pi)}^{\otimes_n}(\ud \boldsymbol{\theta}) \beta^{\otimes_n}(\ud \boldsymbol{\varphi}) \nonumber \\
&=& 3 \Lambda_b \int_{[0, \pi]^n}\int_{(0, 2\pi)^n}\int_{\Omega} (T1) \unifSu u_{(0, 2\pi)}^{\otimes_n}(\ud \boldsymbol{\theta}) \beta^{\otimes_n}(\ud \boldsymbol{\varphi}) \ . \nonumber
\end{eqnarray}
Whence, thanks to (\ref{eq:intS1}), one gets
\begin{eqnarray}
&& \sum_{k=1}^{n} \int_{[0, \pi]^n}\int_{(0, 2\pi)^n}\int_{\Omega} (T2) \unifSu u_{(0, 2\pi)}^{\otimes_n}(\ud \boldsymbol{\theta}) \beta^{\otimes_n}(\ud \boldsymbol{\varphi}) \nonumber \\
&=& 3 \Lambda_b \mathrm{A}_{\lambda}(n, \treen) + 3 \lambda \Lambda_b \int_{[0, \pi]^{n-1}} \left[\sum_{j = 1}^{n} \pi_{j, n}^{4}(\treen, \overline{\boldsymbol{\varphi}})\right] \beta^{\otimes_{n-1}}(\ud \overline{\boldsymbol{\varphi}}) \ . \ \ \ \ \ \ \label{eq:intS2}
\end{eqnarray}
Moreover, for the term $- 2\cos^2\varphi_h \sin^2\varphi_h \pi_{k, n}^{4}(\treen, \Sigma[\treen, k] \overline{\boldsymbol{\varphi}})$ in (\ref{eq:2092}), one has
\begin{eqnarray}
&{}& - \lambda \sum_{k=1}^{n} \int_{[0, \pi]^n} (- 2\cos^2\varphi_h \sin^2\varphi_h) \pi_{k, n}^{4}(\treen, \Sigma[\treen, k] \overline{\boldsymbol{\varphi}}) \beta^{\otimes_n}(\ud \boldsymbol{\varphi}) \nonumber \\
&=& - \lambda \Lambda_b \int_{[0, \pi]^{n-1}} \left[\sum_{j = 1}^{n} \pi_{j, n}^{4}(\treen, \overline{\boldsymbol{\varphi}})\right] \beta^{\otimes_{n-1}}(\ud \overline{\boldsymbol{\varphi}}) \ . \ \ \ \ \ \ \label{eq:intpi4}
\end{eqnarray}
Combining (\ref{eq:intS1})-(\ref{eq:intpi4}) yields
\begin{gather}
\frac{1}{n} \sum_{k = 1}^{n} \mathrm{A}_{\lambda}(n + 1, \treenk) \nonumber \\
= \big{(}1 + \frac{3 \Lambda_b}{n}\big{)} \mathrm{A}_{\lambda}(n, \treen) + \frac{2 \lambda \Lambda_b}{n} \int_{[0, \pi]^n} \left[\sum_{j = 1}^{n} \pi_{j, n}^{4}(\treen, \overline{\boldsymbol{\varphi}})\right] \beta^{\otimes_{n-1}}(\ud \overline{\boldsymbol{\varphi}}) \nonumber \\
+ \frac{1}{n} \sum_{k = 1}^{n} \int_{[0, \pi]^n}\int_{(0, 2\pi)^n}\int_{\Omega}(T3) \unifSu u_{(0, 2\pi)}^{\otimes_n}(\ud \boldsymbol{\theta}) \beta^{\otimes_n}(\ud \boldsymbol{\varphi}) \ . \label{eq:bonynge}
\end{gather}
Then, it remains to consider the term containing $T3$ by showing that there exists a value $\lambda_0 = \lambda_0(\diff^{'})$ such that
\begin{eqnarray}
&{}& 2 \lambda \Lambda_b \int_{[0, \pi]^n} \left[\sum_{j = 1}^{n} \pi_{j, n}^{4}(\treen, \overline{\boldsymbol{\varphi}})\right] \beta^{\otimes_{n-1}}(\ud \overline{\boldsymbol{\varphi}}) \nonumber \\
&+& \sum_{k = 1}^{n} \int_{[0, \pi]^n}\int_{(0, 2\pi)^n}\int_{\Omega}(T3) \unifSu u_{(0, 2\pi)}^{\otimes_n}(\ud \boldsymbol{\theta}) \beta^{\otimes_n}(\ud \boldsymbol{\varphi}) \leq 0 \label{eq:braces}
\end{eqnarray}
for every $\lambda \geq \lambda_0$, $n \geq 2$, $\treen$ in $\mathbb{T}(n)$ and $s = 1, 2, 3$. In fact, the LHS can be written as
\begin{eqnarray}
&& \sum_{k = 1}^{n} \int_{[0, \pi]^n} \pi_{k, n}^{4}(\treen, \Sigma[\treen, k] \overline{\boldsymbol{\varphi}}) \Big{\{} 2 \lambda \Lambda_b + \int_{(0, 2\pi)^n}\int_{\Omega} \nonumber \\
&& \cos^2\varphi_h \sin^2\varphi_h \Big{|} \diff^{'} \big{[} \mathbf{e}_{s}^{t} \mathrm{B}(\ub) \mathrm{O}_{k, n}\big{(}\treen, \Sigma[\treen, k] \overline{\boldsymbol{\varphi}}, \Sigma[\treen, k] \overline{\boldsymbol{\theta}}\big{)} \ \mathrm{K}(\varphi_h, \theta_h) \times \nonumber \\
&\times& \mathrm{O}_{k, n}\big{(}\treen, \Sigma[\treen, k] \overline{\boldsymbol{\varphi}}, \Sigma[\treen, k] \overline{\boldsymbol{\theta}}\big{)}^t \mathrm{B}(\ub)^t \mathbf{e}_s \big{]}\Big{|}^2\unifSu u_{(0, 2\pi)}^{\otimes_n}(\ud \boldsymbol{\theta}) \Big{\}} \beta^{\otimes_n}(\ud \boldsymbol{\varphi}) \nonumber
\end{eqnarray}
where
\begin{gather}
\mathrm{K}(\varphi, \theta) := \nonumber \\
\left( \begin{array}{ccc} 2 \cos^2\theta \cos\varphi \sin\varphi & 2 \cos\theta \sin\theta \cos\varphi \sin\varphi & \cos\theta(\cos^2\varphi - \sin^2\varphi) \\
2 \cos\theta \sin\theta \cos\varphi \sin\varphi & 2 \sin^2\theta \cos\varphi \sin\varphi & \sin\theta(\cos^2\varphi - \sin^2\varphi) \\
\cos\theta(\cos^2\varphi - \sin^2\varphi) & \sin\theta(\cos^2\varphi - \sin^2\varphi) & -2 \cos\varphi \sin\varphi \\
\end{array} \right) \ . \nonumber
\end{gather}
Then, after putting $R = (r_{ij})_{ij} := \mathrm{O}_{k, n} \mathrm{K} \mathrm{O}_{k, n}^t$ and $f_{ij}^{(s)}(\ub) := \big{(}\mathbf{e}_s \cdot \mathrm{B}(\ub)\mathbf{e}_i\big{)} \big{(}\mathbf{e}_s \cdot \mathrm{B}(\ub)\mathbf{e}_j\big{)}$, one notes that
$$
\Big{|} \diff^{'} \sum_{ij} r_{ij} f_{ij}^{(s)}(\ub)\Big{|}^2 \leq 9 \sum_{ij} |r_{ij}|^2 \big{|} \diff^{'} f_{ij}^{(s)}(\ub) \big{|}^2
$$
and that $\max_{ij} |r_{ij}|^2 \leq\ \lnorm \mathrm{O}_{k, n} \rnorm^4 \cdot \lnorm \mathrm{K} \rnorm^2 \leq 36$. Whence,
\begin{eqnarray}
&& \sum_{k = 1}^{n} \int_{[0, \pi]^n} \pi_{k, n}^{4}(\treen, \Sigma[\treen, k] \overline{\boldsymbol{\varphi}}) \Big{\{} 2 \lambda \Lambda_b + \int_{(0, 2\pi)^n}\int_{\Omega} \nonumber \\
&& \cos^2\varphi_h \sin^2\varphi_h \Big{|} \diff^{'} \big{[} \mathbf{e}_{s}^{t} \mathrm{B}(\ub) \mathrm{O}_{k, n}\big{(}\treen, \Sigma[\treen, k] \overline{\boldsymbol{\varphi}}, \Sigma[\treen, k] \overline{\boldsymbol{\theta}}\big{)} \ \mathrm{K}(\varphi_h, \theta_h) \times \nonumber \\
&\times& \mathrm{O}_{k, n}\big{(}\treen, \Sigma[\treen, k] \overline{\boldsymbol{\varphi}}, \Sigma[\treen, k] \overline{\boldsymbol{\theta}}\big{)}^t \mathrm{B}(\ub)^t \mathbf{e}_s \big{]}\Big{|}^2\unifSu u_{(0, 2\pi)}^{\otimes_n}(\ud \boldsymbol{\theta}) \Big{\}} \beta^{\otimes_n}(\ud \boldsymbol{\varphi}) \nonumber \\
&\leq& \sum_{k = 1}^{n} \int_{[0, \pi]^n} \pi_{k, n}^{4}(\treen, \Sigma[\treen, k] \overline{\boldsymbol{\varphi}}) \times \nonumber \\
&\times& \Big{\{} 2 \lambda \Lambda_b + 324 \cos^2\varphi_h \sin^2\varphi_h \int_{\Omega} \sum_{ij} \big{|} \diff^{'} f_{ij}^{(s)}(\ub) \big{|}^2 \unifSu\Big{\}} \nonumber
\end{eqnarray}
and the RHS is zero when $\lambda = \lambda_0(\diff^{'}) := 81\int_{\Omega} \sum_{ij} \big{|} \diff^{'} f_{ij}^{(s)}(\ub) \big{|}^2 \unifSu$, thanks to the fact that $\int_{0}^{\pi} \cos^2\varphi_h \sin^2\varphi_h \beta(\ud \varphi_h) = -\frac{1}{2}\Lambda_b$. Therefore, in view of (\ref{eq:bonynge})-(\ref{eq:braces}),
$$
\frac{1}{n} \sum_{k= 1}^{n} \mathrm{A}_{\lambda_0}(n+1, \treenk) \leq \left(1 + \frac{3 \Lambda_b}{n}\right) \mathrm{A}_{\lambda_0}(n, \treen)
$$
holds for any $n \geq 2$. This inequality entails $a_{\lambda_0}(n+1) \leq \left(1 + \frac{3 \Lambda_b}{n}\right) a_{\lambda_0}(n)$, where $a_{\lambda}(\nu) := \et\Big{[}\mathrm{A}_{\lambda}(\nu, \tau_{\nu})\ \big{|} \ \nu \Big{]}$. At this stage, the same argument developed in \ref{a:gare} shows that
$$
a_{\lambda_0}(n) \leq \frac{2 a_{\lambda_0}(2)}{2 + 3\Lambda_b} \cdot \frac{\Gamma(n + 3\Lambda_b)}{\Gamma(n) \Gamma(2 + 3\Lambda_b)}
$$
holds for every $n \geq 2$, since $2 + 3\Lambda_b > 0$ for any choice of $b$ satisfying (\ref{eq:bsymm})-(\ref{eq:cutoff}). Whence,
\begin{eqnarray}
\et\left[\int_{\Omega_k} \big{|}\diff^{'} \SSS_{k, s}\big{|}^2\unifSu\right] \leq
a_{\lambda_0}(1) e^{-t} + a_{\lambda_0}(2) e^{-t} \frac{e^{(1 + 3\Lambda_b)t} - 1}{1 + 3\Lambda_b} + \lambda_0 e^{\Lambda_b t} \nonumber
\end{eqnarray}
is valid for every $t \geq 0$ with the proviso that $\frac{e^{(1 + 3\Lambda_b)t} - 1}{1 + 3\Lambda_b} := t$ when $\Lambda_b = -1/3$, concluding the proof.

\vspace{3cm}

\noindent \textbf{Acknowledgements} \\

\noindent The authors would like to thank professor Eric Carlen for his constructive and helpful comments, and constant encouragement. They also acknowledge the advice of professor Francesco Bonsante.

\vspace{1cm}

\footnotesize{\textsc{emanuele dolera \\
dipartimento di matematica pura e applicata ``giuseppe vitali'' \\
universit\`a degli studi di modena e reggio emilia \\
via campi 213/b, 41100 modena, italy \\
e-mail:} emanuele.dolera@unimore.it, emanuele.dolera@unipv.it}

\vspace{1cm}

\footnotesize{\textsc{eugenio regazzini \\
dipartimento di matematica ``felice casorati'' \\
universit\`a degli studi di pavia \\
via ferrata 1, 27100 pavia, italy \\
e-mail:} eugenio.regazzini@unipv.it}


\begin{thebibliography}{99}

\bibitem{ark} \textsc{Arkeryd, L.} (1981). Intermolecular forces of infinite range and the Boltzmann equation. \emph{Arch. Rational Mech. Anal}.  $\mathbf{77}$ 11-21.

\bibitem{aub} \textsc{Aubin, T.} (1982). \emph{Nonlinear Analysis on Manifolds. Monge-Amp\`{e}re Equations}. Springer, New York.

\bibitem{bl} \textsc{Bassetti, F.} and \textsc{Ladelli, L.} (2010). Self similar solutions in one-dimensional kinetic models: a probabilistic view. \emph{arXiv:1003.5527}. To appear in \emph{Ann. Appl. Probab}.

\bibitem{blm} \textsc{Bassetti, F., Ladelli, L.} and \textsc{Matthes, D.} (2010). Central limit theorem for a class of one-dimensional kinetic equations. \emph{Probab. Theory Related Fields} $\mathbf{150}$ 77-109.

\bibitem{blr} \textsc{Bassetti, F., Ladelli, L.} and \textsc{Regazzini, E.} (2008). Probabilistic study of the speed of approach to equilibrium for an inelastic Kac model. \emph{J. Statist. Phys.} $\mathbf{133}$ 683-710.

\bibitem{beu} \textsc{Beurling, A.} (1939). Sur les int\'egrales de Fourier absolument convergentes et leur application \`a une transformation fonctionnelle. In \emph{9th Congr. Math. Scandinaves} (Helsinki, 1938) 199-210. Tryekeri, Helsinki. See also \emph{The Collected Works of Arne Beurling. Vol. 2. Harmonic Analysis} (L. Carleson, P. Malliavin, V. Neuberger and J. Wermer, eds.). Birkh\"{a}user, Boston, 1989.

\bibitem{barao} \textsc{Bhattacharya, R.N.} and \textsc{Rao, R.R.}  (1976). \emph{Normal Approximation and Asymptotic Expansions}. Wiley, New York.

\bibitem{bobcer} \textsc{Bobylev, A.V.} and  \textsc{Cercignani, C.} (1999). On the rate of entropy production for the Boltzmann equation. \emph{J. Statist. Phys.} $\mathbf{94}$ 603-618.

\bibitem{bob88} \textsc{Bobylev, A.V.} (1988). The theory of the nonlinear spatially uniform Boltzmann equation for Maxwell molecules. \emph{Mathematical Physics Reviews} $\mathbf{7}$ 111-233.

\bibitem{car} \textsc{Carleman, T.} (1932). Sur la th\'{e}orie de l'equation int\'{e}grodifferentielle de Boltzmann. \emph{Acta Math}. \textbf{60} 91-146.

\bibitem{cc92} \textsc{Carlen, E.A.} and \textsc{Carvalho, M.C.} (1992). Strict entropy production bounds and stability of the rate of convergence to equilibrium for the Boltzmann equation. \emph{J. Statist. Phys.} $\mathbf{67}$ 575-608.

\bibitem{cc94} \textsc{Carlen, E.A.} and \textsc{Carvalho, M.C.} (1994). Entropy production estimates for Boltzmann equation with physically realistic collision kernels. \emph{J. Statist. Phys.} $\mathbf{74}$ 743-782.

\bibitem{ccg0} \textsc{Carlen, E. A., Carvalho, M.C.} and \textsc{Gabetta, E.} (2000). Central limit theorem for Maxwellian molecules and truncation of the Wild expansion. \emph{Comm. Pure Appl. Math}. \textbf{53} 370-397.

\bibitem{ccg5} \textsc{Carlen, E.A., Carvalho, M.C.} and \textsc{Gabetta, E.} (2005). On the relation between rates of relaxation and convergence of Wild sums for solutions of the Kac equation. \emph{J. Funct. Anal}. \textbf{220} 362-387.

\bibitem{ccl} \textsc{Carlen, E.A., Carvalho, M.C.} and \textsc{Loss, M.} (2003). Determination of the spectral gap for Kac's master equation and related stochastic evolution. \emph{Acta Math}. \textbf{191} 1-54.

\bibitem{cgr} \textsc{Carlen, E.A., Gabetta, E.} and \textsc{Regazzini, E.} (2007). On the rate of explosion for infinite energy solutions of the spatially homogeneous Boltzmann equation. \emph{J.
    Statist. Phys.} $\mathbf{129}$ 699-723.

\bibitem{cgrUNI} \textsc{Carlen, E.A., Gabetta, E.} and \textsc{Regazzini, E.} (2008). Probabilistic investigation on the explosion of solutions of the Kac equation with infinite energy initial distribution. \emph{J. Appl. Probab.} $\mathbf{45}$ 95-106.

\bibitem{cgt} \textsc{Carlen, E.A., Gabetta, E.} and \textsc{Toscani, G.} (1999). Propagation
    of smoothness and the rate of exponential convergence to equilibrium for a spatially homogeneous Maxwellian gas. \emph{Comm. Math. Phys.} $\mathbf{199}$ 521-546.

\bibitem{cgl} \textsc{Carlen, E.A., Geronimo, J.S.} and \textsc{Loss, M.} (2008). Determination of the spectral gap in the Kac model for physical momentum and energy-conserving collisions. \emph{SIAM J. Math. Anal.} $\mathbf{40}$ 327-364.

\bibitem{cl} \textsc{Carlen, E.A.} and \textsc{Lu, X.} (2003). Fast and slow convergence to
    equilibrium for Maxwellian molecules via Wild sums. \emph{J. Statist. Phys.} $\mathbf{112}$ 59-134.

\bibitem{carmo} \textsc{do Carmo, M.P.} (1992). \emph{Riemannian Geometry}. Birkh\"{a}user, Boston.

\bibitem{cerS} \textsc{Cercignani, C.} (1988). \emph{The Boltzmann Equation and its Applications}. Springer-Verlag, New York.

\bibitem{cip} \textsc{Cercignani, C., Illner, R.} and \textsc{Pulvirenti, M.} (1994). \emph{The Mathematical Theory of Dilute Gases}. Springer-Verlag, New York.

\bibitem{chte} \textsc{Chow, Y.S.} and \textsc{Teicher, H}. (1997). \emph{Probability Theory. Independence, Interchangeability, Martingales}. $3^{rd}$ ed. Springer-Verlag, New York.

\bibitem{sav} \textsc{Constantine, G.M.} and \textsc{Savits, T.H.} (1996). A multivariate Fa\`a di Bruno formula with applications. \emph{Trans. Amer. Math. Soc.} $\textbf{348}$ 503-520.

\bibitem{devil} \textsc{Desvillettes, L.} and \textsc{Villani, C.} (2005). On the trend to global equilibrium for spatially inhomogeneous kinetic systems: The Boltzmann equation. \emph{Invent. Math.} $\textbf{159}$ 245-316.

\bibitem{dophd} \textsc{Dolera, E.} (2010). Rapidity of convergence to equilibrium of the solution of the Boltzmann equation for Maxwellian molecules. Ph.D. thesis, Universit\`a degli Studi di Pavia.

\bibitem{do} \textsc{Dolera, E.} (2011). On the computation of the spectrum of the linearized Boltzmann collision operator for Maxwellian molecules. \emph{Boll. Unione Mat. Ital. (9)}  $\textbf{4}$ 47-68.

\bibitem{doLom} \textsc{Dolera, E.} (2011). Spatially homogeneous Maxwellian molecules in a neighborhood of the equilibrium. To appear in \emph{Ist. Lombardo Accad. Sci. Lett. Rend. A}.

\bibitem{doBE} \textsc{Dolera, E.} (2012). A new Berry-Esseen-like inequality for randomly weighted sums of independent random variables. \emph{Work in preparation}.

\bibitem{Dbasic} \textsc{Dolera, E.} (2012). On the solution of the homogeneous Boltzmann equation for Maxwellian molecules without cutoff. \emph{Work in preparation}.

\bibitem{dgr} \textsc{Dolera, E., Gabetta, E.} and \textsc{Regazzini, E.} (2009). Reaching the best possible rate of convergence to equilibrium for solutions of Kac's equation via central limit theorem. \emph{Ann. Appl. Probab}. $\textbf{19}$ 186-209.

\bibitem{dore} \textsc{Dolera, E.} and \textsc{Regazzini, E.} (2010). The role of the central limit theorem in discovering sharp rates of convergence to equilibrium for the solution of the Kac equation. \emph{Ann. Appl. Probab}. $\textbf{20}$ 430-461.

\bibitem{drm} \textsc{Drmota, M.} (2009). \emph{Random Trees. An interplay between Combinatorics and Probability}. Springer, Wien.

\bibitem{flr} \textsc{Fortini, S., Ladelli, L.} and \textsc{Regazzini, E.} (1996). A central limit problem for partially exchangeable random variables. \emph{Theory Probab. Appl}. $\textbf{41}$ 224-246.

\bibitem{frgr} \textsc{Fristedt, B.} and \textsc{Gray, L.} (1997). \emph{A Modern Approach to Probability Theory}. Birkh\"{a}user, Boston.

\bibitem{gr6} \textsc{Gabetta, E.} and \textsc{Regazzini, E.} (2006). Some new results for McKean's graphs with applications to Kac's equation. \emph{J. Statist. Phys.} $\textbf{125}$ 947-974.

\bibitem{gr8} \textsc{Gabetta, E.} and \textsc{Regazzini, E.} (2008). Central limit theorem for the solution of the Kac equation. \emph{Ann. Appl. Probab.} $\textbf{18}$ 2320-2336.

\bibitem{gr10} \textsc{Gabetta, E.} and \textsc{Regazzini, E.} (2010). Central limit theorems for the solutions of the Kac equation: Speed of approach to equilibrium in weak metrics. \emph{Probab. Theory Related Fields} $\textbf{146}$ 451-480.

\bibitem{gtw} \textsc{Gabetta, E.}, \textsc{Toscani, G.} and \textsc{Wennberg, B.} (1995). Metrics for propability distributions and the trend to equilibrium for solutions of the Boltzmann equation. \emph{J. Statist. Phys.}. $\textbf{81}$ 901-934.

\bibitem{gri} \textsc{Grigoryan, A.} (2009). \emph{Heat Kernel and Analysis on Manifolds}. American Mathematical Society, Providence.

\bibitem{gru} \textsc{Gr\"{u}nbaum, A.} (1972). Linearization for the Boltzmann equation. \emph{Trans. Amer. Math. Soc.} \textbf{165} 425-449.

\bibitem{hlp} \textsc{Hardy, G.H., Littlewood, J.E.} and \textsc{Polya, G.} (1952). \emph{Inequalities}. $2^{nd}$ ed. Cambridge University Press, Cambridge.

\bibitem{hil} \textsc{Hilbert, D.} (1912). Begr\"{u}ndung der kinetischen Gastheorie. \emph{Math. Ann.} \textbf{72} 562-577.

\bibitem{hir} \textsc{Hirsch, M.W.} (1976). \emph{Differential Topology}. Springer-Verlag, New York.

\bibitem{iktr} \textsc{Ikenberry, E.} and \textsc{Truesdell, C.} (1956). On the pressures and the flux of energy in a gas according to Maxwell's kinetic theory. I. \emph{J. Rational Mech. Anal.} $\textbf{5}$ 1-54.

\bibitem{kac56} \textsc{Kac, M.} (1956). Foundations of kinetic theory. \emph{Proceedings of the Third Berkeley Symposium on Mathematical Statistics and Probability}, vol. III 171-197. University of California Press, Berkeley.

\bibitem{ka} \textsc{Kallenberg, O.} (2002). \emph{Foundations of Modern Probability}. $2^{nd}$ ed. Springer-Verlag, New York.

\bibitem{max} \textsc{Maxwell, J.C.} (1867). On the dynamical theory of gases. \emph{Philos. Trans. Roy. Soc. London Ser. A} \textbf{157} 49-88.

\bibitem{mck6} \textsc{McKean H. P. Jr.} (1966). Speed of approach to equilibrium for Kac's caricature of a Maxwellian gas. \emph{Arch. Rational Mech. Anal.} $\textbf{21}$ 343-367.

\bibitem{mck7} \textsc{McKean, H. P. Jr.} (1967). An exponential formula for solving Boltzmann's equation for a Maxwellian gas. \emph{J. Combinatorial Theory} $\textbf{2}$ 358-382.

\bibitem{mer} \textsc{Merris, R.} (2003). \emph{Combinatorics}. $2^{nd}$ ed. Wiley, New York.

\bibitem{mor} \textsc{Morgenstern, D.} (1954). General existence and uniqueness proof for the spatially homogeneous solutions of the Maxwell-Boltzmann equation in the case of Maxwellian molecules. \emph{Proc. Nat. Acad. Sci. USA} \textbf{40} 719-721.

\bibitem{mor55} \textsc{Morgenstern, D.} (1955). Analytical studies related to the Maxwell-Boltzmann equation. \emph{J. Rational Mech. Anal.} \textbf{4} 533-555.

\bibitem{mouh} \textsc{Mouhot, C.} (2006). Rate of convergence to equilibrium for the spatially homogeneous Boltzmann equation with hard potentials. \emph{Comm. Math. Phys.} $\textbf{261}$ 629-672.

\bibitem{muta} \textsc{Murata, H.} and \textsc{Tanaka, H.} (1974). An inequality for certain functional of multidimensional probability distributions. \emph{Hiroshima Math. J.} $\textbf{4}$ 75-81.

\bibitem{parth} \textsc{Parthasarathy, K.R.} (1967). \emph{Probability Measures on Metric Spaces}. Academic Press, New York. Reprinted in 2005 by AMS Chelsea, Providence.

\bibitem{pe1} \textsc{Petrov, V.V.} (1995). \emph{Limit Theorems of Probability Theory. Sequences of Independent Random Variables}. The Clarendon Press, Oxford University Press, New York.

\bibitem{san} \textsc{Sansone, G.} (1959). \emph{Orthogonal Functions}. Interscience Publishers, New York. Reprinted in 1991 by Dover Publications, New York.

\bibitem{stro} \textsc{Stroock, D.W.} (2011). \emph{Probability theory. An analytic view}. $2^{nd}$ ed. Cambridge University Press, Cambridge.

\bibitem{ta} \textsc{Tanaka, H.} (1978). Probabilistic treatement of the Boltzmann equation of Maxwellian molecules. \emph{Z. Wahrsch. Verw. Gebiete} $\textbf{46}$ 67-105.

\bibitem{tovil} \textsc{Toscani, G.} and \textsc{Villani, C.} (1999). Probability metrics and uniqueness of the solution of the Boltzmann equation for a Maxwell gas. \emph{J. Statist. Phys.} $\textbf{94}$ 619-637.

\bibitem{trumu} \textsc{Truesdell, C.} and \textsc{Muncaster, R.} (1980). \emph{Fundamentals of Maxwell's Kinetic Theory of a Simple Monoatomic Gas}. Academic Press, New York.

\bibitem{vil98} \textsc{Villani, C.} (1998). Fisher information estimates for Boltzmann's collision operator. \emph{J. Math. Pures Appl.} \textbf{77} 821-837.

\bibitem{vil}  \textsc{Villani, C.} (2002). A review of mathematical topics in collisional kinetic theory. \emph{Handbook of Mathematical Fluid Dynamics}, Vol. I, 71-305. (S. Friedlander and D. Serre eds.). North-Holland, Amsterdam.

\bibitem{vilcerc} \textsc{Villani, C.} (2003). Cercignani's conjecture is sometimes true and always almost true. \emph{Comm. Math. Phys.} \textbf{234} 455-490.

\bibitem{wil} \textsc{Wild, E.} (1951). On Boltzmann's equation in kinetic theory of gases. \emph{Proc. Cambridge Philos. Soc.} \textbf{47} 602-609.

\end{thebibliography}
\end{document}